\documentclass[12pt]{article}
\usepackage{graphicx,amssymb,amsfonts,fullpage,latexsym,times}

\newcommand{\myeqref}[1]{Eq.~\ref{#1}}
\newcommand{\myfigref}[1]{Figure~\ref{#1}}
\newcommand{\mysecref}[1]{Sec.~\ref{#1}}
\newcommand{\comment}[1]{}
\let\hat\widehat
\let\tilde\widetilde
\newtheorem{lemma}{Lemma}
\newtheorem{theorem}{Theorem}
\newtheorem{remark}{Remark}
\newtheorem{example}{Example}

\usepackage{natbib}
\usepackage[english]{babel}
\usepackage{natbib}

\newcommand{\Aeps}[1]{A_{\varepsilon}}
\newcommand{\Meps}[1]{M_{\varepsilon}}

\newcommand{\bbR}{\mathbb{R}}
\newcommand{\bbK}{\mathbb{K}}
\newcommand{\bbL}{\mathbb{L}}
\newcommand{\bbM}{\mathbb{M}}
\newcommand{\bbA}{\mathbb{A}}

\begin{document}

\title{Spectral Connectivity Analysis}

%\begin{center}
%{\bf \Huge Spectral Connectivity Analysis}\\
%\end{center}
%\begin{center}
%{\bf \large Ann B. Lee and Larry Wasserman}\\
%{\bf \large Carnegie Mellon University}\\
%{\bf \large \today}
%\end{center}

\vskip 1.5in
\author{Ann B. Lee and Larry Wasserman\\ \\
Department of Statistics\\
 Carnegie Mellon University\\
 Pittsburgh, USA}

%\date{}
\maketitle

%\vspace{2cm}

%\begin{quote}
\begin{abstract}
Spectral kernel methods are techniques
for transforming data into a coordinate system that
efficiently reveals the geometric structure---
in particular, the ``connectivity''---of the data.
These methods depend on certain tuning parameters.
We analyze the dependence of the method on these tuning parameters.
We focus on one particular technique---diffusion maps---but our analysis
can be used for other methods as well.
We identify the population quantities implicitly being estimated,
we explain how these methods relate to classical kernel smoothing and
we define an appropriate risk function for analyzing the estimators.
We also show that, in some cases, fast rates of convergence
are possible even in high dimensions.
\end{abstract}
%\end{quote}

\vspace{1cm}


\vspace{1cm}

{\noindent \bf Address for correspondence:}

\noindent Larry Wasserman, Department of Statistics, Carnegie Mellon
University, 5000 Forbes Avenue, Pittsburgh, PA 15213, USA. E-mail:
larry@stat.cmu.edu \\

\vspace{1cm}

\noindent
Research supported by
NSF grant DMS-0707059 and
ONR grant N00014-08-1-0673.

\newpage

\tableofcontents
\newpage

\section{Introduction}

There has been growing interest in spectral kernel
methods such as spectral clustering~\citep{Luxburg:2007}, Laplacian
maps~\citep{BelkinNiyogi03}, Hessian maps~\citep{DonohoGrimes03}, 
and
locally linear embeddings~\citep{RoweisSaul00}. The main idea
behind these methods is that the geometry of a data set can be 
analyzed
using certain operators and their corresponding
eigenfunctions. These eigenfunctions describe the main variability of
the data and often provide an efficient parameterization of the
data.  

\begin{figure}
\begin{center}
\includegraphics[width=3in,angle=-0]{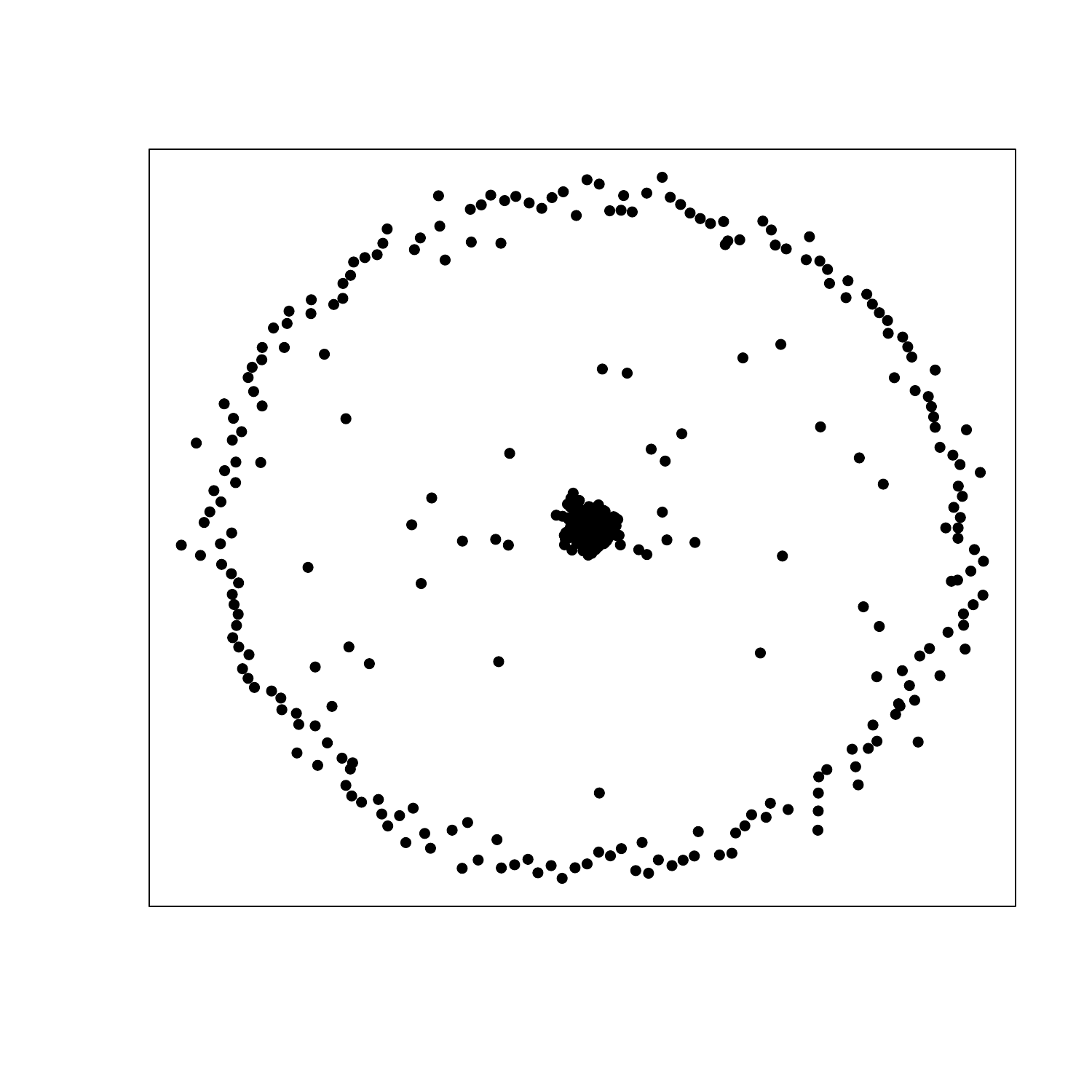}
\includegraphics[width=3in,angle=-0]{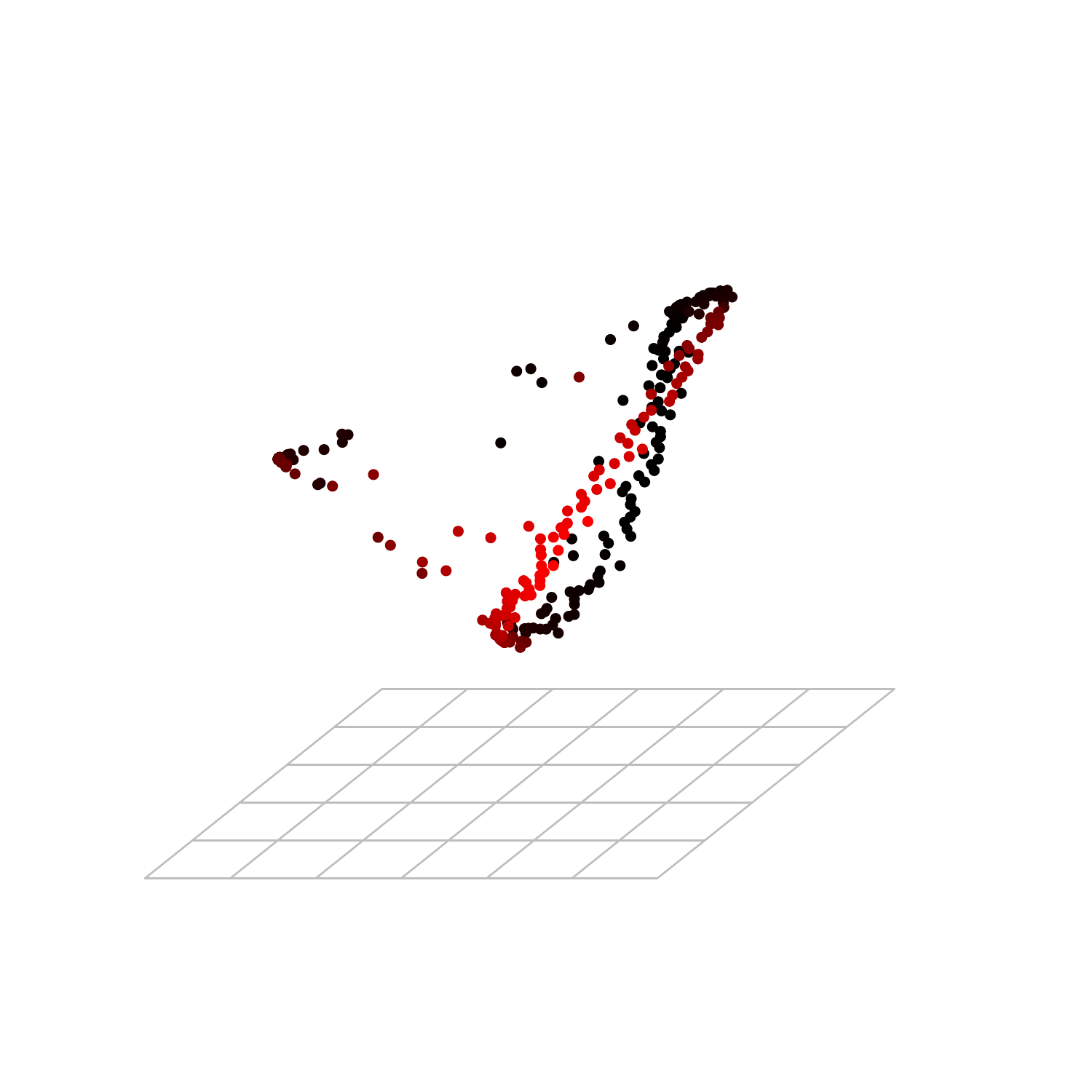}
\end{center}
\vspace{-2cm}
\caption{\footnotesize Synthetic data in original and diffusion coordinates}
\label{fig::rings}
\end{figure}

\begin{figure}
\begin{center}
\includegraphics[width=3in]{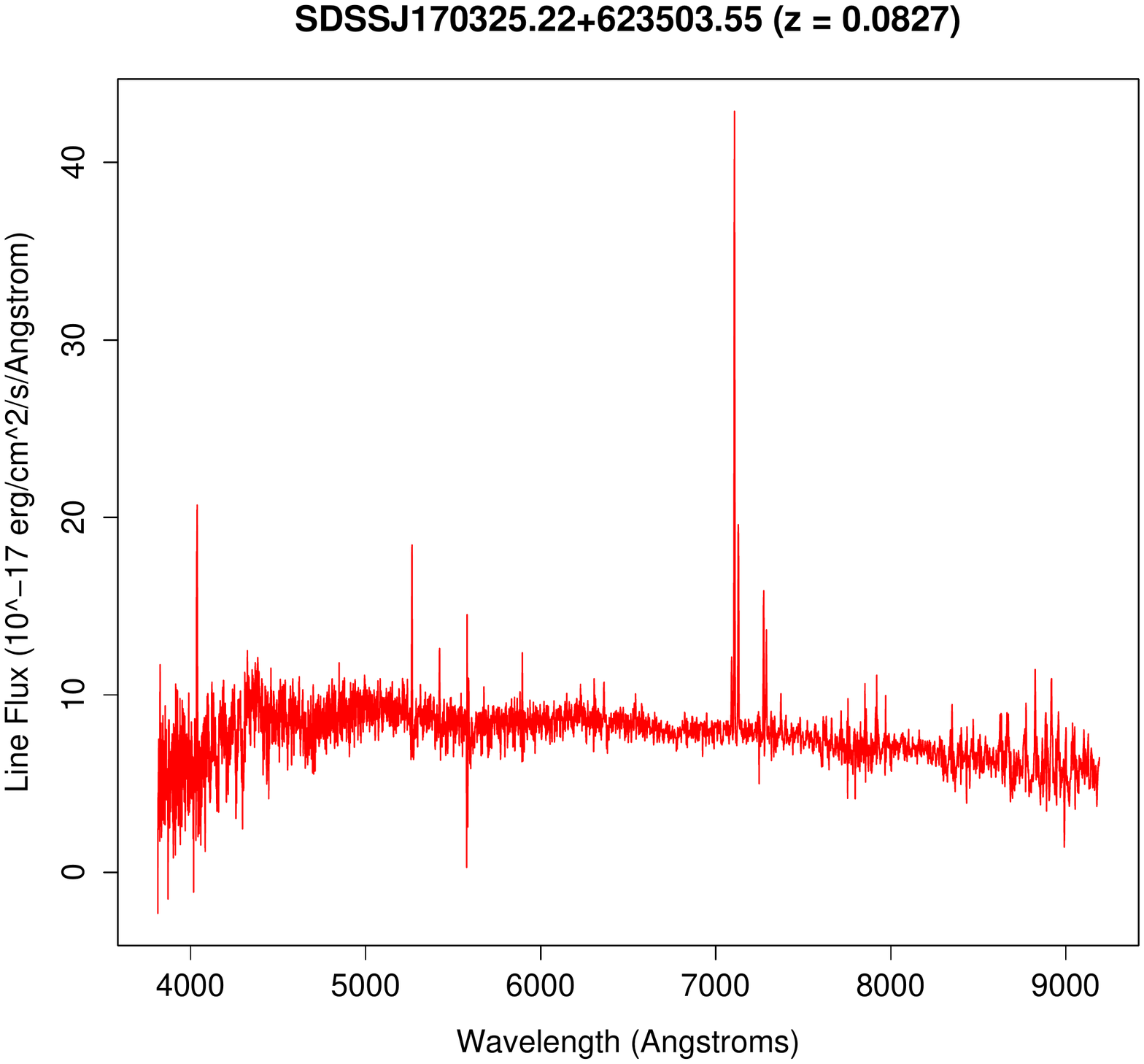}
\includegraphics[width=3in]{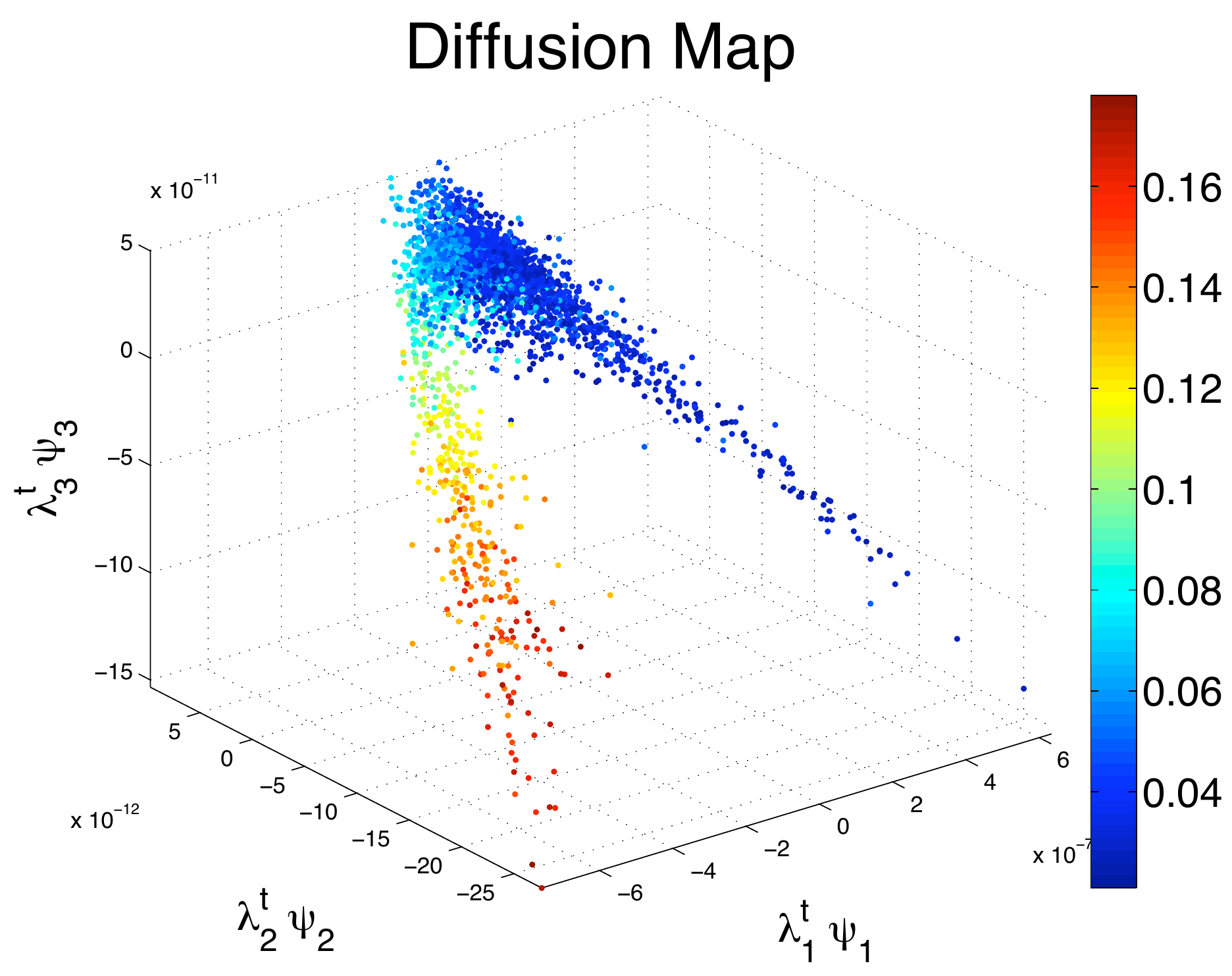}
\end{center}
\vspace{-.5cm}
\caption{\footnotesize Left: Flux versus wavelength for a typical Sloan 
Digital Sky Survey (SDSS) %~\footnote{www.sdss.org} 
galaxy spectrum. Right: Embedding of a sample of 2,793 SDSS galaxy spectra using
        the first 3 diffusion map coordinates.  The color codes for 
        redshift. The reparameterization shows a clear correspondence with variations in redshift,
         even though redshift was not taken into account in the construction. 
          (Reproduced from~\cite{Richards:EtAl:2008})
}
\label{fig::redshift}
\end{figure}

Figure \ref{fig::rings} shows an example.
The left plot is a synthetic dataset
consisting of a ring, a blob, and some uniform noise.
The right plot shows the data
in a new parameterization computed using
the methods described in this paper.
In this representation the data take the form of a cone.
The data can be much simpler to deal with in the new parameterization.
For example, a linear plane will easily separate the two clusters
in this parameterization.
In high-dimensional cases
the reparameterization leads to dimension reduction as well. 
\myfigref{fig::redshift} shows an application to astronomy data. 
Each point in the low-dimensional embedding to the right represents
 a galaxy spectrum (a function that measures 
photon flux at more than 3000 different wavelengths). The results indicate that by analyzing 
only a few dominant eigenfunctions of this highly complex data set, one can 
capture the variability in redshift (a quantity related to the distance of a galaxy from the observer) 
very well.

More generally, the central goal of spectral kernel methods can be described as follows:
\begin{quote}
{\sf Find a transformation
$Z= \Psi(X)$ such that the structure of the distribution $P_Z$ is simpler
than the structure of the distribution $P_X$
while preserving key geometric properties of $P_X$.}
\end{quote}
``Simpler'' 
can mean lower dimensional but can be intepreted much more broadly 
as we shall see.

These new methods of data reparameterization are more flexible than 
traditional
methods such as principal component analysis, clustering and
kernel smoothing.  Applications of these methods include: manifold
learning, 
\citep{Bickel:Levina:04},
fast internet web searches~\citep{Page:EtAl:98},
semi-supervised learning for regression and
classification~\citep{SzummerJaakkola01,Lafferty:Wasserman:07},
inference of arbitrarily shaped clusters, etc. The added flexibility
however comes at a price: there are tuning parameters,
such as a kernel bandwidth $\varepsilon$, and the dimension $q$ 
of the embedding
that need to be
chosen and these parameters often interact in a complicated way.
The first step in understanding these tuning parameters
is to identify the
population quantity these methods are actually estimating, then define 
an
appropriate loss function.  

We restrict our discussion
to Laplacian-based methods, though the analysis generalizes to other
spectral kernel methods. Several authors, including
\cite{Coifman:Lafon:06}, \cite{Belkin:Niyogi:05},
\cite{Hein:EtAl:05} and~\cite{Singer:06}, and 
\cite{Gine:Koltchinskii:2006} have studied the convergence
of the empirical graph Laplacian to the Laplace-Beltrami operator of a
smooth manifold as the sample size $n \rightarrow \infty$ and the
kernel bandwidth $\varepsilon \rightarrow 0$. In all these studies,
the data are assumed to lie exactly on a Riemannian submanifold
in the ambient space $\mathbb{R}^p$. Although the theoretical
framework is appealing, there
are several concerns with this approach: 
(i) distributions are rarely supported exactly on a manifold,
(ii) even in cases where the manifold assumption
is approximately reasonable, the bias-variance calculations do not 
actually take
into account stochastic variations about a perfect manifold, 
(iii) the
calculations give no information on how the parameters in the model
(such as for example the number of eigenvectors in the embedding)
depend on the sample size $n$ and the dimension $p$ when noise is
present. 

We drop the manifold assumption and instead
consider data that are drawn from some general underlying
distribution. Recently, other work has taken a similar
approach. For example, \cite{Luxburg:EtAl:2008} study the consistency
of spectral clustering. For a fixed kernel
bandwidth $\varepsilon$ and in the limit of the sample size $n
\rightarrow \infty$, the authors show that the eigenvectors of the
graph Laplacian converge to the eigenvectors of certain limit
operators. In this paper, we allow $\varepsilon$ to go to 0.
 
\newpage

The goals of the paper are to:
\begin{enumerate}
\item identify the population quantities being implicitly estimated in 
Laplacian-based spectral methods,
\item explain how these methods relate to classical kernel smoothing 
methods, 
\item find the appropriate risk and propose an approach to choosing 
the tuning parameters. 
\end{enumerate}
We show that spectral methods are closely related to classical kernel
smoothing.
This link provides
insight into the problem of parameter estimation in Laplacian
eigenmaps and spectral clustering. 
The real power in spectral methods is 
that they find structure 
in the data.
In particular, they perform
connectivity learning, with
data reduction and manifold learning
being special cases.

Laplacian-based kernel methods essentially use the same smoothing
operators as in traditional nonparametric statistics but the end goal
is not smoothing. These new kernel methods exploit the fact that the
eigenvalues and eigenvectors of local smoothing operators provide
information on the underlying geometry of the
data. 

In this paper, we describe a version of Laplacian-based spectral
methods, called {\em diffusion maps}. These techniques capture 
multiscale
structure in data by propagating local neighborhood information
through a Markov process. Spectral geometry and 
higher-order connectivity are two new concepts in data analysis. In
this paper, we show how these ideas can be incorporated into a
traditional statistical framework, and how this connection extends
classical techniques to a whole range of new applications.
We refer to the resulting method as Spectral Connectivity Analysis 
(SCA).

\section{Review of Spectral Dimension Reduction Methods}

The goal of dimensionality reduction is to find a function $\Psi$ that
maps our data $X$ from a space $\mathcal{X}$ to a new space
$\mathcal{Z}$ where their description is considered to be simpler. 
Some of the methods naturally lead to an
eigen-problem. Below we give some examples.

\subsection{Principal Component Analysis and Multidimensional 
Scaling}\label{sec:PCA}

Principal component mapping is a simple and popular method for data
reduction. In principal component analysis (PCA), one attempts to fit
a globally linear model to the data. If $S$ is a set, define
\begin{equation}
R(S)=\mathbb{E} \| X - \pi_S X\|^2
\label{eq:projection_risk}
\end{equation}
where $\pi_S X$ is the projection of $X$ onto
$S$. 
Finding $ {\rm argmin}_{S\in {\cal C}} R(S) $, where ${\cal C}$ is the
set of all $q$-dimensional planes, gives a solution that corresponds
to the first $q$ eigenvectors of the covariance matrix of $X$.

In principal coordinate analysis, the projections $\pi_S x = (z_1,
\ldots, z_q)$ on these eigenvectors are used as coordinates of the
data. This method of reparameterization is also known as classical or
metric multidimensional scaling (MDS).  The goal here is to find a
lower-dimensional embedding of the data that best preserves pairwise
Euclidean distances.  Assume that $X$ and $Y$ are covariates in
$\mathbb{R}^p$. One way to measure the discrepancy between the
original configuration and its embedding is to compute
$$
R(\Psi)=\mathbb{E} \left( d(X,Y)^2 - \|\Psi(X)-\Psi(Y)\|^2 \right)
=\int \left(d(x,y)^2 - \|\Psi(x)-\Psi(y)\|^2 \right)dP(x) dP(y) \ ,
$$
where $d(x,y)^2= \| x-y\|^2$. One can show that amongst all linear
projections $ \Psi=\pi_S$ onto $q$-dimensional subspaces of
$\mathbb{R}^p$, this quantity is minimized when the data are projected
onto their first $q$ principal components~\citep{Mardia:1980}. Thus,
there is a close connection between principal component analysis,
which returns the span of a hyperplane, and classical MDS or
``principal coordinate analysis'', which returns the new
parameterization $\Psi(x)=(z_1, \ldots, z_q)$.

The duality between PCA and MDS is also directly apparent in empirical
computations: Let $\mathbb{X}$ be an $n \times p$ data matrix, where
the rows are observations $x_i \in \mathbb{R}^p$ centered so that
$\frac{1}{n}\sum_{i=1}^n x_i =0$. The solution to PCA is then given by
the principal eigenvectors $\{v_\ell\}$ of the $p \times p$ sample
covariance matrix $\mathbb{S}=\frac{1}{n}\mathbb{X}^T\mathbb{X}$. 
The
solution to the MDS problem, on the other hand, is given by the
rescaled eigenvectors of the $n \times n$ Gram or (the positive
semi-definite) inner product matrix
$\mathbb{K}=\mathbb{X}\mathbb{X}^T$, where element
$\mathbb{K}(i,j)=\langle x_i, x_j \rangle$.
If $\{\lambda_\ell,
u_\ell\}$ are the principal eigenvalues and eigenvectors of
$\mathbb{K}$, then $\Psi(x_i)=(\lambda_1^{1/2}
u_1(i),\lambda_2^{1/2} u_2(i),\ldots)$.

\subsection{Non-Linear Methods}

For complex data, a linear model may not be adequate.
There are a large number of
non-linear data reduction methods; some of these are direct
generalizations of the PCA projection method. For example, local
PCA~\citep{Kambhatla:Leen:1997} partitions the data space into
different regions and fits a hyperplane to the data in each
partition. In principal curves~\citep{Hastie:Stuetzle:1989}, the goal
is to minimize a risk of the same form as in
Equation~\ref{eq:projection_risk}, but with $S$ representing some
class of smooth curves or surfaces.

Among non-linear extensions of PCA and MDS, we also have kernel
PCA~\citep{Scholkopf98} which applies PCA to data $\Phi(X)$ in a
higher (possibly infinite) dimensional ``feature space''. The kernel
PCA method never explicitly computes the map $\Phi$, but instead
expresses all calculations in terms of inner products $k(x,y)=\langle
\Phi(x),\Phi(y)\rangle$ where the ``kernel'' $k$ is a symmetric and
positive semi-definite function. Common choices include the Gaussian
kernel $k(x,y)=\exp \left(- \frac{\|x-y\|^2}{4 \varepsilon}\right)$ and the
polynomial kernel $k(x,y)=\langle x,y\rangle^r$, where $r=1$
corresponds to the linear case in~\mysecref{sec:PCA}. As shown
in~\cite{Bengio:EtAl:2004}, the low-dimensional embeddings $\Psi(x)$ used by eigenmap
and spectral clustering methods are equivalent to the
projections 
(of $\Phi(x)$ on the principal axes in feature space) computed by the
kernel PCA method.
 
In this paper, we study diffusion maps, a particular spectral
embedding technique.
Because of the close connection between 
MDS,
kernel PCA and eigenmap techniques, our analysis can be used for 
other
methods a well. Below we start by providing some background on
spectral dimension reduction methods from a more traditional
graph-theoretic perspective. In the next section we begin our main
analysis.

\subsubsection{Laplacian eigenmaps and other locality-preserving spectral methods}

Most spectral methods take a data-analytic rather than a
probabilistic approach to dimension reduction. The usual strategy
is to construct an adjacency graph on a given data set and then find
the optimal clustering or parameterization of the data that minimizes
some empirical locality-preserving objective function on the graph.

For a data set with $n$ observations, we define a graph $G=(V,E)$,
where the vertex set $V=\{1,\ldots,n\}$ denotes the observations, and
the edge set $E$ represents connections between pairs of 
observations.
Typically, the graph is also associated with a weight matrix
$\mathbb{K}$ that reflects the ``edge masses'' or strengths of the
edge connections. A common choice for data in Euclidean space is to
start with a Gaussian kernel: Define $\mathbb{K}(u,v) = \exp \left(-
\frac{\|x_u-x_v\|^2}{4 \varepsilon}\right)$ for all data pairs $(x_u,
x_v)$ with $(u,v) \in E$, and only include cases where the weights
$\mathbb{K}(u,v)$ are above some threshold $\delta$ in the definition
of the edge set $E$ .

Consider now a one-dimensional map $f: V \rightarrow \mathbb{R}$ that assigns a real
value to each vertex; we will later generalize to the multidimensional
case.  Many spectral embedding techniques are locality-preserving;
e.g. locally linear embedding, Laplacian eigenmaps, Hessian 
eigenmaps,
local tangent space alignment, etc. These methods are special cases of
kernel PCA, and all aim at minimizing distortions of the form
\begin{equation} 
Q(f) = \sum_{v \in V} Q_v(f) \label{eq:obj_local}
\end{equation} 
under the constraints that $Q_M(f)=1$. Typically, $Q_v(f)$ is a 
symmetric positive semi-definite quadratic form that measures 
local variations of $f$ around vertex $v$, and $Q_M(f)$ is a quadratic 
form that acts as a normalization for $f$.  
For Laplacian eigenmaps, for example, the neighborhood structure of
$G$ is described in terms of the graph Laplacian
matrix $$\bbL=\bbM-\bbK,$$ where
$\bbM=\rm{diag}(\rho_1,\ldots,\rho_n)$ is a diagonal matrix with
$\rho_u=\sum_v \mathbb{K}(u,v)$ for the ``node mass'' or degree of
vertex $u$. The goal is to find the map $f$ that minimizes the
weighted local distortion
\begin{equation}
	Q(f) = f^T \bbL f =  \sum_{(u,v) \in E} \bbK(u,v) \left(f(u)-f(v)\right)^2  
\
\geq \ 0 ,  \label{eq:objective_function}
\end{equation}
under the constraints that
$$
Q_M(f)= f^t \bbM f = \sum_{v \in V} m_v f(v)^2=1
$$
and (to avoid the trivial solution of a constant function) $f^T \bbM
1=0$.  Minimizing the distortion in (\ref{eq:objective_function})
forces $f(u)$ and $f(v)$ to be close if $\bbK(u,v)$ is large. From
standard linear algebra it follows that the optimal embedding is given
by the eigenvector of the generalized eigenvalue problem
\begin{equation}
\bbL f = \mu \bbM f \label{eq:Laplacian_eigenproblem}
\end{equation}
with the smallest non-zero
eigenvalue.

We can easily extend the discussion to higher dimensions.
Let $f_1, \ldots, f_{q}$ be the $q$ first non-trivial eigenvectors of
(\ref{eq:Laplacian_eigenproblem}), normalized so that 
$f_{i}^T M f_j= \delta_{ij}$, 
where $\delta_{ij}$ is Kronecker's delta function. The
map $f: V \rightarrow \bbR^q$, where
\begin{equation}
 f = (f_1, \ldots, f_{q}) \  ,
 \label{eq:Laplacian_eigenmap}
\end{equation}
is the Laplacian eigenmap~\citep{BelkinNiyogi03} of
$G$ in $q$ dimensions. 
It is optimal in the sense that it provides the $q$-dimensional
embedding that minimizes
\begin{equation}
\sum_{(u,v)\in E} \bbK(u,v) \|f(u)-f(v)\|^2 = \sum_{i=1}^{q} f_i^T \bbL f_i 
\label{eq:k-dimensional_embedding}
\end{equation}
 in the subspace orthogonal to $\bbM 1$, under the constraints that  
$f_{i}^T
\bbM f_j = \delta_{ij}$ for $i,j=1,\ldots,q$.

If the data points $x_u$ lie on a Riemannian manifold $\mathcal{M}$,
and $f: \mathcal{M} \rightarrow \bbR$ is a twice differentiable
function, then the expression in \myeqref{eq:objective_function} is
the discrete analogue on graphs of $\int_{\mathcal{M}}
\|\nabla_{\mathcal{M}} f\|^2 = - \int_{\mathcal{M}}
(\triangle_\mathcal{M} f)f$, where $\nabla_{\mathcal{M}}$ and
$\triangle_{\mathcal{M}}$, respectively, are the gradient and
Laplace-Beltrami operators on the manifold. The solution of $\arg
\min_{\|f\|=1} \int_{\mathcal{M}} \|\nabla_{\mathcal{M}} f\|^2$ is
given by the eigenvectors of the Laplace-Beltrami operator
$\triangle_{\mathcal{M}}$.  To give a theoretical justification for
Laplacian-based spectral methods, several authors have derived results
for the convergence of the graph Laplacian of a point cloud to the
Laplace-Beltrami operator under the manifold assumption; see
\cite{Belkin:Niyogi:05,Coifman:Lafon:06,Singer:06,Gine:Koltchinskii:2006}.

\subsubsection{Laplacian-based methods with an explicit metric}

Diffusion mapping is an MDS technique that belongs to the family of
Laplacian-based spectral methods. The original scheme was introduced
in the thesis work by~\cite{Lafon:2004} and
in~\cite{PNAS1,PNAS2}. See also independent work
by~\cite{FoussEtAl:05} for a similar technique called Euclidean commute time (ECT)
maps. In this paper, we will describe a slightly modified version of
diffusion maps that appeared 
in~\citep{Coifman:Lafon:06,LafonLee2006}~\footnote{See http://www.stat.cmu.edu/$\sim$annlee/software.htm for example code in Matlab and R.}.

The starting point of the diffusion framework is to introduce a
distance metric that reflects the higher-order connectivity of the
data. This is effectively done by defining a diffusion process or
random walk on the data.

As before, we here describe a graph approach where the nodes of the
graph represent the observations in the data set. Assuming
non-negative weights $\bbK$ and a degree matrix $\bbM$, we define a
row-stochastic matrix $\bbA=\bbM^{-1} \bbK$. We then imagine a 
random
walk on the graph $G=(V,E)$ where $\bbA$ is the transition matrix, and
element $\bbA(u,v)$ corresponds to the probability of reaching node
$v$ from $u$ in one step. Now if $\bbA^m$ is the $m^{\rm th}$ matrix power
of $\bbA$, then element $\bbA^m(u,v)$ can be interpreted as the
probability of transition from $u$ to $v$ in $m$ steps.  By increasing
$m$, we are running the Markov chain forward in time, thereby
describing larger scale structures in the data set. Under certain
conditions on $\bbK$, the Markov chain has a unique stationary
distribution $s(v) = \rho(v)/{\sum_{u \in V} \rho(u)}$.

We define the diffusion distance between nodes $u$ and $v$ as a
weighted $L^2$ distance between the two distributions
$\bbA^m(u,\cdot)$ and $\bbA^m(v,\cdot)$,
$$ D_m(u,v)^2 = \sum_{k \in V} \frac{\left(\bbA^m(u,k)-\bbA^m(v,k)
\right)^2}{s(k)}.$$
This quantity captures the higher-order connectivity of the data at a
scale $m$ and is very robust to noise 
since it integrates 
multiple-step, multiple-path connections between points. The distance
$D_m(u,v)^2$ is small when $A^m(u,v)$ is large, or when there are 
many paths between nodes $u$ and $v$ in the graph.
 
As in multidimensional scaling, the ultimate goal is to find an
embedding of the data where Euclidean distances reflect similarities
between points. In classical MDS, one attempts to preserve the
original Euclidean distances $d^2(u,v)= \| x_u-x_v\|^2$ between
points. In diffusion maps, the goal is to approximate diffusion
distances $D_m^2(u,v)$.  One can show (see appendix) that the
optimal embedding in $q$ dimensions is given by a ``diffusion
map'' $\Psi_m$, where the coordinates of the data are the (rescaled) 
right 
eigenvectors of the Markov matrix $\bbA$. In fact, assuming the kernel
matrix $\bbK$ is positive semi-definite, we have that
$$
v \in V \ \mapsto \ \Psi_m(v) = (\lambda_1^{m} \psi_1(v), \lambda_2^{m}
\psi_2(v),\ldots, \lambda_q^{m} \psi_q(v)) \in \bbR^q  \ ,
$$ 
where $\{\psi_{\ell}\}_{\ell \geq 0}$ are the principal eigenvectors
of $\bbA$ and the eigenvalues $\lambda_0=1 \geq \lambda_1 \geq 
\ldots
0$. This solution is, up to a rescaling, the same as the solution of
Laplacian eigenmaps and spectral clustering, 
since
$$ \ \ \bbL \psi = \mu \bbM \psi  \  \  \Leftrightarrow \  \  \bbA \psi = 
\lambda \psi \ .$$
for $\lambda = 1-\mu$ and $\bbL=\bbM-\bbK$. The diffusion framework
provides a link between Laplacian-based spectral methods, MDS and kernel
PCA, and can also be generalized to multiscale
geometries~\citep{PNAS2,CoifmanMauro05}.

\begin{remark}
  The link to MDS and kernel PCA is even more explicit in the original
  version of diffusion maps~\citep{PNAS1}, which is based on the symmetric (positive
  semi-definite) kernel matrix $\widetilde{\bbA}^m = \bbM^{1/2} \bbA^m
  \bbM^{-1/2} = \mathbb{I} - \bbM^{-1/2} \bbL \bbM^{-1/2}$,
  and the metric $ D_m^2(u,v) = \widetilde{\bbA}^m(u,u) +
  \widetilde{\bbA}^m(v,v) -2 \widetilde{\bbA}^m(u,v)$ induced by this
  kernel. In classical MDS and linear PCA, the analogue is a positive
  semi-definite kernel matrix $\bbK$, where $\bbK(u,v)=\langle x_u, x_v
  \rangle$, and Euclidean distances $d^2(u,v)=\|x_u-x_v\|^2 =
  \bbK(u,u) + \bbK(v,v) -2 \bbK(u,v)$. In both cases, the data are 
parameterized by the 
  rescaled principal eigenvectors $(\lambda_1^{1/2}
\phi_1,\lambda_2^{1/2} \phi_2,\ldots)$ of the kernel matrix associated 
with the metric.
\end{remark}

\section{Diffusion Maps}

The diffusion map creates a distribution-sensitive
reparameterization. We will study the method under the assumption
that the data are drawn from an underlying distribution.
We begin by introducing a Markov chain that plays
an important role in the definition of the diffusion map.

\subsection{A Discrete-Time Markov Chain}

{\bf Definitions.}
Suppose that the data $X_1,  \ldots, X_n$
are drawn from some underlying distribution $P$
with compact support
${\cal X}\subset\mathbb{R}^d$.
We assume $P$ has a density 
$p$ with respect to Lebesgue measure $\mu$.
Let  
\begin{equation}
k_{\varepsilon}(x,y) =  
\frac{1}{(4\pi \varepsilon)^{d/2}} \exp \left(-\frac{\|x-y\|^2}{4 \varepsilon} 
\right)
\label{eq:heat_kernel}
\end{equation}
denote the Gaussian kernel\footnote{
Other kernels can be used. For simplicity, we will focus
on the Gaussian kernel which is also the Green's function of the heat equation in $\mathbb{R}^d$.}
with bandwidth $h=\sqrt{2\varepsilon}$.
We write the bandwidth in terms of $\varepsilon$ instead of $h$
because $\varepsilon$ is more 
natural for our purposes.
Consider the Markov chain with transition 
kernel
$\Omega_\varepsilon(x,\cdot)$
defined by
\begin{equation}
\Omega_\varepsilon(x,A) = \mathbb{P}(x \to A) =
\frac{\int_A k_\varepsilon(x,y) dP(y)}{\int k_\varepsilon(x,y) dP(y)} =
\frac{\int_A k_\varepsilon(x,y) dP(y)}{p_\varepsilon(x)}
\end{equation}
where $p_\varepsilon(x) = \int k_\varepsilon(x,y)dP(y)$.

Starting at $x$, this chain moves to points $y$ close to $x$, giving 
preference
to points with high density $p(y)$.
In a sense, this chain measures the connectivity of 
the sample space relative to $p$.
The stationary distribution $S_\varepsilon$ 
is given by
$$
S_\varepsilon (A) =
\frac{\int_A p_\varepsilon(x) dP(x)}{\int p_\varepsilon(x) dP(x)}
$$
and
$$
S_\varepsilon(A) \to \frac{\int_A p(x) dP(x)}{\int p(x) dP(x)}\ \ \ {\rm as}\ 
\varepsilon \to 0.
$$

Define
the densities
\begin{eqnarray*}
\omega_\varepsilon(x,y) &=&
\frac{d \Omega_\varepsilon}{d\mu}(x,y)=\frac{k_{\varepsilon}(x,y)p(y)}
{p_\epsilon(x)}\\
a_\varepsilon(x,y) &=&
\frac{d \Omega_\varepsilon}{dP}(x,y)=\frac{k_{\varepsilon}(x,y)}{p_
\epsilon(x)}.
\end{eqnarray*}
The {\em diffusion operator}
$A_\varepsilon$---which maps a function $f$ to a new function
$A_\varepsilon f$---
is defined by
\begin{equation}
\label{eq:A_eps}
A_\varepsilon f(x) = \int a_\varepsilon(x,y) f(y) dP(y) =
\frac{\int k_\varepsilon(x,y)f(y) dP(y)}{\int k_\varepsilon(x,y)dP(y)}.
\end{equation}

\begin{figure}
\begin{center}
\includegraphics{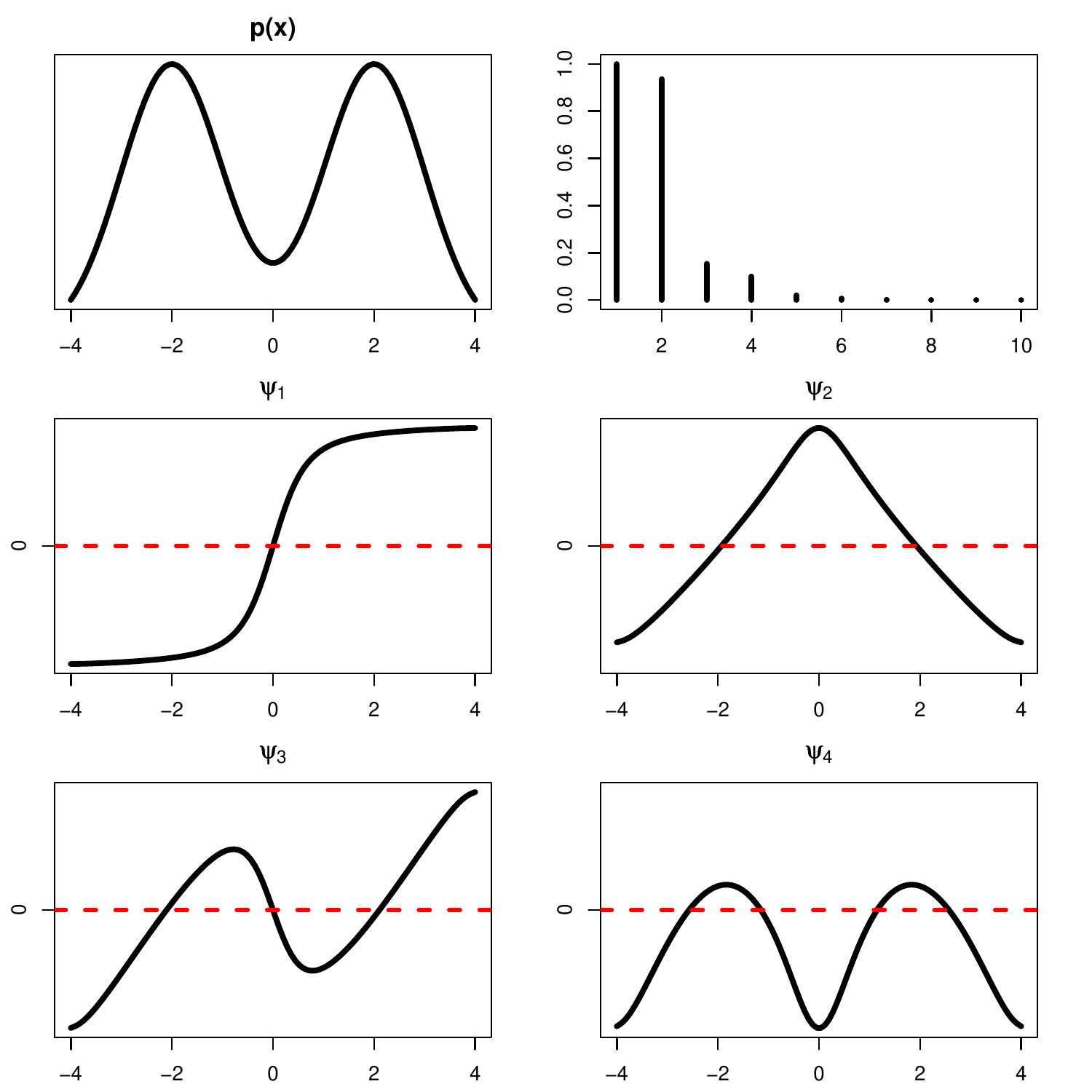}
\end{center}
\caption{\footnotesize A mixture of two Gaussians.
Density, eigenvalues, and first four eigenfunctions.}
\label{fig::two1}
\end{figure}

We normalize the eigenfunctions 
$\{\psi_{\varepsilon,0},\psi_{\varepsilon,1},\ldots \}$
of $A_\varepsilon$ by
$$
\int \psi_{\varepsilon,\ell}^2(x) s_\varepsilon(x) dP(x)  = 1, 
$$
where 
$$
s_\varepsilon(x) = \frac{p_\varepsilon(x)}{\int p_\varepsilon(y) dP(y)}
$$ 
is the density
of the stationary distribution
with respect to $P$. The first eigenfunction 
of the operator $A_\varepsilon$ is
$\psi_{\varepsilon,0}(x) = 1$ with eigenvalue 
$\lambda_{\varepsilon,0} = 1$.
In general, the eigenfunctions 
have the following interpretation: $\psi_{\varepsilon,j}$ is the smoothest function 
relative to $p$, subject to being orthogonal to $\psi_{\varepsilon,i}$, $i < j$.
The eigenfunctions form an efficient basis
for expressing smoothness, relative to $p$.
If a distribution has a few well defined clusters
then the first few eigenfunctions tend to behave
like indicator functions
(or combinations of indicator functions) for those clusters.
The rest of the eigenfunctions provide smooth basis
functions within each cluster.
These smooth functions are Fourier-like.
Indeed, the uniform distribution on the circle yields
the usual Fourier basis. 
Figure \ref{fig::two1} shows a density which is a mixture of two Gaussians.
Also shown are the eigenvalues and the first 4 eigenfunctions
which illustrate these features.

Denote the $m$-step transition measure 
by $\Omega_{\varepsilon,m}(x,\cdot)$.
Let 
$A_{\varepsilon,m}$ be the corresponding diffusion operator
which can be written as
$$
A_{\varepsilon,m} f(x) = \int a_{\varepsilon,m}(x,y) f(y) dP(y)
$$
where $a_{\varepsilon,m}(x,y)= d \Omega_{\varepsilon,m}/dP$.

Define the empirical operator $\hat{A}_\varepsilon$ by
\begin{equation}
\hat{A}_\varepsilon f(x) = 
 \frac{\sum_{i=1}^n k_\varepsilon(x,X_i) f(X_i)}{\sum_{i=1}^n k_
\varepsilon(x,X_i)}=
\int \hat{a}_\varepsilon(x,y) f(y) d\hat{P}_n(y)
\end{equation}
where
$\hat{P}_n$
denotes the empirical distribution,
$\hat{a}_\varepsilon(x,y)={k_\varepsilon(x,y)}/{\hat{p}_\varepsilon(x)}$  
and
\begin{equation}
\hat{p}_\varepsilon(x)=\int k_\varepsilon(x,y) d\hat{P}_n(y) = \frac{1}{n}
\sum_{i=1}^n k_\varepsilon(x,X_i)
\end{equation}
is the kernel density estimator.
Let $\hat{A}_{\varepsilon,m}$ be the 
corresponding $m$-step operator.
Let $\hat\psi_{\varepsilon,\ell}$ denote the eigenvectors
of the matrix $\mathbb{A}_\varepsilon$
where
$\mathbb{A}_\varepsilon(j,k)=
k_\varepsilon(X_j,X_k)/\hat{p}_\varepsilon(X_j)$.
These eigenvectors are estimates
of $\psi_\ell$ at the observed values
$X_1,\ldots, X_n$.
The function $\psi_\ell(x)$ can be estimated 
at values of $x$ not corresponding to one of the $X_i$'s
by kernel smoothing as follows.
The eigenfunction-eigenvalue equation
$\lambda_{\varepsilon,\ell} \psi_{\varepsilon,\ell} =
A_{\varepsilon} \psi_{\varepsilon,\ell}$
can be rearranged as
\begin{equation}
\psi_{\varepsilon,\ell}(x) =
\frac{A_\varepsilon \psi_{\varepsilon,\ell}}{\lambda_{\varepsilon,\ell}}=
\frac{\int k_\varepsilon(x,y) \psi_{\varepsilon,\ell}(y) dP(y)}
     {\lambda_{\varepsilon,\ell}\int k_\varepsilon(x,y) dP(y)}
\end{equation}
suggesting the estimate
\begin{equation}\label{eq::nystrom}
\hat\psi_{\varepsilon,\ell}(x)=
\frac{\sum_i k_\varepsilon(x,X_i) \hat\psi_{\varepsilon,\ell}(X_i)}
     {\hat\lambda_{\varepsilon,\ell}\sum_i k_\varepsilon(x,X_i)}
\end{equation}
which is known in the 
applied mathematics literature as the
Nystr\"{o}m approximation.

\vspace{1cm}
 
{\bf Interpretation.} 
The diffusion operators 
are averaging operators.
Equation (\ref{eq:A_eps}) arises in nonparametric regression.
If we are given regression data
$Y_i= f(X_i) + \epsilon_i$,
$i=1,\ldots, n$,
then the kernel regression estimator of $f$ is
\begin{equation}\label{eq::nadaray}
\hat{f}(x)=\frac{\frac{1}{n}\sum_{i=1}^n 
Y_i k_\varepsilon(x,X_i)}{\frac{1}{n}\sum_{i=1}^n k_\varepsilon(x,X_i)}.
\end{equation}
Replacing the sample averages in (\ref{eq::nadaray})
with their population averages yields
(\ref{eq:A_eps}).
One may then wonder: in what way spectral
smoothing is different from traditional nonparametric smoothing? 
There are at least three differences:

\begin{enumerate}
\item Estimating $A_\varepsilon$ is an unsupervised problem, 
that is, there are no responses $Y_i$.
(But see Section \ref{section::discussion} for applications to supervised 
problems.)
\item In spectral methods, smoothing is not the end goal. The main
  objective is finding structure in the data. The eigenvalues and
  eigenvectors of $\hat{A}_\varepsilon$ provide information on the
  intrinsic geometry of the data and can be used to parameterize the
  data.
\item In spectral smoothing, we are interested in
  $\hat{A}_{\varepsilon,m}$ for $m\geq 1$.  The value $m=1$ leads to a 
local analysis of
  the nearest-neighbor structure --- this part is equivalent to
  classical smoothing. Powers $m>1$, however, takes higher-order
  structure into account.
  \end{enumerate}

The concept of connectivity is new in
nonparametric statistics and is perhaps best explained in terms of
stochastic processes.  
Introduce the forward Markov operator
\begin{equation}
M_\varepsilon g(x) = \int_{\mathcal{X}} a_\varepsilon(y,x) g(y) dP(y) 
\label{eq:M_eps}
\end{equation}
and its $m$-step version
$M_{\varepsilon,m}$. 
The first eigenfunction of $M_\varepsilon$ is
$\varphi_{\varepsilon,0}(x)= s_\varepsilon(x)$,
the density of the stationary distribution. 
In general,
$$
\varphi_{\varepsilon,\ell} = s_\varepsilon(x)\psi_{\varepsilon,\ell}(x).
$$
The averaging operator $A$ and the Markov operator $M$ and (and 
hence
also the iterates $A_{\varepsilon,m}$ and $M_{\varepsilon,m}$) are
adjoint under the inner product $ \langle f,g
\rangle=\int_{\mathcal{X}} f(x)g(x) dP(x)$, i.e. $ \langle
A_\varepsilon f, g \rangle =\langle f, M_\varepsilon g\rangle$.
By comparing (\ref{eq:heat_kernel}) and the heat kernel of a 
continuous-time diffusion process (see equation (3.28)  in~
\cite{Grigoryan:06}), we 
identify the time step of the discrete system as 
$t=m\varepsilon$. 

The Markov operator $M_\varepsilon = A^{*}_\varepsilon$ maps measures into
measures.
That is,
let
$L^1_P(\mathcal{X}) = \{ g:\ g(y) \geq 0, \int g(y) dP(y) =1\}$.
Then
$g \in L^1_P(\mathcal{X})$ 
implies that
$M_{\varepsilon,m} g \in L^1_P(\mathcal{X})$.
In particular, if $\varphi$ is the probability
density at time $t=0$, then
$M_{\varepsilon,m} \varphi$ is the probability
density after $m$ steps.
The averaging operator $A_\varepsilon$ maps observables into
observables. Its action is to compute conditional
expectations. If $f \in L^{\infty}_P(\mathcal{X})$ is the test
function (observable) at $t=0$, then 
$A_{\varepsilon,m} f\in L^{\infty}_P(\mathcal{X})$ 
is the average of the function after $m$
steps, i.e. at a time comparable to $t=m \varepsilon$ for a continuous
time system.

\subsection{Continuous Time}

Under appropriate regularity conditions,
the eigenfunctions 
$\{\psi_{\varepsilon,\ell}\}$ 
converge
to a set of functions
$\{\psi_{\ell}\}$ as $\varepsilon\to 0$.
These limiting eigenfunctions
correspond to some operator.
In this section we identify this operator.
The key is to consider the Markov chain
with infinitesimal transitions.
In physics,
local infinitesimal transitions 
of a system lead to global macroscopic descriptions by integration. 
Here we use the same tools 
(infinitesimal operators, generators, exponential maps, etc) to extend 
short-time 
transitions to larger times.

Define the operator 
\begin{equation}
G_\varepsilon f(x) = 
\frac{1}{\varepsilon} 
\left(\int_{\mathcal{X}} a_\varepsilon(x,y) f(y) dP(y) - f(x) \right)  \ .
\label{eq:G_eps}
\end{equation}
Assume that the limit 
\begin{equation}
\label{eq:generator}
\mathbf{G} f =   
\lim_{\varepsilon\to 0} G_\varepsilon f =
\lim_{\varepsilon \rightarrow 0} \frac{A_\varepsilon f- f}{\varepsilon}
\end{equation}
exists for all functions $f$
in some appropriately defined space
of functions ${\cal F}$.
The operator $\mathbf{G}$ is known as
the {\em infinitesimal generator}. 
A Taylor expansion shows that
\begin{equation}
\mathbf{G} = -\triangle+\frac{\nabla p}{p}
\end{equation}
for smooth functions
where
$\triangle$ is the Laplacian and $\nabla$ is the gradient.
Indeed,
$G_\varepsilon f = -\triangle f +\frac{\nabla p}{p} + O(\varepsilon)$
which is precisely the bias for kernel regression.

\begin{remark}
In kernel regression smoothing, the term
$\nabla p/p$ is considered an undesirable extra bias, called
design bias (\cite{Fan:1993}).
In regression it is removed by using local linear smoothing
which is asymptotically equivalent
to replacing the Gaussian kernel $k_\varepsilon$
with a bias-reducing kernel $k_\varepsilon^*$.
In this case,
$\mathbf{G} = - \Delta$.
\end{remark}
 
For $\ell >0$ define
\begin{equation}
\nu_{\varepsilon,\ell}^2 = \frac{1-\lambda_{\varepsilon,\ell}}{\varepsilon}
\ \ \ {\rm and}\ \ \ 
\nu_{\ell}^2 = \lim_{\varepsilon\to 0}
\nu_{\varepsilon,\ell}^2.
\end{equation}
The eigenvalues and eigenvectors of $G_\varepsilon$ are
$-\nu_{\varepsilon,\ell}^2$ and $\psi_{\varepsilon,\ell}$
while
the eigenvalues and eigenvectors of the generator $\mathbf{G}$
are $-\nu_{\ell}^2$ and $\psi_{\ell}$.
Also, $\psi_{\varepsilon,\ell} \approx \psi_\ell$.

Let $\mathbf{A}_{t} = \lim_{\varepsilon \rightarrow 0}
A_{\varepsilon,t/\varepsilon}$. From (\ref{eq:G_eps}) and
(\ref{eq:generator}), it follows that
\begin{equation}
\mathbf{A}_{t} \equiv 
\lim_{\varepsilon \rightarrow 0} A_{\varepsilon,t/\varepsilon} = 
\lim_{\varepsilon \rightarrow 0} (I+\varepsilon G_\varepsilon)^{t/
\varepsilon} = 
\lim_{\varepsilon \rightarrow 0} (I+\varepsilon \mathbf{G})^{t/\varepsilon} 
= 
e^{\mathbf{G}t}.
\label{eq:heat_op}
\end{equation}
The family $\{\mathbf{A}_{t}\}_{t\geq0}$ defines a continuous semigroup 
of 
operators~\citep{Lasota:Mackey:94}.
The notation is summarized in Table \ref{table::notation}.

%\begin{table}
%\begin{center}
%{\caption{Summary of notation}\label{table::notation}}
%\begin{tabular}{lll} \hline
\begin{table}
{\caption{Summary of notation}\label{table::notation}}
%\hrule
\begin{center}
\begin{tabular}{lll}
\hline
Operator & Eigenfunctions & Eigenvalues\\ \hline
\noalign{\vskip 8pt}
$A_\varepsilon f(\cdot) = 
\frac{\int k_\varepsilon (\cdot,y) f(y) dP(y)}{\int k_\varepsilon (\cdot,y) 
dP(y)}$ &
$\psi_{\varepsilon,\ell}$ &
\phantom{-}$\lambda_{\varepsilon,\ell}$ \\
\noalign{\vskip 8pt}
$\mathbf{G} = \lim_{\varepsilon\to 0} \frac{A_\varepsilon -I}{\varepsilon}$ 
&
$\psi_\ell$ &
$-\nu_\ell^2 =  \lim_{\varepsilon\to 0} \frac{\lambda_{\varepsilon,\ell}-1}
{\varepsilon}$\\
\noalign{\vskip 8pt}
$\mathbf{A}_t = e^{ t \mathbf{G}} =
\sum_{\ell=0}^\infty e^{-\nu_\ell^2 t} \Pi_\ell $ &
$\psi_{\ell}$ &
\phantom{-}$e^{-t\nu_\ell^2} = \lim_{\varepsilon \rightarrow 0} \lambda_{\varepsilon,\ell}^{t/\varepsilon}$\\
\phantom{$\mathbf{A}_t$} $= \lim_{\varepsilon\to 0}A_{\epsilon,t/
\varepsilon}$ & & \\
\hline
\end{tabular}
\end{center}
%\caption{Summary of notation}
%\vspace{.1cm}
%\hrule
%\label{table::notation}
\end{table}

One of our goals is to find the bandwidth $\varepsilon$
so that $\hat{A}_{\varepsilon,t/\varepsilon}$
is a good estimate of
$\mathbf{A}_{t}$.
We show that this is a well-defined problem. Related work on
manifold learning, on the other hand, only discusses the convergence
properties of the graph Laplacian to the Laplace-Beltrami operator,
i.e. the generators of the diffusion. Estimating the generator
$\mathbf{G}$, however, does not answer questions regarding the 
optimal
choice of the number of eigenvectors, the number of groups in spectral
clustering etc.

We can express the diffusion in terms of its eigenfunctions.
Mercer's theorem gives the biorthogonal decomposition
\begin{eqnarray}
a_\varepsilon(x,y) &=& 
\sum_{\ell \geq 0} 
\lambda_{\varepsilon,\ell}  
\psi_{\varepsilon,\ell}(x) \varphi_{\varepsilon,\ell}(y) , 
\label{eq:approx_heatkernel} \\
a_{\varepsilon,t/\varepsilon}(x,y) 
&=& 
\sum_{\ell \geq 0} \lambda_{\varepsilon,\ell}^{t/\varepsilon}  
\psi_{\varepsilon,\ell}(x) \varphi_{\varepsilon,\ell}(y) 
\end{eqnarray}
where  $\psi_{\varepsilon,\ell}$ are the 
eigenvectors of $A_\varepsilon$, and 
$\varphi_{\varepsilon,\ell}$ 
are the eigenvectors of its adjoint $M_\varepsilon$. 
The details are given in Appendix~\ref{appendix:diffusion}. 
From (\ref{eq:G_eps}), it follows that 
the eigenvalues $\lambda_{\varepsilon,\ell}=1- \varepsilon
\nu_{\varepsilon,\ell}^2$. 
The averaging operator 
$A_\varepsilon$ and its generator $G_\varepsilon$ have the same 
eigenvectors.
Inserting (\ref{eq:approx_heatkernel}) into 
(\ref{eq:A_eps}) 
and recalling
that $\varphi_{\varepsilon,\ell}(x) = s_\varepsilon(x) \psi_{\varepsilon,\ell}(x)$,
gives
\begin{eqnarray*}
 A_\varepsilon f(x) &=& 
\sum_{\ell \geq 0} \lambda_{\varepsilon,\ell}  \psi_{\varepsilon,\ell}(x) 
\int_{\mathcal{X}} \varphi_{\varepsilon,\ell}(y)  f(y) dP(y)\\
&=&
\sum_{\ell \geq 0} \lambda_{\varepsilon,\ell}  \psi_{\varepsilon,\ell}(x) 
\int_{\mathcal{X}} \psi_{\varepsilon,\ell}(y)  f(y) s_\varepsilon(y) dP(y)\\
&=&
\sum_{\ell \geq 0} \lambda_{\varepsilon,\ell}  \psi_{\varepsilon,\ell}(x) 
\langle \psi_{\varepsilon,\ell}, f\rangle_\varepsilon =
\sum_{\ell \geq 0} \lambda_{\varepsilon,\ell} \ \Pi_{\varepsilon,\ell} f (x)
\end{eqnarray*}
where
$\langle f,g\rangle_\varepsilon \equiv
\int_{\mathcal{X}}   f(y)g(y) s_\varepsilon(y) dP(y)$
and
$\Pi_{\varepsilon,\ell}$ is the weighted orthogonal projector on the 
eigenspace 
spanned by $\psi_{\varepsilon,\ell}$.
Thus, 
\begin{equation}
A_{\varepsilon,t/\varepsilon} =  
\sum_{\ell \geq 0}  \lambda_{\varepsilon,\ell}^{t/\varepsilon} 
\Pi_{\varepsilon,\ell}.
\end{equation}
Similarly, assuming the limit in 
(\ref{eq:heat_op}) exists,
\begin{equation}
\mathbf{A}_{t} =  \sum_{\ell \geq 0}  e^{-\nu_{\ell}^2t} \Pi_{\ell} 
\label{eq:At_proj}
\end{equation}
where  $\Pi_{\ell}$ is the weighted orthogonal projector on the 
eigenspace corresponding to the 
eigenfunction $\psi_{\ell}$ of $\mathbf{G}$. 
Weyl's theorem (\cite{Stewart:1991}) gives
\begin{eqnarray}
\sup_\ell  |   e^{-\nu_{\ell}^2t} -\lambda_{\varepsilon,\ell}^{t/\varepsilon} | 
\leq \|A_{\varepsilon,t/\varepsilon}-e^{\mathbf{G}t}\|= t\varepsilon + 
O(\varepsilon^2)   ,\\
\lim_{\varepsilon \rightarrow 0} \lambda_{\varepsilon,\ell}^{t/\varepsilon} 
=  
e^{-\nu_{\ell}^2t} , \ \ \ 
\lim_{\varepsilon \rightarrow 0}\Pi_{\varepsilon,\ell} = \Pi_{\ell} . 
\nonumber \\
\end{eqnarray}

Note that
$\{\psi_\ell\}$ is an orthonormal basis with respect to the inner product
$$
\langle f , g \rangle_\varepsilon =
\int f(x) g(x) s_\varepsilon(x) dP(x)
$$
while
$\{\varphi_\ell\}$ is an orthonormal basis with respect to the inner 
product
$$
\langle f , g \rangle_{1/\varepsilon} =
\int \frac{f(x) g(x)}{s_\varepsilon(x)} dP(x).
$$

Equation (\ref{eq:At_proj}) implies that to estimate 
the action of the limiting operator $\mathbf{A}_{t}$ at 
a given time $t>0$, we need the dominant eigenvalues and 
eigenvectors of the generator $\mathbf{G}$. 
Finally, we also define the limiting transition density
\begin{equation}
\mathbf{a}_t(x,y) = \lim_{\varepsilon\to 0} a_{\varepsilon,t/\varepsilon}
(x,y).
\end{equation}
As $t\to 0$, $\mathbf{a}_t(x,y)$ converges to a point mass at $x$;
as $t\to \infty$, $\mathbf{a}_t(x,y)$ converges to $p(y)$.

\begin{remark} 
There is an important difference between estimating $\mathbf{A}_t$
and $\mathbf{G}$: the diffusion operator $\mathbf{A}_t$ is a compact
operator, while the generator $\mathbf{G}$ is not even a bounded
operator. Consider, for example, the Laplacian on a circle $S^1$
(Rosenberg, 1997). The eigenfunctions of $\mathbf{G}$ and
$\mathbf{A}_t$ are here the Fourier basis functions $e^{i\ell x}$ where
$\ell=0,\pm1, \pm 2,\ldots$. The heat operator $e^{-t \Delta}$ is a
compact operator. Its eigenvalues are $e^{-\ell^2 t} (t>0)$ which are
clearly bounded above and go to zero. The Laplace-Beltrami operator
$\Delta$, on the other hand, has eigenvalues $n^2$ which are
unbounded.  \label{remark:A_vs_G}
\end{remark}

We will consider some examples in Section \ref{section::examples}
but let us first illustrate the definitions for a one-dimensional distribution with
multiscale structure.

\begin{figure}
\vspace{-1cm}
\begin{center}
\includegraphics[width=3in]{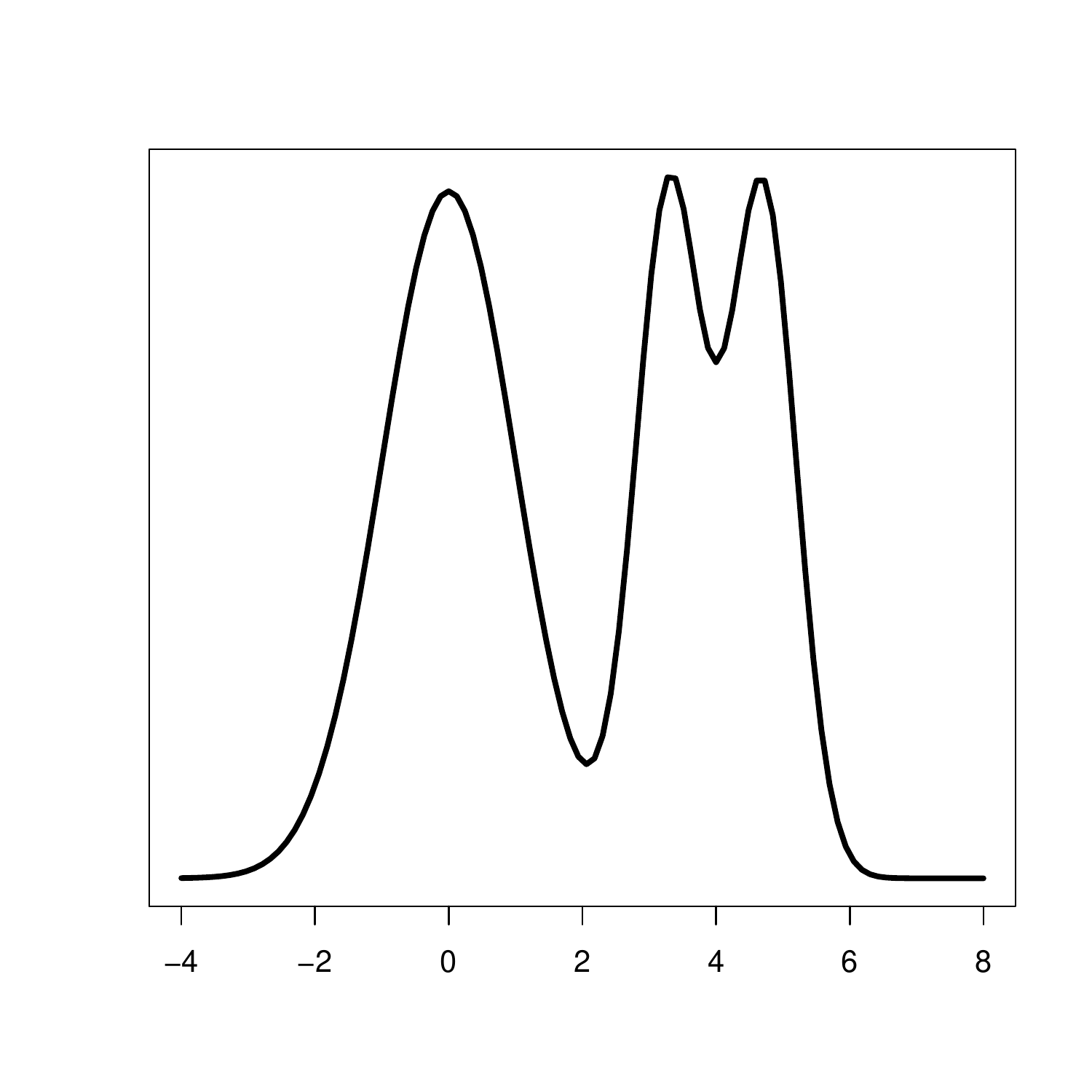}
\end{center}
\vspace{-1cm}
\caption{\footnotesize The density $p$ for Example \ref{example::mix}.}
\label{fig::mixdens}
\end{figure}

\begin{figure}
\begin{center}
\includegraphics[width=4in]{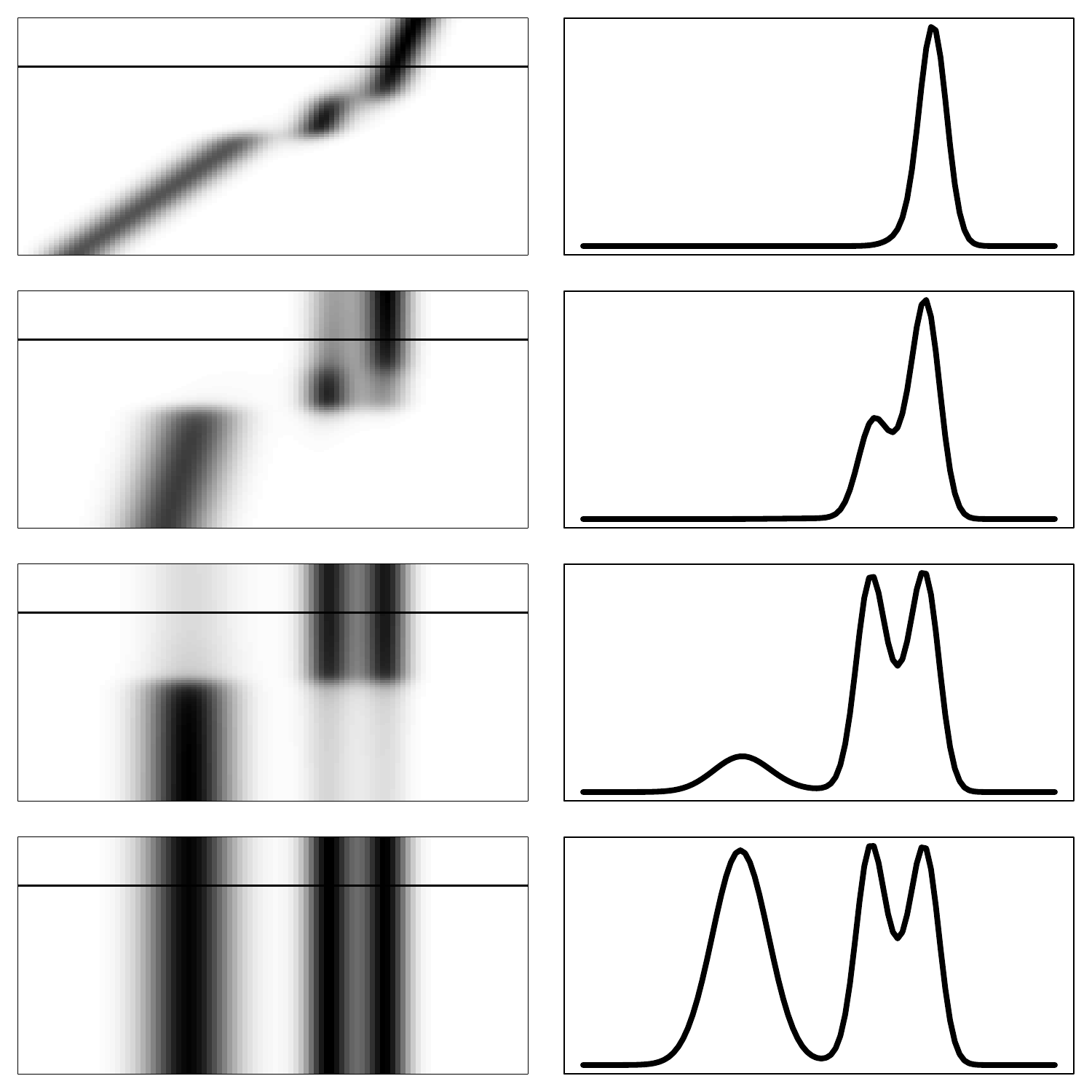}
\end{center}
\caption{\footnotesize Example \ref{example::mix}.
Left column: $\omega_t(x,y)$ for
$t = .1, 1, 10 , 1000$.
Right column, $\omega_t(x,y)$ for a fixed $x$.}
\label{fig::m1}
\end{figure}

\begin{example}
\label{example::mix}
Suppose that 
$P$ is a mixture of three Gaussians.
Figure \ref{fig::mixdens} shows the density $p$.
The left column of Figure \ref{fig::m1}
shows $\Omega_t$ for increasing $t$.
The right column shows 
a fixed row of $\Omega_t$, namely $\omega_t(x,\cdot)$ for a fixed $x$
indicated by the horizontal line.
The density
$\omega_t(x,\cdot)$ starts out concentrated near $x$. 
As $t$ increases, it begins to spread out. It becomes bimodal at $t=1$
indicating 
that the two closer clusters have merged. Eventually, the density has three modes (indicating a single cluster)
at $t=10$, 
and then resembles $p$ when $t=1000$
since  $\omega_t(x,\cdot)\to p^2(\cdot)/\int p^2(u)du$ as $t \rightarrow \infty$.
%Figure \ref{fig::mixdenseigen} shows the first 6 eigenfunctions.
\end{example}

\subsection{Comparing $\varepsilon$ and $t$}

The parameters $t$ and $\varepsilon$ 
are both related to smoothing
but they are quite different.
The parameter $t$ is part of the population quantity
being estimated and controls the scale of the analysis. Hence, the choice 
of $t$ is often determined by the problem at hand. The parameter $\varepsilon$ is a smoothing parameter for 
estimating the population quantity from data. As $n \rightarrow \infty$, we let $\varepsilon_n \rightarrow 0$ 
for more accurate estimates. The following two examples illustrate the differences of smoothing in data 
when using $\varepsilon$ or $t$.

\begin{figure}
\begin{center}
\includegraphics[width=3in]{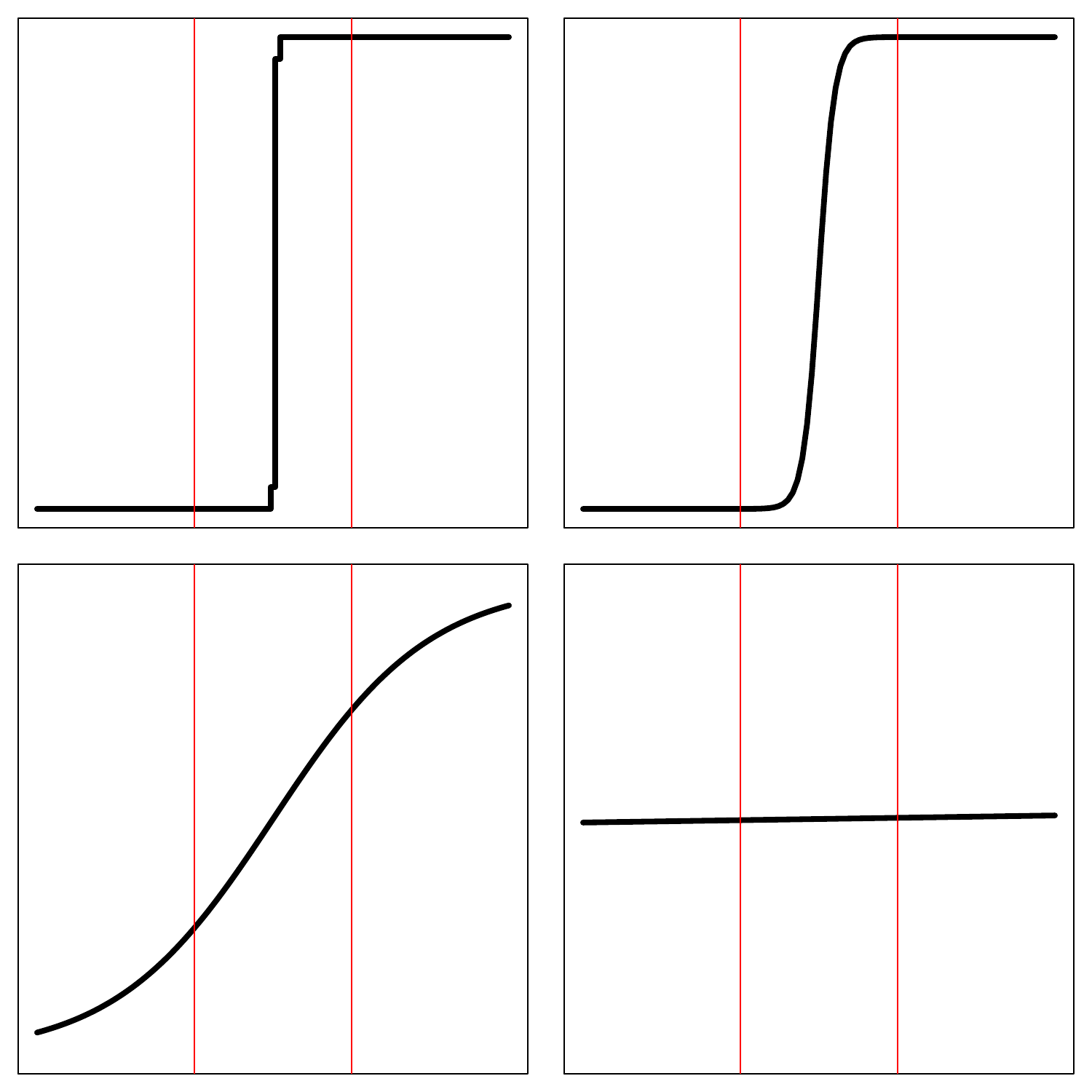}
\end{center}
\caption{\footnotesize $A_\varepsilon f$ for increasing values of $\varepsilon$.
The function $\mathbf{A}_t f$ corresponds to the top left plot
and does not change as $t$ changes.}
\label{fig::twopoint}
\end{figure}

\begin{example}
Consider a fixed test function $f$.
Define
$$
{\cal A} =
\Biggl\{ g=A_\varepsilon f:\ \ 0 \leq \varepsilon \leq \infty \Biggr\}
\ \ \ {\rm and}\ \ \ 
{\cal A}^* =
\Biggl\{ g=\mathbf{A}_t f:\ \ 0 \leq t \leq \infty \Biggr\}.
$$
Let 
$$
P = \frac{1}{2}\delta_{0} + \frac{1}{2}\delta_{1}
$$
where
$\delta_{c}$
denotes a point mass distribution at
$c$.
If $f$ is any continuous functions then
both
$A_\varepsilon$ and
$\mathbf{A}_t$ depend only on the two values
$f(0)$
and
$f(1)$
which we will assume are distinct.

Now
$$
A_\varepsilon f (x) =
\frac{ k_\varepsilon(x,0) f(0) + k_\varepsilon(x,1) f(1)}
     { k_\varepsilon(x,0)  + k_\varepsilon(x,1)}.
$$
In particular
$$
A_0 f (x) =
\left\{
\begin{array}{ll}
f(0) & x < 1/2\\
f(1) & x > 1/2.
\end{array}
\right.
$$
and
$A_\infty f(x) = c$ for all $x$
where
$c = (f(0) + f(1))/2$.
For $0 < \varepsilon <\infty$, 
$A_\varepsilon f (x)$ is a smooth monotone function;
see Figure \ref{fig::twopoint}.

In contrast,
$$
\mathbf{A}_t f(x) = 
\left\{
\begin{array}{ll}
f(0) & x < 1/2\\
f(1) & x > 1/2
\end{array}
\right.
$$
for all values of $t$.
In other words,
$\mathbf{A}_t f(x) = A_0 f(x)$ for all $t$.
The reason is that
$\mathbf{A}_t$ has two eigenfunctions:
$\psi_0(x) =1$ and
$\psi_1(x) = I(x > 1/2) - I(x < 1/2)$
(assuming the normalization
$\int \psi^2(x) dP(x) = 1$.)
The eigenvalues are
$\lambda_0 = \lambda_1 =1$.
Hence,
$\nu_0 = \nu_1 =0$ and so
$$
\mathbf{A}_t = \Pi_0 + \Pi_1
$$
where $\Pi_0$ projects onto $\psi_0$ and 
$\Pi_1$ projects onto $\psi_1$.
The step function behavior of
$\mathbf{A}_t$ reflects the lack of
connectivity
of $P$.
\end{example}

\begin{example}
Assume that the distribution is supported along two parallel lines of 
length $\pi$ at $v=0$ and $v=1$, respectively. 
The probability measure is
$$
P = \frac{1}{2}U_0 + \frac{1}{2} U_1
$$
where $U_0$ is uniform on $\{ (0,x):\ 0 \leq x \leq \pi\}$ and
$U_1$ is uniform on $\{ (1,x):\ 0 \leq x \leq \pi\}$.
Consider a fixed 
test function $f$. We have that
\begin{eqnarray*}
A_\varepsilon f (x) &=& \int \ell_{\varepsilon}(x,y) f(y) dy \\
\mathbf{A}_t f(x) &=& \int \mathbf{\ell}_{t}(x,y) f(y) dy \ ,
\end{eqnarray*}
where the weights $\ell_{\varepsilon}(x,y) = 
\frac{k_\varepsilon(x,y)p(y)}{p_\varepsilon(x)}$ and $\ell_{t}(x,y)=\mathbf{a}_t(x,y)p(y) = 
\lim_{\varepsilon\to 0} a_{\varepsilon,t/\varepsilon}(x,y) p(y)$ \ .

Let $x=(0,0)$ and $y=(u,v)$. \myfigref{fig:ex_eps} shows how the 
weights $\ell_{\varepsilon}(x,y)$ change with the parameter $\varepsilon$. 
When $\varepsilon$ is small, $A_\varepsilon f(x)$ will only depend on the values of $f$ 
close to the origin along the line at $v=0$. However, with increasing  $\varepsilon$, 
smoothing will also involve function values further from the origin, 
including values along the parallel line at $v=1$, as indicated by the red dashed curves in the figure.

\begin{figure}
\begin{center}
\includegraphics[width=4in]{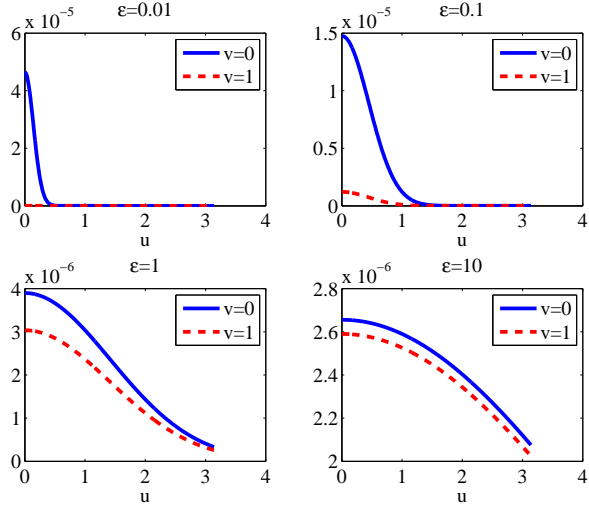}
\end{center}
\vspace{-1.5in}
\caption{\footnotesize $\ell_{\varepsilon}(x,y)$ for $x=(0,0)$, $y=(u,v)$ 
and $\varepsilon=0.01, 0.1, 1, 10$.}
\label{fig:ex_eps}
\end{figure}

In contrast, for $x=(0,0)$, $ \mathbf{A}_t f(x)$ only depends on values 
of $f$ in the same connected set as $x$, i.e. function values along the line at $v=0$, 
regardless of $t$.  \myfigref{fig:ex_t} illustrates how the weights $\ell_{t}(x,y)$ 
change as the parameter $t$ increases. Smoothing by $t$ reflects 
the {\em connectivity} of the data. In particular, there is no 
mixing of values of $f$ from disconnected sets.
\begin{figure}
\begin{center}
\includegraphics[width=4in]{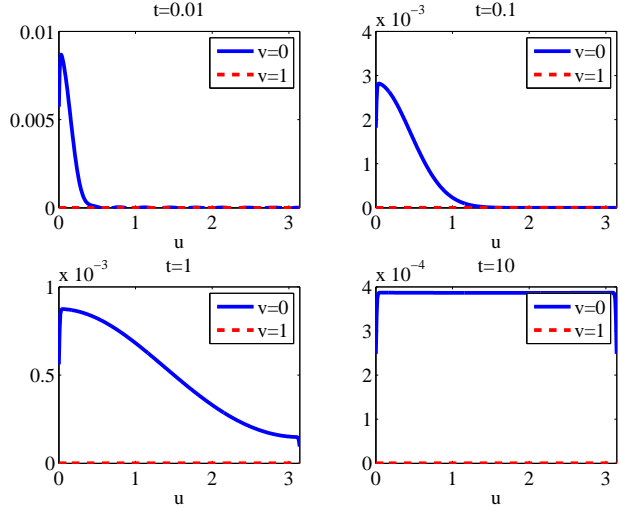}
\end{center}
\vspace{-1.5in}
\caption{\footnotesize $\ell_{t}(x,y)$ for $x=(0,0)$, $y=(u,v)$ and $t=0.01, 0.1, 1, 10$.}
\label{fig:ex_t}
\end{figure}

\end{example}

\newpage

\section{Diffusion Distance}

The diffusion distance is another quantity that captures
the underlying geometry.

\subsection{Definition}

For an $m$-step Markov chain, the diffusion distances are defined by
$$
D_{\varepsilon,m}^2(x,z) =
\int \frac{( a_{\varepsilon,m}(x,u) - a_{\varepsilon,m}(z,u))^2}{s_
\varepsilon(u)} dP(u)
$$
for $m=1,2,\ldots$.
It can be shown (see appendix) that
\begin{equation}
D_{\varepsilon,m}^2(x,z)=\sum_{\ell \geq 0} \lambda_{\varepsilon,
\ell}^{2m}
(\psi_{\varepsilon,\ell}(x) - \psi_{\varepsilon,\ell}(z))^2 .
\end{equation}
Following the same arguments as 
before
we deduce
that the corresponding population quantity is
\begin{equation}
\mathbf{D}_{t}^2(x,z)=\sum_{\ell \geq 0} e^{- 2\nu_{\ell}^2t}
( \psi_{\ell}(x) -  \psi_{\ell}(z))^2 .
\end{equation}

Now we compare diffusion distance to
two other distances that have been used recently:
geodesic distance and density distance.

\subsection{Geodesic Distance}

The geodesic distance, or the shortest path, is a very intuitive way of measuring 
the distance between two points in a set. Some manifold learning algorithms, 
such as Isomap~\citep{Tenenbaum:2000}, rely on being able to estimate the 
geodesic distance on a manifold given data in $\mathbb{R}^p$. 
The idea is to construct a graph $G$ on pairs of points at a
distance less than a given threshold $\delta$, and define a graph distance 
$$
d_G(A,B) = \min_\pi \left(\| x_0-x_1\| + \ldots + \|x_{m-1}-x_m\|\right)
$$
where $\pi=(x_0,\ldots,x_m)$ varies over all paths along the edges of $G$ 
connecting the points $A=x_0$ and $B=x_m$. Multidimensional scaling is then used 
to find a low-dimensional embedding of the data that best preserves these distances.  

Under the assumption that the data lie exactly on a smooth manifold 
$\mathcal{M}$, \cite{Bernstein:EtAl:00} have shown that the 
graph distance $d_G(A,B)$ converges to the geodesic manifold 
metric $$d_{\mathcal M}(A,B)=\inf \{\mathrm{length}(\gamma)\},$$ 
where $\gamma$ varies over the set of smooth arcs connecting $A$ and $B$ in 
$\mathcal{M}$. Beyond this ideal situation, little is known about the statistical 
properties of the graph distance. Here we show by two examples  
that the geodesic (graph) distance is {\em inconsistent} 
if the support of the distribution is not exactly on a manifold.

Consider a one-dimensional spiral in a plane:
$$
\left\{
\begin{array}{ccc}
x &=& t^a \cos(bt) \\
y &=& t^a \sin(bt)
\end{array} 
\right.
$$
where $a=0.8$ and $b=10$. The geodesic manifold distance $d_{\mathcal M}(A,B)$ 
between two reference points A 
and B with $t=\pi/2b$ and $t=5\pi/2b$, respectively, is 3.46. The corresponding Euclidean distance is $0.60$. 

\begin{figure}[htb]
\begin{center}
\includegraphics[width=4in]{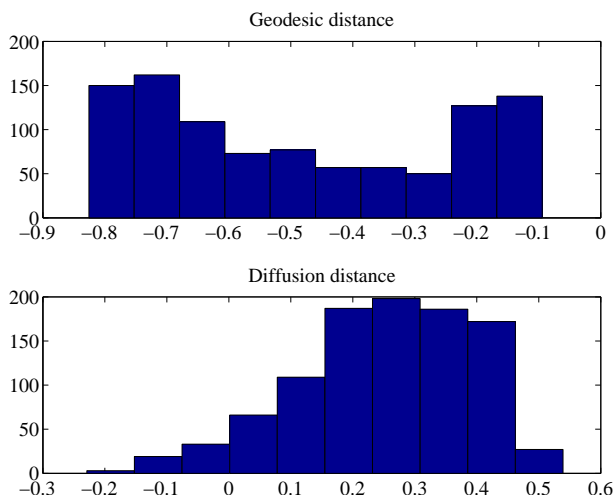}
\end{center}
\vspace{-1.5in}
\caption{\footnotesize Sensitivity to noise. Distribution of the geodesic (top) and diffusion (bottom) distances 
for a noisy spiral. Each histogram has been rescaled so as to show the relative change 
from the noiseless case. }
\label{fig:spiral_robustness}
\end{figure}

\begin{figure}[htb]
\begin{center}
\begin{tabular}{cc}
\includegraphics[width=2.8in]{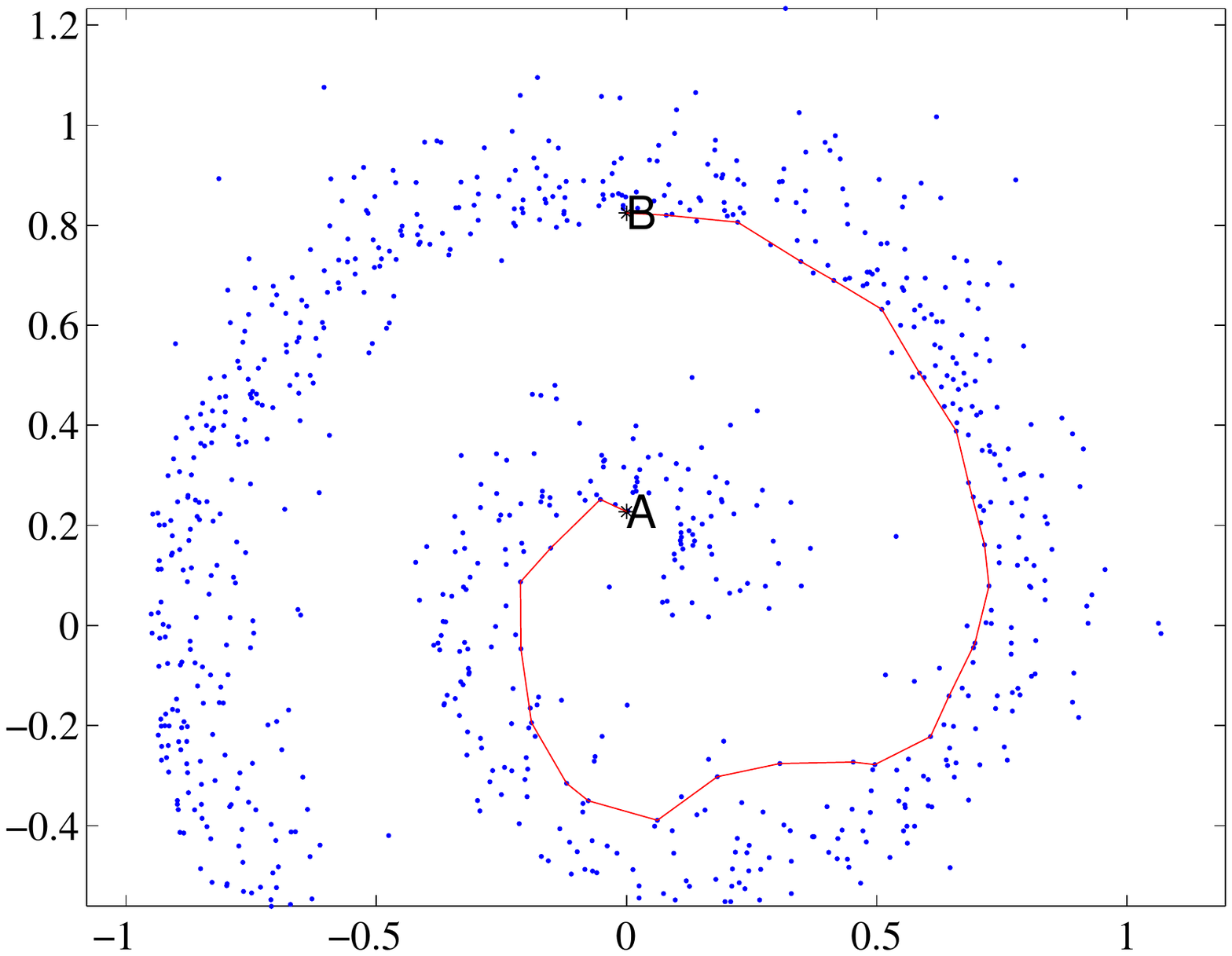}&\includegraphics[width=2.8in]{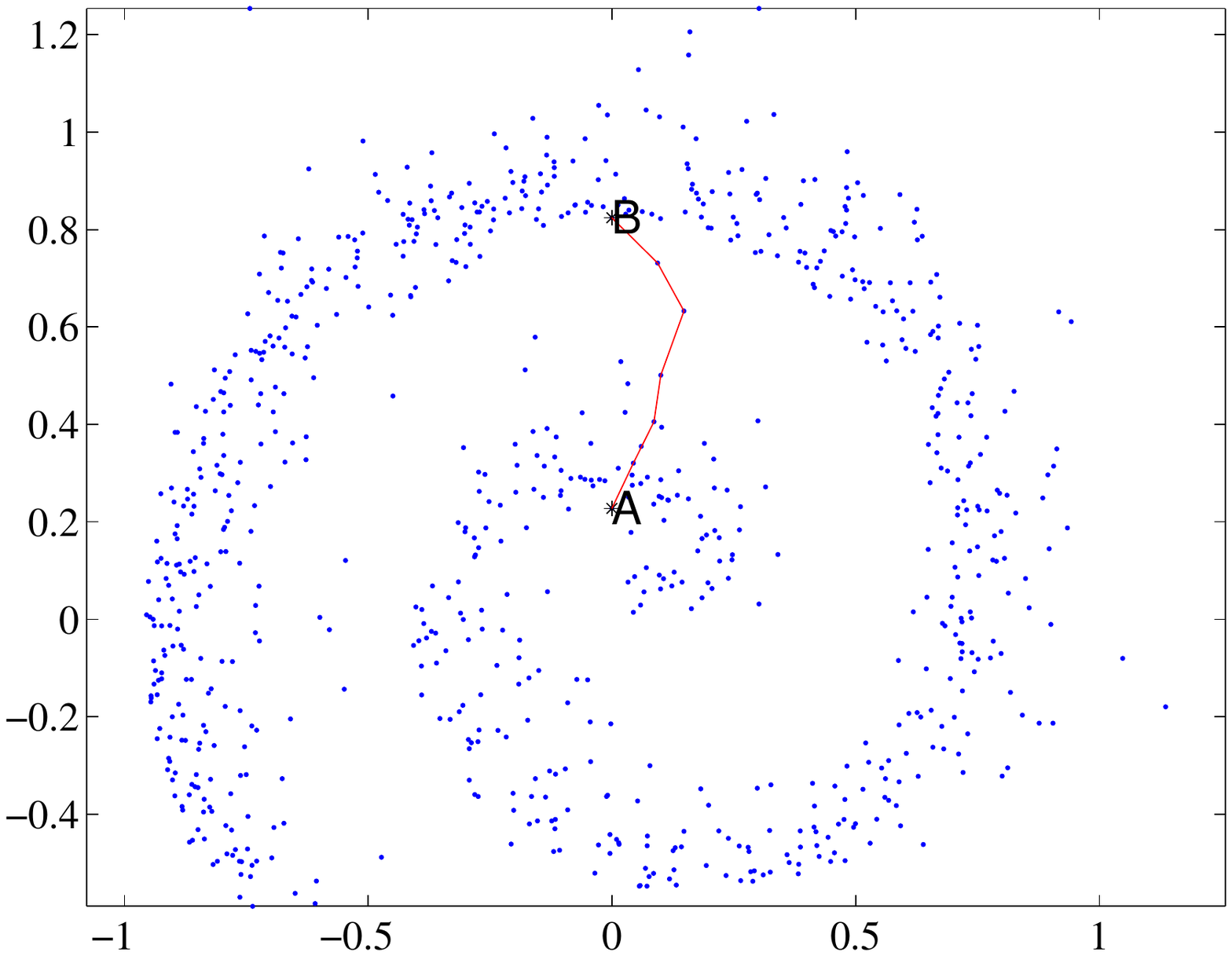}
\end{tabular}
\end{center}
\vspace{-1in}
\caption{\footnotesize Two realizations of a noisy spiral. The solid line 
represents the shortest path between two reference points $A$ and $B$ 
in a graph constructed on the data.}
\label{fig:spiral_paths}
\end{figure}

\begin{example}
{\bf (Sensitivity to noise)} We first generate 1000 instances of the spiral without noise,
(that is, the data fall exactly on the spiral)
and then 1000 instances of the spiral with exponential noise with mean 
parameter $\beta=0.09$ added to both $x$ and $y$. For each realization of the spiral, 
we construct a graph by connecting 
all pairs of points at a distance less than a threshold $\tau$. 
\myfigref{fig:spiral_robustness} shows histograms of the relative change in the 
geodesic graph distance (top) and the diffusion distance (bottom) when the data 
are perturbed. (The value 0 corresponds 
to no change from the average distance in the noiseless cases). 
For the geodesic distance, we have a bimodal distribution with a large variance. 
The mode near $-0.15$ corresponds to cases where the shortest path between 
$A$ and $B$ approximately follows the branch of the spiral; 
see~\myfigref{fig:spiral_paths} (left) for an example. The second mode around $-0.75$ 
occurs because some realizations of the noise give rise to shortcuts, 
which can dramatically reduce the length of the shortest path; 
see~\myfigref{fig:spiral_paths} (right) for an example. The diffusion distance, on the other hand, 
is not sensitive to small random perturbations of the data, because 
unlike the geodesic distance, it represents an average quantity. Shortcuts due to noise 
have little weight in the computation, as the number of such paths is much smaller 
than the number of paths following the shape of the spiral. This is also what our 
experiment confirms: \myfigref{fig:spiral_robustness} (bottom) shows a unimodal distribution 
with about half the variance as for the geodesic distance. 
In our experiment, the sample size $n=800$ and the neighborhood size $\tau=0.15$. 
To be able to directly compare the two methods and use the same parameters, we have 
for the diffusion distance calculation digressed from the Gaussian kernel and instead 
defined an adjacency matrix with only zeros or ones, 
corresponding to the absence or presence of an edge, respectively, in the graph construction.
\end{example}

\begin{figure}[htb]
\begin{center}
\includegraphics[width=4in]{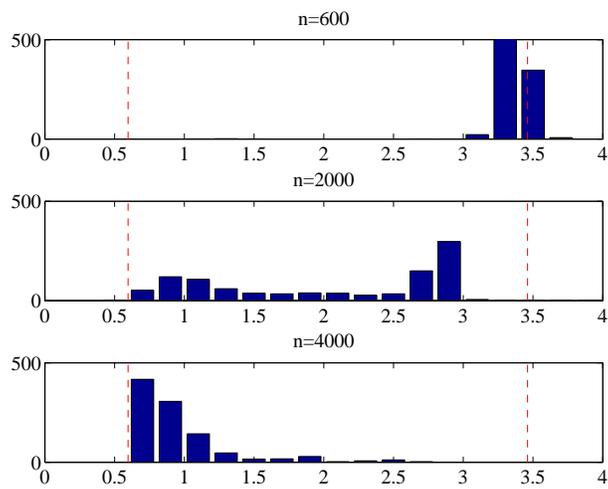}
\end{center}
\vspace{-1.5in}
\caption{\footnotesize Inconsistency of the geodesic graph distance. Distribution of the 
geodesic distance for (top to bottom) 
sample sizes $n=600, 2000$ and $4000$. The dashed vertical lines indicate the 
Euclidean distance in the ambient space and the geodesic distance of the ideal manifold. }
\label{fig:spiral_consistency}
\end{figure}

\begin{example} ({\bf Consistency})
For a distribution not supported exactly on a manifold, 
the problem with shortcuts gets worse as the sample size increases. 
This is illustrated by our next experiment 
where the noise level $\beta=0.09$ and the neighborhood size $\tau=0.1$ are fixed, 
and the sample size $n=600, 2000$ and $4000$. \myfigref{fig:spiral_consistency} 
shows that for a small enough sample size, the graph estimates are close to the 
theoretical value $d_{\mathcal M}=3.46$. For intermediate sample sizes, 
we have a range of estimates between the Euclidean distance $\|x_A-x_B\|=0.6$ 
and $d_{\mathcal M}$. As $n$ increases, shortcuts are more likely to occur, 
with the graph distance eventually converging to the Euclidean distance in the ambient space. 
\end{example}

\subsection{Density Sensitive Metrics}

In certain machine learning methods, such as semisupervised learning,
it is useful to define a density sensitive distance for which $x$ and $y$ 
are close
just when there is a high density path connecting $x$ and $y$.
This is precisely what diffusion distances do.
	Another such metric is \citep{BousquetCH03}
$$
d(x,y) = \inf_q \int_x^y \frac{ds}{p(q(s))}
$$
where the infimum is over all smooth paths 
(parameterized by path length)
connecting $x$ and $y$.
The two metrics have similar goals but
$\mathbf{D}_t(x,y)$ is more robust and easier to estimate.
Indeed, to find $d(x,y)$ one has to examine all paths connecting $x$ 
and $y$.

\clearpage

\section{Estimation}
\label{sec:estimation}

Now we study the properties of
$\hat{A}_{\varepsilon,t/\varepsilon}$
as an estimator of $\mathbf{A}_t$.
Let $\Pi_{\varepsilon,\ell}$ be the 
%subspace 
orthogonal projector onto the subspace spanned by $
\psi_{\varepsilon,\ell}$
and let
$\Pi_{\ell}$ be the projector onto the subspace spanned by $\psi_{\ell}$.
Consider the following operators:
\begin{center}
\begin{tabular}{lcllcl}
$A_t(\varepsilon,P)$ &$\equiv$& $A_{\varepsilon,t/\varepsilon}$   
  $=$  $\sum_{\ell=0}^\infty 
\lambda_{\varepsilon,\ell}^{t/\varepsilon}\ \Pi_{\varepsilon,\ell}$, \ \ \ &
$A_t(\varepsilon,q,P)$  &          
$=$ & $\sum_{\ell=0}^q \lambda_{\varepsilon,\ell}^{t/\varepsilon}\ 
\Pi_{\varepsilon,\ell}$,\\
\noalign{\vskip 5pt}
$A_t(\varepsilon,q,\hat{P}_n)$ & $=$ & 
$\sum_{\ell=0}^{q}
\hat{\lambda}_{\varepsilon,\ell}^{t/\varepsilon}\hat{\Pi}_{\varepsilon,\ell}$, 
\ \ \ &
$\mathbf{A}_t$ & $=$ & $\sum_{\ell \geq 0}e^{-\nu_{\ell}^2t} \ \Pi_{\ell}$, 
\\
\end{tabular}
\end{center}
where $\psi_{\varepsilon,\ell}$ and
$\lambda_{\varepsilon,\ell}$ denote the eigenfunctions and
eigenvalues of $A_\varepsilon$, and $\hat\psi_{\varepsilon,\ell}$ and
$\hat\lambda_{\varepsilon,\ell}$ are the 
eigenfunctions and eigenvalues of the data-based operator $\hat{A}_
\varepsilon$. 
Two estimators of $\mathbf{A}_t$ are
the truncated estimator $A_t(\varepsilon,q,\hat{P}_n)$ 
and the non-truncated estimator
$A_t(\varepsilon,\hat{P}_n)  \equiv e^{t (\hat{A}_\varepsilon - I)/
\varepsilon}$.
In practice, truncation is important since it
corresponds to choosing a dimension for the reparameterized data.

\subsection{Estimating the Diffusion Operator $\mathbf{A}_t$ }

\begin{figure}
\hspace{-2cm}
\includegraphics[bb=40 520 80 660]{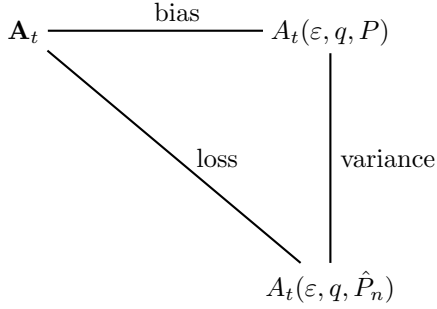}
\vspace{-1cm}
\caption{\footnotesize Decomposition of the loss into bias and variance.}
\label{fig::risk}
\end{figure}

Given data with a sample size $n$, 
we estimate $\mathbf{A}_t$ using a finite number $q$ of eigenfunctions 
and a kernel 
bandwidth $\varepsilon>0$. 
\comment{$\Pi_{\varepsilon,\ell}\to \Pi_\ell$ 
and 
$\lambda_{\varepsilon,\ell}^{t/\varepsilon}\to e^{-\nu_\ell^2 t}$ as $
\varepsilon \to 0$.
We call
$\mathbf{A}_t$ the {\em diffusion operator}.
Note that $\mathbf{A}_0 = A$.}
We define the loss function as
\begin{equation}\label{eq::loss}
 L_n(\varepsilon, q,t) =
\|\mathbf{A}_t - A_t(\varepsilon, q, \hat{P}_n)\|
\end{equation}
where
$\|B\| = \sup_{f\in {\cal F}} 
\|Bf\|_2/\|f\|_2$ and $\|f\|_2 = \sqrt{\int_{\mathcal{X}} f^2(x)  dP(x)}$
where ${\cal F}$ is the set of 
uniformly bounded, three times differentiable functions
with uniformly bounded derivatives
whose gradients vanish at the boundary.
By decomposing $L_n$ into a bias-like and variance-like term 
(\myfigref{fig::risk}), we derive the following result
for the estimate based on truncation.
Define
\begin{equation}
\rho(t)=\sum_{\ell=1}^\infty e^{-\nu_\ell^2 t}.
\end{equation}

\begin{theorem}\label{thm::main}
Suppose that $P$ has compact support,
and has bounded density $p$ such that
$\inf_x p(x) > 0$ and $\sup_x p(x) < \infty$.
Let $\varepsilon_n\to 0$ and $n \varepsilon_n^{d/2}/\log(1/\varepsilon_n) \to \infty$.
Then
\begin{equation}\label{eq::main}
L_n(\varepsilon_n,q,t) = 
\rho(t)
\left(O_P\left(\sqrt{\frac{ \log(1/\varepsilon_n)}{n \varepsilon_n^{(d+4)/2}}}\right) +
O(\varepsilon_n)\right)\ + \ 
O(1)\sum_{q+1}^\infty e^{-\nu_\ell^2 t}.
\end{equation}
The optimal choice of $\varepsilon_n$ is
$\varepsilon_n \asymp (\log n /n)^{2/(d+8)}$ in which case
\begin{equation}
L_n(\varepsilon_n,q,t) = 
\rho(t)\times
O_P\left(\frac{\log n}{n}\right)^{2/(d+8)} +
O(1)\sum_{q+1}^\infty e^{-\nu_\ell^2 t}.
\end{equation}
\end{theorem}

We also have the following result which does not use truncation.

\begin{theorem}
\label{theorem::alt}
Define
$$
A_t(\varepsilon,\hat{P}_n)  = e^{t (\hat{A}_{\varepsilon_n} - I)/
\varepsilon_n}.
$$
Then,
\begin{equation}
\|\mathbf{A}_t - A_t(\varepsilon_n,\hat{P}_n) \| =
\left(O_P\left(\sqrt{\frac{ \log(1/\varepsilon_n)}{n \varepsilon_n^{(d+4)/2}}}\right) +
O(\varepsilon_n)\right)
\times  \rho(t)
\end{equation}
The optimal $\varepsilon_n$ is
$\varepsilon_n \asymp (\log n /n)^{2/(d+8)}$.
With this choice,
$$
\|\mathbf{A}_t - A_t(\varepsilon,\hat{P}_n) \| = O_P\left(\frac{\log n}{n}\right)^{2/(d+8)}
\times  \rho(t).
$$
\end{theorem}

Let us now make some remarks on the interpretation of these reults.

\begin{enumerate}
\item The terms $O(\varepsilon_n)$ and
$\sum_{q+1}^\infty e^{-\nu_\ell^2 t}$ correspond to bias.
The term
$O_P\left(\sqrt{\frac{ \log(1/\varepsilon_n)}{n \varepsilon_n^{(d+4)/2}}}\right)$
corresponds to the square root of the variance.
\item The rate $n^{-2/(d+8)}$ is slow.
Indeed, the variance term
$1/(n\varepsilon_n^{(d+4)/2})$
is the usual rate for estimating the second derivative of a regression function
which is a notoriously difficult problem.
This suggests that estimating $\mathbf{A}_t$ accurately is quite difficult.
\item 
We also have that
$$
\|G_\varepsilon - \mathbf{G}\| = 
O_P\left(\sqrt{\frac{ \log(1/\varepsilon_n)}{n \varepsilon_n^{(d+4)/2}}}
\right) +
O(\varepsilon)
$$
and the first term 
is slower than the rate 
$1/\sqrt{n \varepsilon_n^{(d+2)/2}}$
in
Gin\'e and Koltchinskii (2006) and Singer (2006).
This is because they assume a uniform distribution. The slower rate comes from the term
$p_\varepsilon(x) - \hat{p}_\varepsilon(x)$ which cannot be ignored when $p$ is unknown.
\item 
If the distribution is supported on a manifold of dimension
$r < d$ then $\varepsilon^{(d+4)/2}$ becomes
$\varepsilon^{(r+4)/2}$.
In between full support and manifold support, one can create distributions
that are concentrated near manifolds.
That is, one first draws $X_i$ from a distribution supported on
a lower dimensional manifold, then adds noise to $X_i$.
This corresponds to full support again unless one lets the variance
of the noise decrease with $n$.
In that case, the exponent of $\varepsilon$ can be between $r$ and $d$.
\item Combining the above results with the result from 
\cite{Zwald:Blanchard:2006}, we have that
$$
\|\psi_{\ell} - \hat\psi_{\varepsilon_n,\ell}\| = 
\left(O_P\left(\sqrt{\frac{ \log(1/\varepsilon_n)}{n \varepsilon_n^{(d+4)/2}}}\right) +
O(\varepsilon_n)\right) \times
\frac{1}{\min_{0\leq j \leq \ell} (\nu_j^2 - \nu_{j-1}^2)}.
$$
\item The function
$\rho(t)$
is decreasing in $t$.
Hence for large $t$,
the rate of convergence can be arbitrarily fast,
even for large $d$.
In particular, for
$t \geq \rho^{-1}( n^{-(d+4)/(2(d+8))})$
the loss has the parametric rate
$O_P(\sqrt{\log n/n})$.
\item An estimate of the diffusion distance is
$$
\hat{D}_t^2(x,y) = \sum_{\ell=0}^\infty
\hat\lambda_{\varepsilon,\ell}^{2t/\varepsilon}
(\hat\psi_{\varepsilon,\ell}(x) - \hat\psi_{\varepsilon,\ell}(y))^2 .
$$
The approximation properties are similar to those of
$\hat{A}_t$.
\item The dimension reduction parameter $q$ should be chosen as small as
possible while keeping the last term 
in (\ref{eq::main})
no bigger than the first term.
This is illustrated below.
\end{enumerate}

\begin{example}
Suppose that
$\nu_\ell = \ell^\beta$ for some $\beta > 1/2$.
Then
\begin{eqnarray*}
L_n(\varepsilon_n,q,t) &=& 
\frac{C_1 }{t^{1/(2\beta)}}
O_P\left(\frac{\log n}{n^{2/(d+8)}}\right) +
C_2 e^{ - t q^{2\beta}}.
\end{eqnarray*}
The smallest $q_n$ such that
the last term in (\ref{eq::main})
does not dominate is
$$
q_n \asymp 
\left(\frac{ \frac{1}{2\beta}\log t + \frac{2}{d+8}\log n}{t}\right)^{1/(2\beta)}
$$
and we get
$$
L_n(\varepsilon_n,q,t) =
O_P\left( \frac{1 }{t^{1/(2\beta)}} \frac{\log n}{n^{2/(d+8)}}\right).
$$
\end{example}

\subsection{Nodal Domains and Low Noise}

An eigenfunction $\psi_\ell$ partitions the sample space into regions
where $\psi_\ell$ has constant sign.
This partition is called the
{\em nodal domain} of $\psi_\ell$.
In some sense, the nodal domain represents the basic
structural information in the eigenfuction.
In many applications, such as spectral clustering,
it is sufficient to estimate the nodal domain rather than
$\psi_\ell$.
We will 
show that fast rates are sometimes available for 
estimating the nodal domain even when the eigenfunctions are hard
to estimate.
This explains why spectral methods can be very successful despite the slow
rates of convergence that we and others have obtained.

Formally, the nodal domain of $\psi_\ell$ is
$N_\ell = \{C_1, \ldots, C_k\}$
where the sets $C_1,\ldots, C_k$ partition
the sample space and
the sign of $\psi_\ell$ is constant over each 
partition element $C_j$.
Thus, estimating the nodal domain corresponds
to estimating
$H_\ell(x) = {\rm sign}(\psi_\ell(x))$.
\footnote{If $\psi$ is an eigenfunction then so is $-\psi$.
We implicitly assume that the sign ambiguity
of the eigenfunction has been removed.}

Recently, in the literature on classification,
there has been a surge of research on the
so-called ``low noise'' case.
If the data have a low probability of being close
to the decision boundary, then very fast
rates of convergence are possible even in high dimensions.
This theory explains the success of classification techniques
in high dimensional problems.
In this section we show that a similar phenomema
applies to spectral smoothing when estimating the nodal domain.

\begin{figure}
\includegraphics{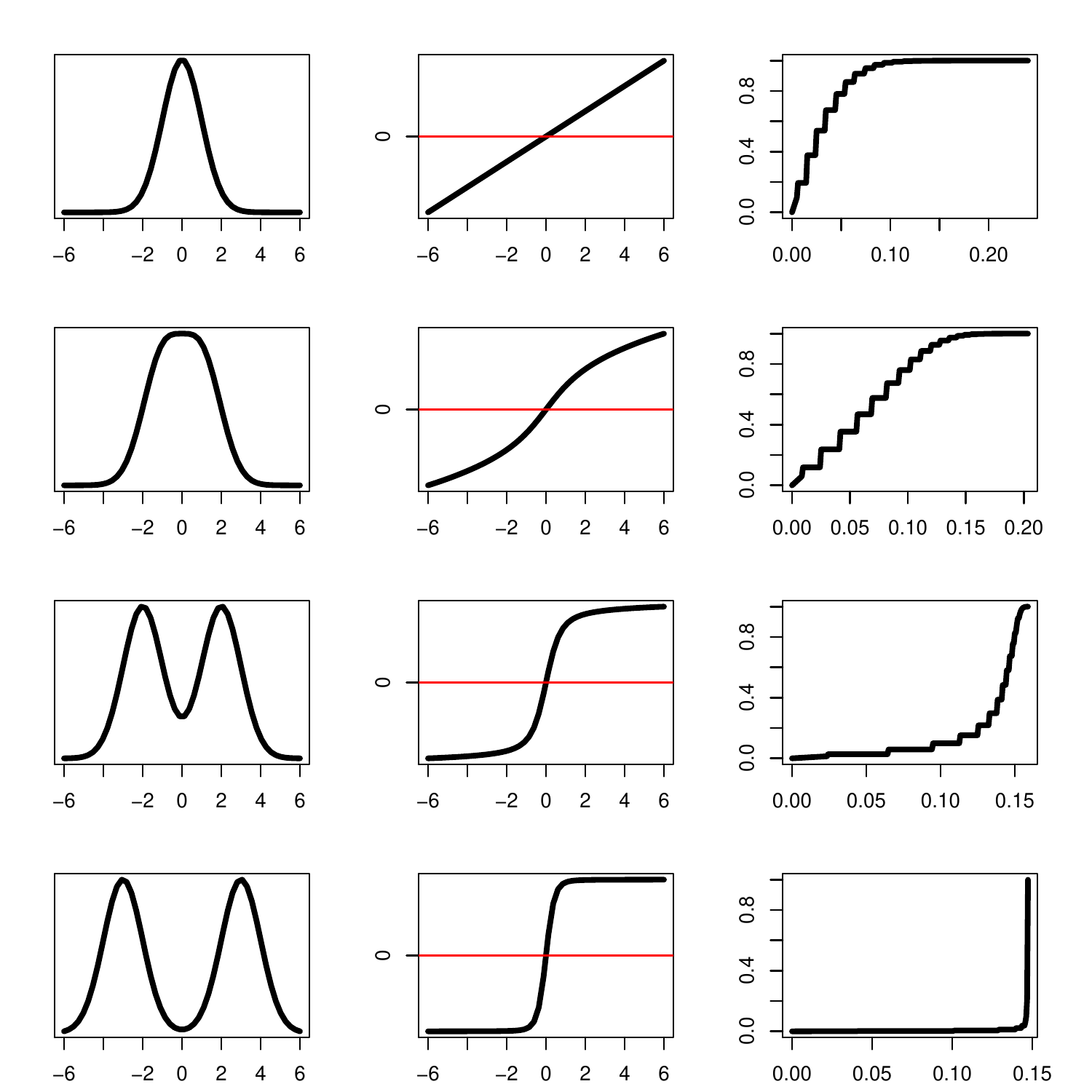}
\caption{\footnotesize
Each row corresponds to a mixture of two Gaussians.
The first column of plots shows the densities of these distributions.
The second column shows $\psi_1$.
The third column shows
$\mathbb{P}( 0 < | \psi_1(X)| \leq \delta)$ as a function of $\delta$.}
\label{fig::lownoise}
\end{figure}

Inspired by the definition of low noise in 
\cite{Mammen99smoothdiscrimination}, \cite{Audibert:Tsybakov:2007}, and \cite{Kohler:Krzyzak:07},
we say that $P$ has noise exponent $\alpha$ 
if there exists $C>0$ such that
\begin{equation}\label{eq::lownoise}
\mathbb{P}( 0 < | \psi_1(X)| \leq \delta) \leq C \delta^\alpha.
\end{equation}
We are focusing here on $\psi_1$ although extensions to
other eigenfunctions are immediate.
Figure \ref{fig::lownoise} shows 4 distributions.
Each is a mixture of two Gaussians.
The first column of plots shows the densities of these 4 distributions.
The second column shows $\psi_1$.
The third column shows
$\mathbb{P}( 0 < | \psi_1(X)| \leq \delta)$.
Generally, as clusters become well separated,
$\psi_1$ behaves like a step function and
$\mathbb{P}( 0 < | \psi_1(X)| \leq \delta)$ puts less and less mass near 0
which corresponds to $\alpha$ being large.

\begin{theorem}\label{theorem::lownoise}
Let
$H(x) = {\rm sign}(\psi_1(x))$ and
$\hat{H}(x) = {\rm sign}(\hat\psi_1(x))$.
Suppose that (\ref{eq::lownoise}) holds.
Set
$\varepsilon_n = n^{-2/(4\alpha + d + 8)}$.
Then, 
\begin{equation}
\mathbb{P}( H(X) \neq \hat{H}(X))\leq
n^{ - \frac{2\alpha}{4\alpha + 8 + d}}
\end{equation}
where $X\sim P$.
\end{theorem}

Note that, as $\alpha\to\infty$ the rate tends to
the parametric rate $n^{-1/2}$.

\subsection{Choosing a Bandwidth}

The theory we have developed gives insight into the behavior of the 
methods.
But we are still left with the need for a practical
method for choosing $\varepsilon$.
Given the similarity with kernel smoothing,
it is natural to use methods from density estimation
to choose $\varepsilon$.
In density estimation it is common to use the
loss function
$\int (p(x) - \hat{p}_\varepsilon(x))^2 dx$
which is equivalent, up to a constant, to
$$
{\cal L}(\varepsilon) = \int \hat{p}_\varepsilon^2(x) dx - 
2 \int \hat{p}_\varepsilon(x) p(x) dx.
$$
A common method to approximate this loss
is the cross-validation score
$$
\hat{\cal L}(\varepsilon) =
\int \hat{p}_\varepsilon^2 (x) dx - \frac{2}{n}\sum_{i=1}^n 
\hat{p}_{\varepsilon,i}(X_i)
$$
where
$\hat{p}_{\varepsilon,i}$
is the same as
$\hat{p}_{\varepsilon}$
except that $X_i$ is omitted.
It is well-known that
$\hat{\cal L}(\varepsilon)$
is a nearly unbiased estimate of
${\cal L}(\varepsilon)$.
One then chooses $\tilde{\varepsilon}_n$ to minimize
$\hat{\cal L}(\varepsilon)$.

The optimal $\varepsilon_n^*$ from 
our earlier result is (up to log factors)
$O(n^{-2/(d+8)})$ but the optimal bandwidth 
$\varepsilon_n^0$
for minimizing
${\cal L}$ is
$O(n^{-2/(d+4)})$.
Hence,
$\varepsilon_n^*/\varepsilon_n^0 \asymp n^{8/((d+4)(d+8))}$.
This suggests that density cross-validation
is not appropriate for our purposes.

Indeed, there appears to be no unbiased risk estimator
for this problem.
In fact, estimating the risk is difficult in most problems
that are not prediction problems.
As usual in nonparametric inference,
the problem is that estimating bias is harder than
the original estimation problem.
Instead, we take a more modest view of simply trying to find the smallest
$\varepsilon$ such that the resulting variability is tolerable.
In other words,
we choose the smallest $\varepsilon$ that
leads to stable estimates of the eigenstructure
(similar to the approach for choosing the
number of clusters in~\cite{Lange:EtAl:04}).
There are several ways to do this as we now explain.

\vspace{.5cm}

{\bf Eigen-Stability.}
Define
$\overline{\psi}_{\varepsilon,\ell}(x) = \mathbb{E} (\hat\psi_{\varepsilon,\ell}(x))$.
Although
$\overline{\psi}_{\varepsilon,\ell} \neq \psi_\ell$,
they do have a similar shape.
We propose to choose $\varepsilon$ by
finding the smallest $\varepsilon$ for which
$\overline{\psi}_{\varepsilon,\ell}$ can be estimated with
a tolerable variance.
To this end we define
\begin{equation}
{\rm SNR}(\varepsilon) =
\sqrt{
\frac{ ||\overline{\psi}_{\varepsilon,\ell}||_2^2}
    { \mathbb{E}||\hat\psi_\ell - \overline{\psi}_{\varepsilon,\ell}||_2^2}
    }
\end{equation}
which we will refer to as the signal-to-noise ratio.
When $\varepsilon$ is small,
the denominator will dominate and
${\rm SNR}(\varepsilon) \approx 0$.
Conversely, when 
$\varepsilon$ is large,
the denominator tends to 0 so that
${\rm SNR}(\varepsilon)$ gets very large.
We want to find $\varepsilon_0$ such that
$$
\varepsilon_0 = \inf \Biggl\{ \varepsilon :\ {\rm SNR}(\varepsilon) \geq K_n \Biggr\}
$$
for some $K_n \geq 1$.%$K$ such as $K=1$.

We can estimate ${\rm SNR}$ as follows.
We compute $B$ bootstrap replications
$$
\hat\psi_{\varepsilon,\ell}^{(1)}, \ldots, \hat\psi_{\varepsilon,\ell}^{(B)}.
$$
We then take
\begin{equation}
\hat{\rm SNR}(\varepsilon) =
%\sqrt{  \frac{||\overline{\psi}_{\varepsilon,\ell}^*||^2_2 - \xi^2}{\xi^2}}
\sqrt{  \frac{ \left(||\overline{\psi}_{\varepsilon,\ell}^*||^2_2 - \xi^2 \right)_+}{\xi^2}}
\end{equation}
where $c_+=\max\{c,0\}$,
% $\left(||\overline{\psi}_{\varepsilon,\ell}^*||^2_2 - \xi^2 \right)_+ 
%=\max\{ ||\overline{\psi}_{\varepsilon,\ell}^*||^2_2 - \xi^2 ,0\}$ and
$$
\xi^2 = \frac{1}{B}\sum_{b=1}^B || \hat\psi_{\varepsilon,\ell}^{(b)} - \overline{\psi}_{\varepsilon,\ell}^* ||^2_2
$$
and
$\overline{\psi}_{\varepsilon,\ell}^* = B^{-1}\sum_{b=1}^B \hat\psi_{\varepsilon,\ell}^{(b)}$.
Note that we subtract $\xi^2$ from the numerator
to make the numerator approximately an unbiased estimator of
$||\overline{\psi}_{\varepsilon,\ell}||^2$.
Then we use
$$
\hat\varepsilon = \min\bigl\{\varepsilon:\ \hat{\rm SNR}(\varepsilon) \geq K_n \bigr\}.
$$
We illustrate the method in Section \ref{section::examples}.
%Now
%$\varepsilon_0 =O(n^{-2/(d+4)})$
%which is smaller than the optimal $\varepsilon$
%which is $O(n^{-2/(d+8)})$.
For $K_n = C n^{2/(d+8)}$, where $C$ is a constant, the optimal $\varepsilon$
is $O(n^{-2/(d+8)})$.
To see this, 
write
$$
\hat\psi_{\varepsilon,\ell}(x) = \psi_\ell(x) + b(x) + \xi(x)
$$
where $b(x)$ denotes the bias and
$\xi(x) = \hat\psi_{\varepsilon,\ell}(x) - \psi_\ell(x) - b(x)$
is the random component.
Then
$$
{\rm SNR}^2(\varepsilon)  =
\frac{ || \psi_\ell(x) + b(x) ||^2}{\mathbb{E} ||\xi||^2} =
\frac{O(1)}{O_P\left(\frac{1}{n \varepsilon^{(d+4)/2}}\right)}.
$$
Setting this equal to $K_n^2$ yields
$\varepsilon_0 =O(n^{-2/(d+8)})$.
%a constant yields
%$\varepsilon_0 =O(n^{-2/(d+4)})$.

The same bootstrap idea can be applied to estimating the nodal domain.
In this case we define
\begin{equation}
\hat{\rm SNR}(\varepsilon) =
\sqrt{  \frac{\left(||\overline{H}_{\varepsilon,\ell}^*||^2_2 - \xi^2\right)_+}{\xi^2}}
\end{equation}
where
$$
\xi^2 = \frac{1}{B}\sum_{b=1}^B || \hat{H}_{\varepsilon,\ell}^{(b)} - \overline{H}_{\varepsilon,\ell}^* ||^2_2
$$
and
$\overline{H}_{\varepsilon,\ell}^* = B^{-1}\sum_{b=1}^B \hat{H}_{\varepsilon,\ell}^{(b)}$.

\vspace{.5cm}

{\bf Neighborhood Size Stability.}
Another way to control the variability is to ensure
that the number of points involved in the local averages does not get too small.
For a given $\varepsilon$ let
$N = \{N_1,\ldots, N_n\}$
where $N_i = \# \{X_j:\ \|X_i-X_j\| \leq \sqrt{2\epsilon}\}$.
One can informally examine the histogram of $N$
for various $\varepsilon$.
A rule for selecting $\varepsilon$ is
$$
\hat\varepsilon = \min\bigl\{\varepsilon:\ {\rm median}\{N_1,\ldots, N_n\} \geq k \bigr\}.
$$
We illustrate the method in Section \ref{section::examples}.

An alternative, suggested by
\cite{Luxburg:2007},
is to choose the smallest $\varepsilon$
that makes the resulting graph well-connected.
This leads to
$\varepsilon = O((\log n/n)^{1/d})$.
More specifically,
\cite{Luxburg:2007}
 suggests to
``... choose $\varepsilon$ as the length of the longest edge
in a minimal spanning tree of the fully connected graph
on the data points.''

\vspace{.5cm}

\section{Examples}
\label{section::examples}

\subsection{Two Gaussians}

\begin{figure}
\begin{center}
\includegraphics[width=6in]{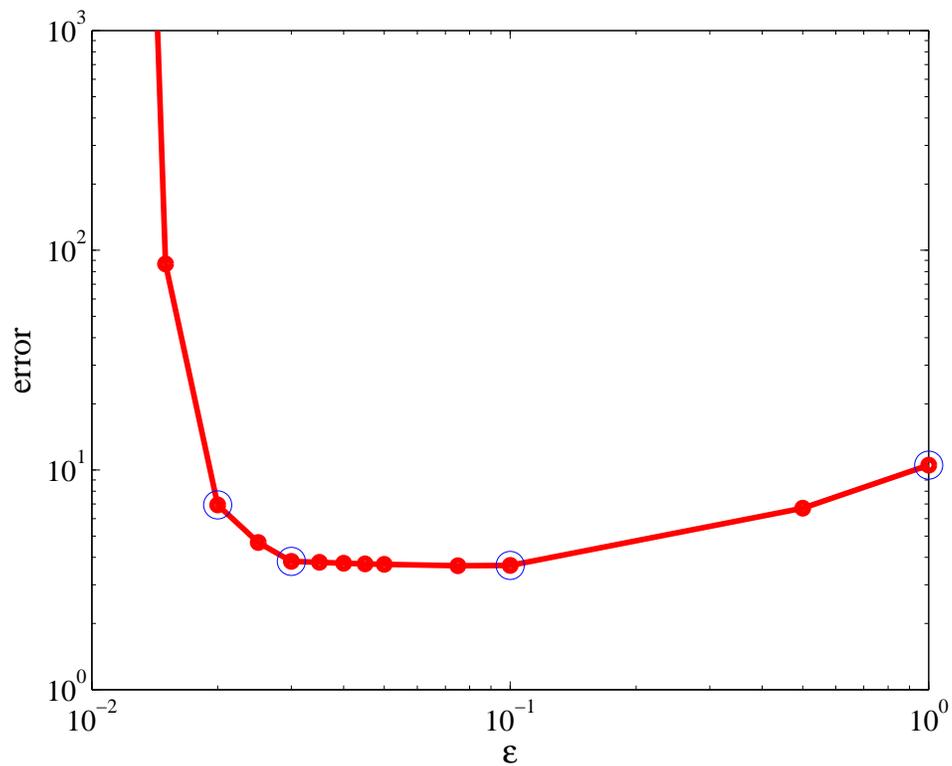}
\end{center}
\vspace{-5cm}
\caption{\footnotesize The error $\|\psi_1 - \hat{\psi}_{\varepsilon,1} \|$ in the estimate of the 
first eigenvector as a function of $\varepsilon$. For each $\varepsilon$ (red dots), 
an average is taken over approximately $200$
independent simulations with $n=1000$ points 
from a mixture distribution with two Gaussians. \myfigref{fig::eigvec1} shows 
the estimated mean and variance of  
$\hat{\psi}_{\varepsilon,1}$ for $\varepsilon = 0.02, 0.03,  0.1, 1$ (blue circles)}
\label{fig::error_rate}
\end{figure}

\begin{figure}
%\vspace{-4cm}
\begin{center}
\includegraphics[width=4in,bb = 73 210 544 595]{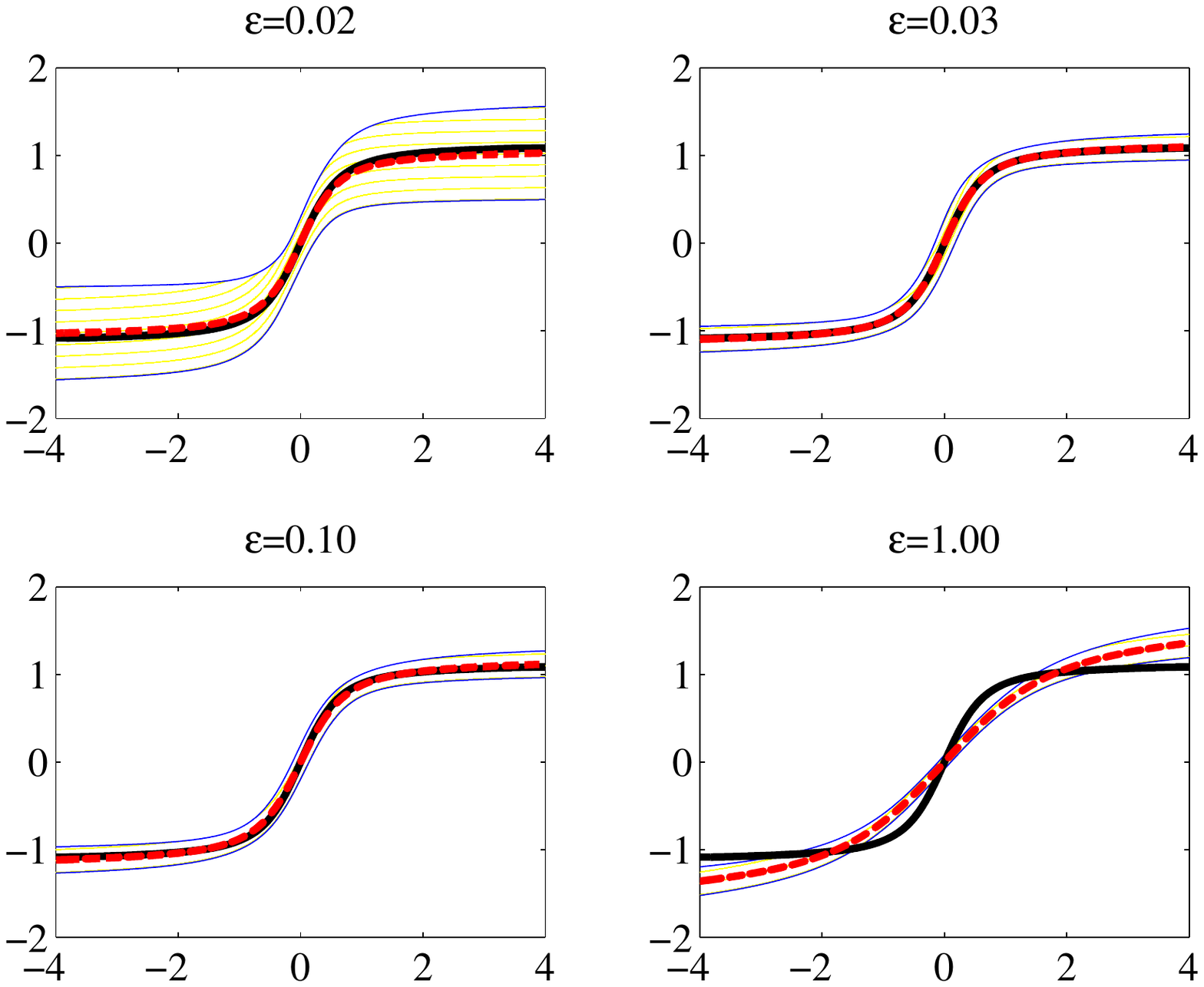}
\end{center}
%\vspace{-4cm}
\caption{ \footnotesize The first eigenvector
$\hat{\psi}_{\varepsilon,1}$ for $\varepsilon = 0.02, 0.03,  0.1, 1$ and $n=1000$. 
The red dashed curves with 
shaded regions indicate the mean value $\pm$ two standard deviations for approximately $300$
independent simulations. The black solid curves show ${\psi}_{\varepsilon,1}$ 
as $\varepsilon \rightarrow 0$.}
\label{fig::eigvec1}
\end{figure}

\begin{figure}
%\vspace{-4cm}
\begin{center}
\includegraphics[width=4in,bb = 73 210 544 595]{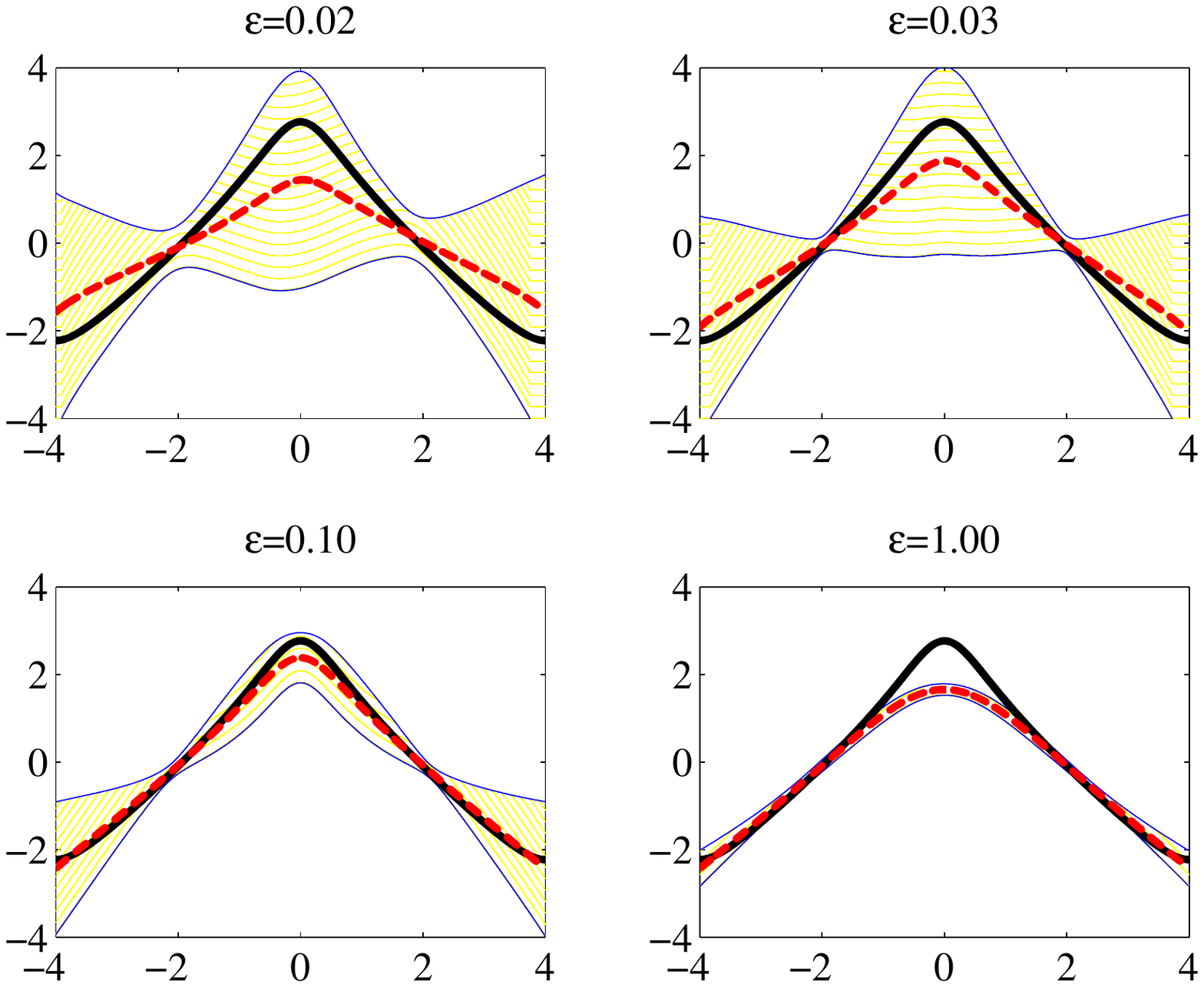}
\end{center}
%\vspace{-4cm}
\caption{\footnotesize The second eigenvector
$\hat{\psi}_{\varepsilon,2}$ for $\varepsilon = 0.02, 0.03,  0.1, 1$ and $n=1000$. 
The red dashed curves with shaded regions indicate the mean value $\pm$ two standard deviations for approximately $300$ 
independent simulations. The black solid curves show ${\psi}_{\varepsilon,2}$ 
as $\varepsilon \rightarrow 0$.}
\label{fig::eigvec2}
\end{figure}

\begin{figure}
\begin{center}
\includegraphics[width=4in,bb = 73 210 544 595]{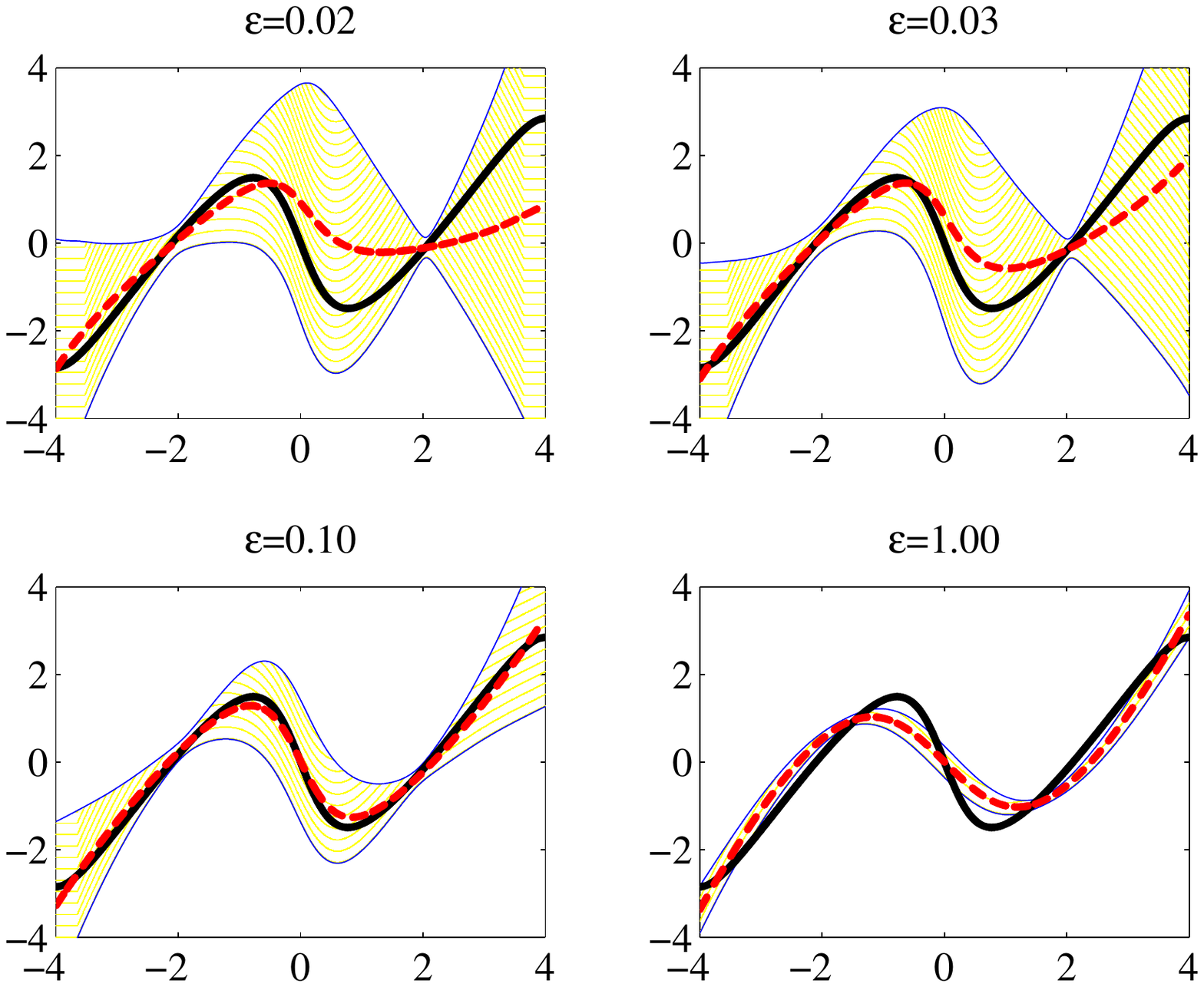}
\end{center}
\caption{\footnotesize The third eigenvector
$\hat{\psi}_{\varepsilon,3}$ for $\varepsilon = 0.02, 0.03,  0.1, 1$ and $n=1000$. 
The red dashed curves with shaded regions indicate the mean value $\pm$ two standard deviations for approximately $300$
independent simulations. The black solid curves show ${\psi}_{\varepsilon,3}$ 
as $\varepsilon \rightarrow 0$.}
\label{fig::eigvec3}
\end{figure}

\begin{figure}
\begin{center}
\includegraphics[width=4in,bb = 73 210 544 595]{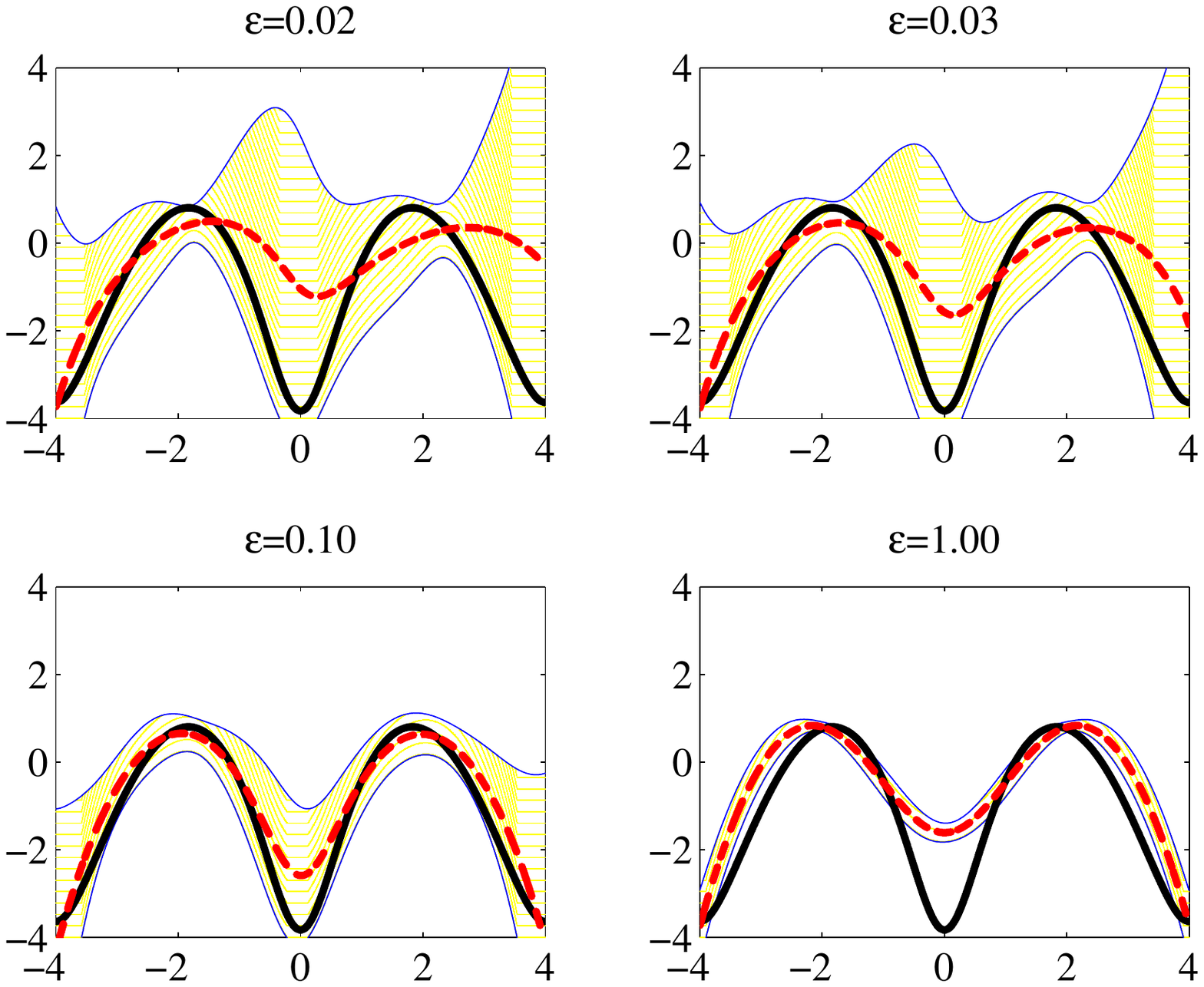}
\end{center}
\caption{\footnotesize The fourth eigenvector
$\hat{\psi}_{\varepsilon,4}$ for $\varepsilon = 0.02, 0.03,  0.1, 1$ and $n=1000$. 
The red dashed curves with shaded regions indicate the mean value $\pm$ two standard deviations for
approximately $300$
independent simulations. The black solid curves show ${\psi}_{\varepsilon,4}$ 
as $\varepsilon \rightarrow 0$.}
\label{fig::eigvec4}
\end{figure}

Let
$$
p(x) = \frac{1}{2} \phi(x;-2,1) + \frac{1}{2} \phi(x;2,1)
$$
where $\phi(x;\mu,\sigma)$ denotes
a Normal density with mean $\mu$
and variance $\sigma^2$.
\myfigref{fig::error_rate} shows the error $\|\psi_1 -
\hat{\psi}_{\varepsilon,1} \|$ as a function of $\varepsilon$ for a
sample of size $n=1000$. The results are averaged over
approximately~\footnote{We discard simulations where
$\widehat{\lambda}_1=\widehat{\lambda}_0=1$ for
$\varepsilon=0.02$. }  $200$ independent draws. A minimal error
occurs for a range of different values of $\varepsilon$ between $0.03$
and $0.1$. The variance dominates the error in the small $\varepsilon$
region ($\varepsilon<0.03$) , while the bias dominates in the large
$\varepsilon$ region ($\varepsilon > 0.1$). These results are
consistent with \myfigref{fig::eigvec1}, which shows the estimated
mean and variance of the first eigenvector
$\hat{\psi}_{\varepsilon,1}$ for a few selected values of
$\varepsilon$ ($\varepsilon = 0.02, 0.03, 0.1, 1$), marked with blue
circles in
\myfigref{fig::error_rate}. Figures~\ref{fig::eigvec2}-\ref{fig::eigvec4}
show similar results for the second, third and fourth eigenvectors
%$\hat{\psi}_{\varepsilon,2}, \hat{\psi}_{\varepsilon,3}, \hat{\psi}_{\varepsilon,4}$. 
$\psi_2, \psi_3, \psi_4$.
Note that even in cases where the error in the estimates of the eigenvectors is large, 
the variance around the cross-over points (where the eigenvectors switch signs) 
can be small.

\begin{figure}
\begin{center}
\includegraphics[width=3.2in]{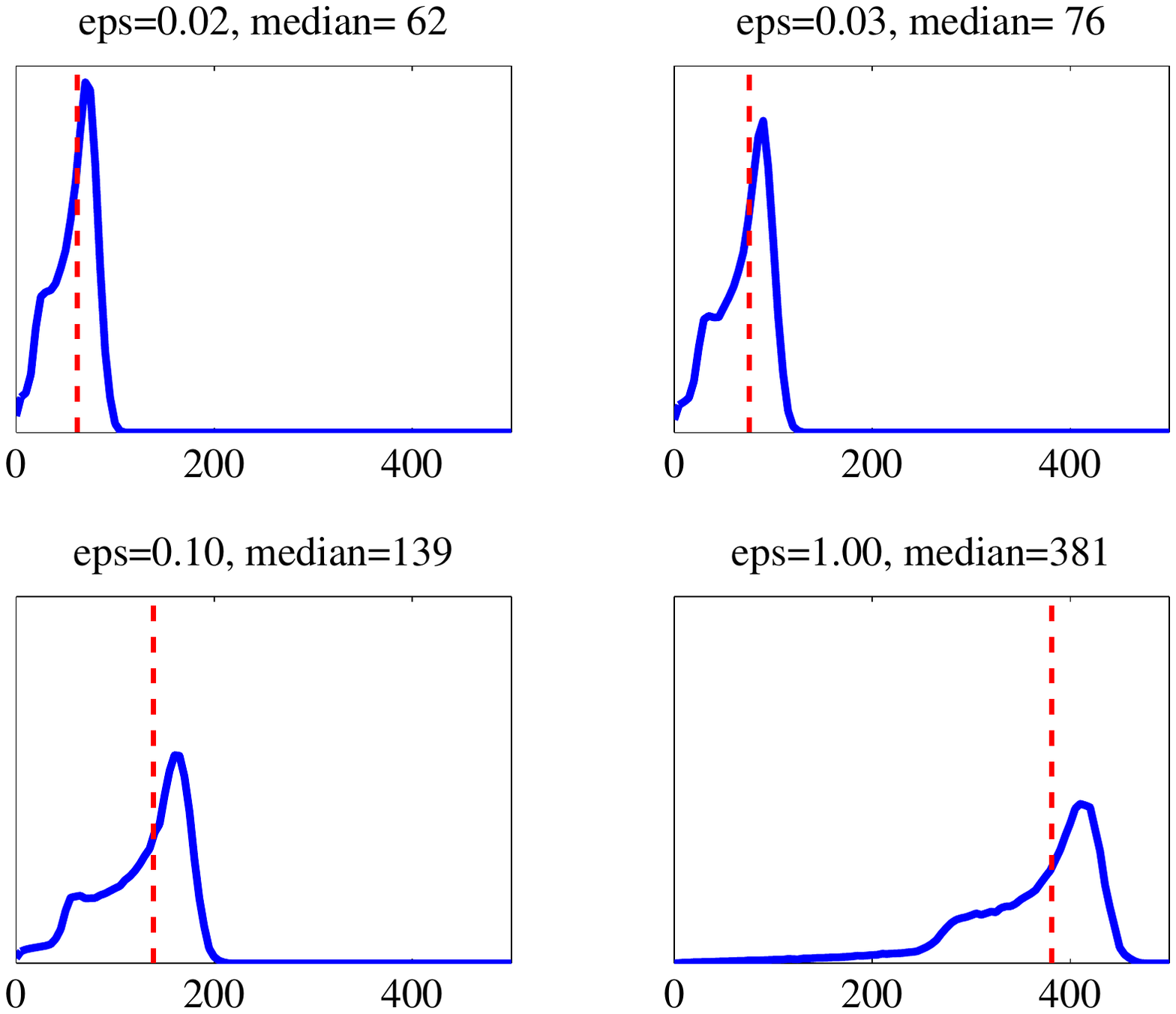}
\includegraphics[width=3.2in]{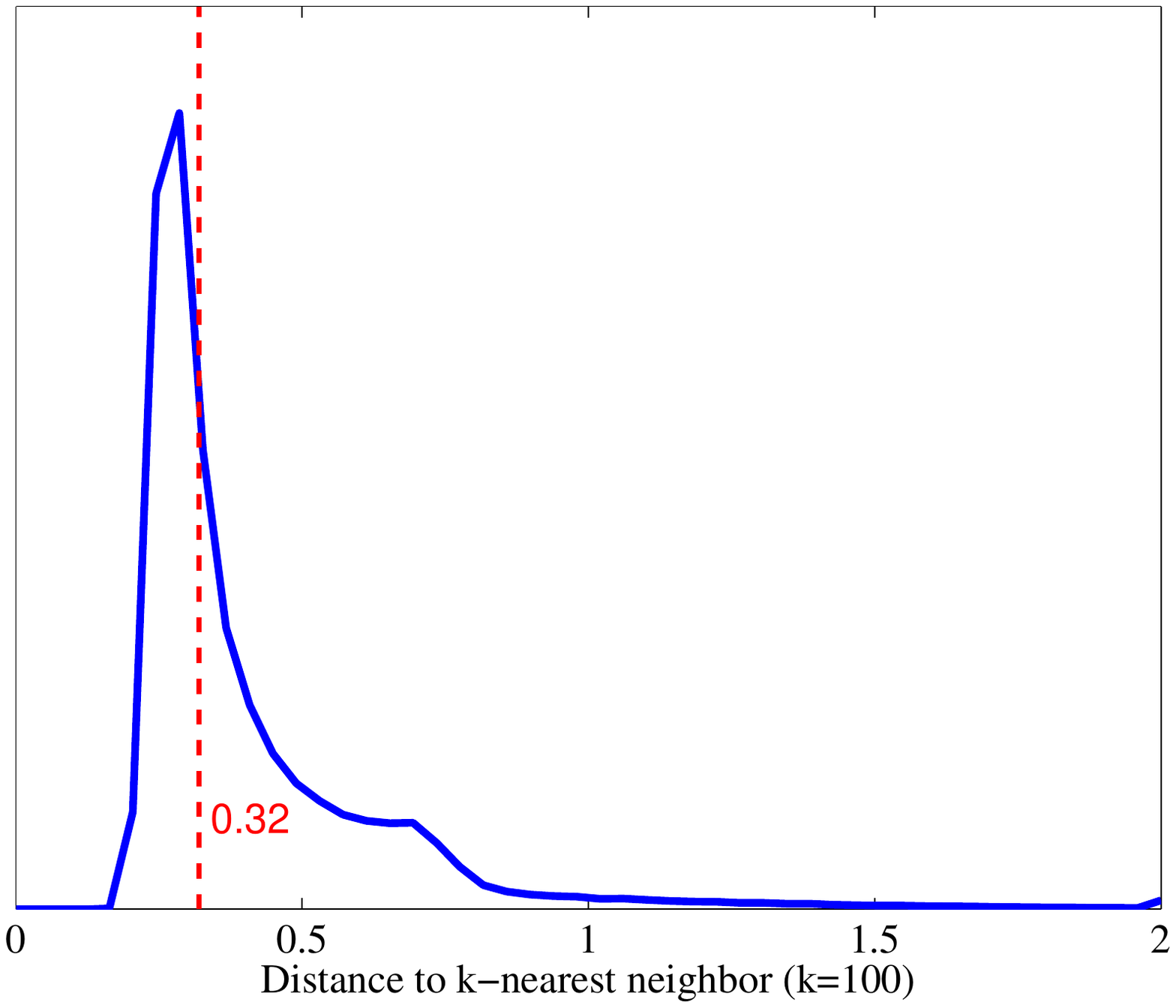}
\end{center}
\vspace{-3cm}
\caption{\footnotesize Left: Histogram of $N_i = \# \{X_j:\ \|X_i-X_j\| \leq \sqrt{2 \epsilon}\}$ for 
$\varepsilon = 0.02, 0.03,  0.1, 1$ and $n=1000$. The vertical dashed lines indicate the 
median values. 
Right: Histogram of the distance to the $k$-nearest neighbor for $k=100$. 
The median value $0.32$ (vertical dashed line) roughly corresponds 
to $\varepsilon=0.32^2/2 = 0.05$.
All results are averaged over $500$ independent simulations.}
\label{fig::neighborhood}
\end{figure}

\myfigref{fig::neighborhood} (left) shows a histogram of $N_i = \#
\{X_j:\ \|X_i-X_j\| \leq \sqrt{2 \varepsilon}\}$ for $\varepsilon =
0.02, 0.03, 0.1, 1$ and $n=1000$. All results are averaged over $500$
independent simulations. The vertical dashed lines indicate the median
values. For this particular example, we know that the error is small
when $\varepsilon$ is between $0.03$ and $0.1$. This corresponds to
${\rm median}\{N_1,\ldots, N_n\}$ being around
$100$. \myfigref{fig::neighborhood} (right) shows a histogram of the
distance to the $k$-nearest neighbor for $k=100$. The median value
$0.32$ (see vertical dashed line) roughly corresponds to the tuning
parameter $\varepsilon=0.32^2/2 = 0.05$.

\begin{figure}
\begin{center}
\includegraphics[width=4in]{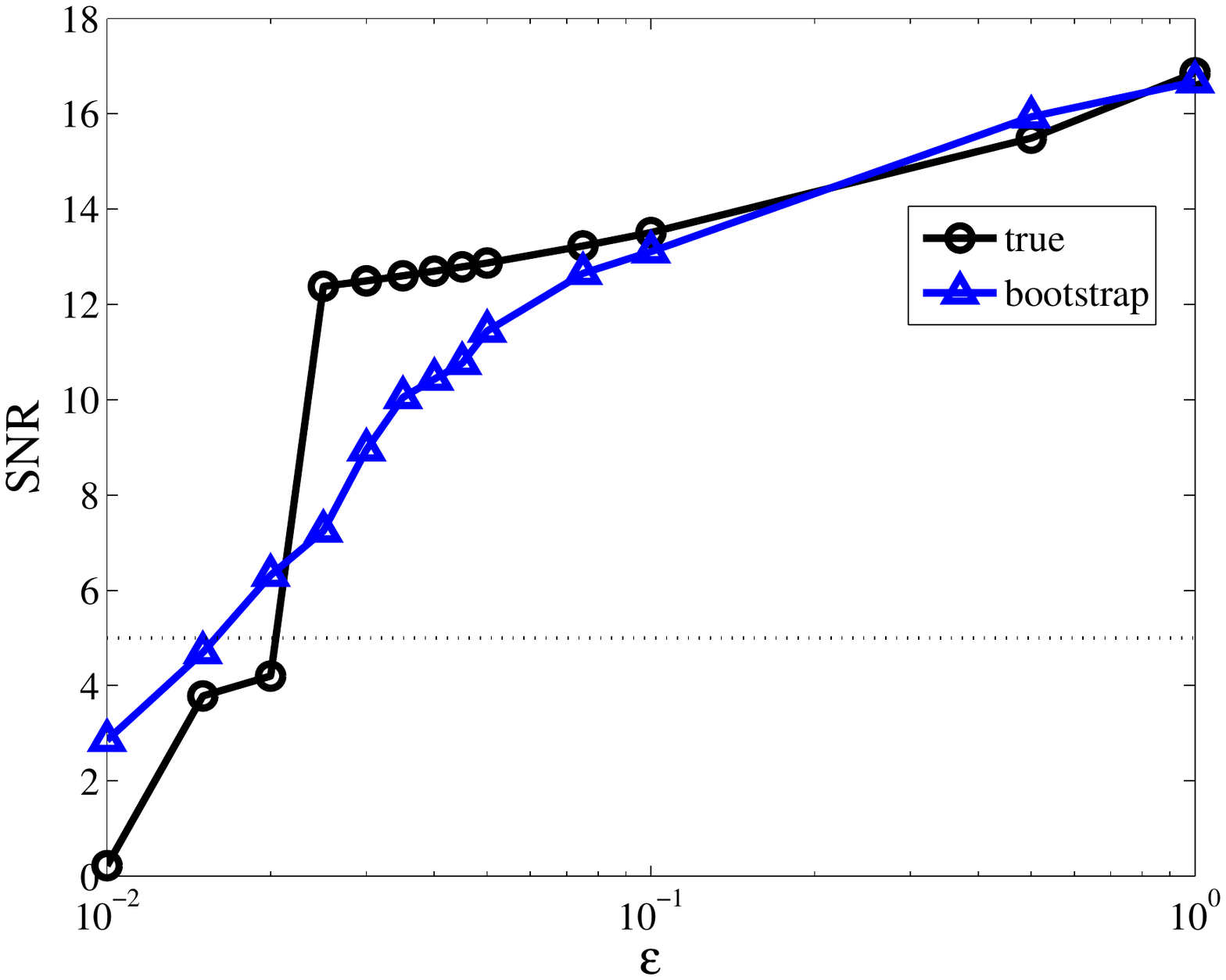}
\end{center}
\vspace{-3cm}
\caption{\footnotesize True signal-to-noise ratio estimated by simulation (rings)
and mean of the bootstrap estimated signal-to-noise ratio (triangles)
as a function of $\varepsilon$.
}
\label{fig::SNR1}
\end{figure}

\begin{figure}
\begin{center}
\includegraphics[width=3.2in]{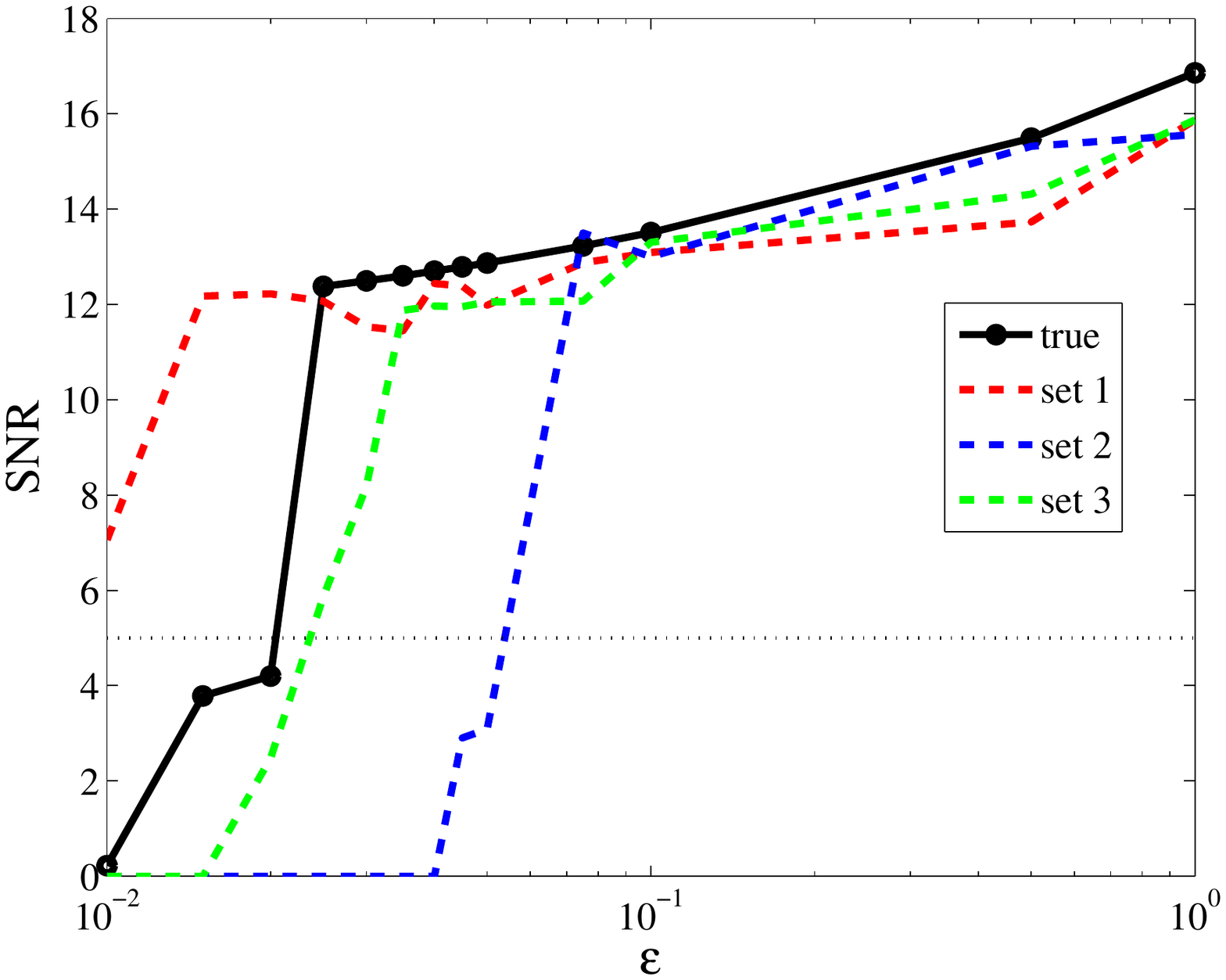}
\includegraphics[width=3.2in]{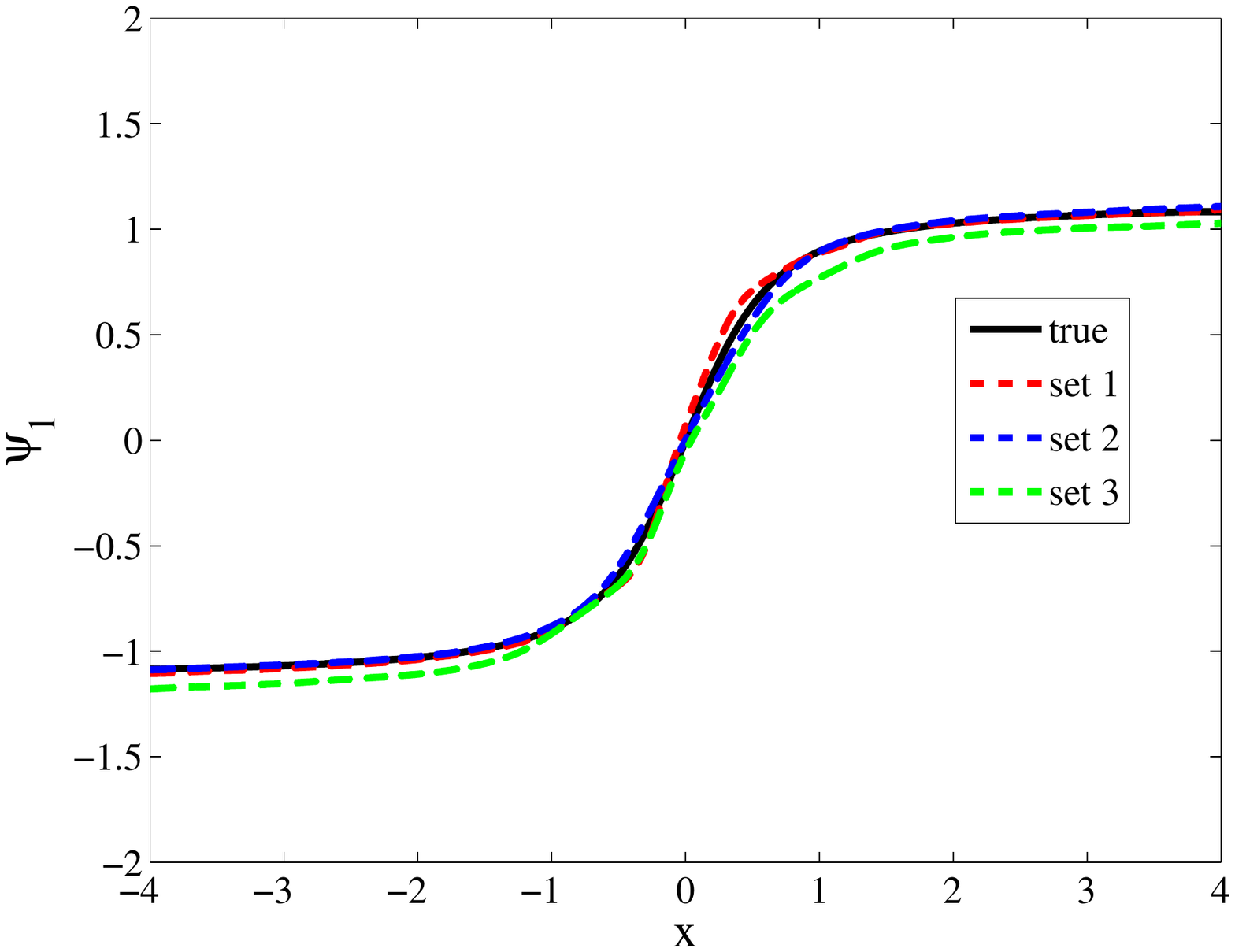}
\end{center}
\vspace{-3cm}
\caption{\footnotesize Left: Signal-to-noise ratio, as a function of  $\varepsilon$,  
estimated by simulation (solid black line), and 
by the bootstrap for three different data sets (dashed lines). 
Right:  $\psi_1$ (solid black line) and resulting bootstrap 
estimates of $\psi_1$ using SNR = 5  (dashed lines).
}
\label{fig::SNR2}
\end{figure}

{\bf Choosing the Bandwidth Using SNR.}
\myfigref{fig::SNR1} (line with circles) shows the signal-to-noise ratio for $\psi_1$ with $n=1000$, 
estimated by simulation. 
For each simulation, we also computed the 
bootstrap estimate of SNR and averaged this over the simulations.
The result is the line with triangles.
The dashed lines in \myfigref{fig::SNR2} represent bootstrap 
estimates of SNR for three typical data sets.
%each resampled $B=300$ times. 
The resulting 
$\hat\psi_{\widehat\varepsilon,1}$ 
using $SNR=5$ are shown to the right. 
For all three data sets, the bootstrap estimates of $\psi_1$ (dashed lines)
almost overlap the true eigenvector (solid line).

\begin{figure}
\begin{center}
\includegraphics[width=3.5in]{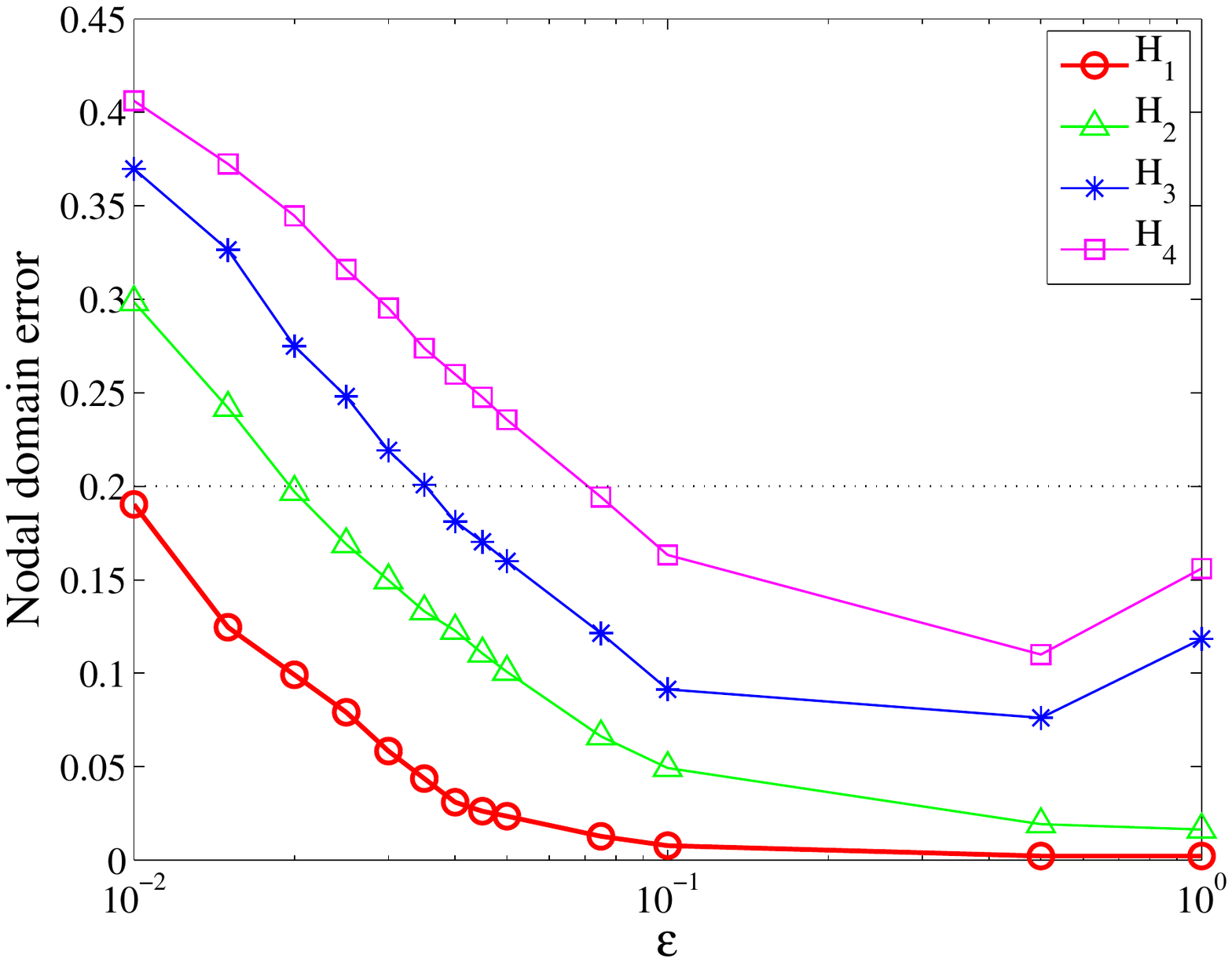}
\end{center}
\vspace{-3cm}
\caption{\footnotesize The nodal domain error $\mathbb{P}(\widehat{H}_{\ell}(X)\neq H_{\ell}(X))$
as a function of $\varepsilon$ for $\ell=1,2,3,4$.}
\label{fig::twonodal}
\end{figure}

{\bf Estimating the Nodal Domain.}
Now consider estimating
$H_{\ell}(x) = {\rm sign}(\psi_{\ell}(x))$.
\myfigref{fig::twonodal}
shows the nodal domain error for $H_{\ell}$ when $\ell=1,2,3,4$,
estimated by simulation. We see that the error is relatively small and 
stable over $\varepsilon$. As predicted by our results, large $\varepsilon$ can lead to very low error.
We can use the instability measure 
$\Xi(\varepsilon,\ell) = \mathbb{P}(\widehat{H}_{\ell}(X)\neq H_{\ell}(X))$,
where $\widehat{H}_{\ell}(x) = {\rm sign}(\widehat \psi_{\varepsilon,\ell}(x))$,
to choose $\varepsilon$ and $q$.
For example, find the smallest $\varepsilon$ and the largest 
number $q$ of eigenvectors such that
$\hat\Xi(\varepsilon,\ell)\leq \alpha$
for all $\ell \leq q$. (In this case, $\alpha=0.2$ 
approximately corresponds to $\varepsilon=0.075$ and $q=4$.)
We should caution the reader, however, that stability-based ideas have drawbacks.
In clustering, for example, \cite{Ben-David:EtAl:06} showed that
choosing the number of clusters based on stability can lead to poor clusters.

\subsection{Words}

The last example is an application of SCA to text data mining. The example
shows how one can measure the semantic association of words using diffusion distances, 
and how one can organize and form representative ``meta-words'' by eigenanalysis 
and quantization of the diffusion operator.

The data consist of $p=1161$ Science News articles. To encode the text, we extract 
$n=1004$ words (see~\cite{LafonLee2006} for details) and form a 
document-word information matrix. The mutual information between 
document $x$ and word $y$ is defined as
 $$I_{x,\,y}=\log \left( \frac{f_{x,\,y}}{\sum_{\xi} f_{\xi,\,y} \sum_{\eta} f_{\xi,\eta}} \right) \ ,$$
where $f_{x,y}=c_{x,y}/n$, and $c_{x,y}$ is the number of times word $y$ 
appears in document $x$. Let $$e_y = [I_{1,\,y}, I_{2,\,y}, \ldots I_{p,\,y}] \ .$$ be a 
p-dimensional feature vector for word $y$.

Our goal is to reduce both the dimension $p$ and the number of variables $n$, 
while preserving the main connectivity structure of the data. 
In addition, we seek a parameterization of the words 
that reflect how similar they are in meaning. Diffusion maps and diffusion coarse-graining 
 (quantization) offer a natural framework for achieving these objectives.

Define the weight matrix $\mathbb{K}(i,j) = \exp \left( - \frac{\|e_i-e_j\|^2}{4 \varepsilon} \right)$ 
for a graph with $n$ nodes. Let $\mathbb{A}_{\varepsilon,m}$ be the corresponding 
$m$-step transition matrix with eigenvalues 
$\lambda_{\ell}^m$ and eigenvectors $\psi_{\ell}$. Using the bootstrap, we estimate the 
SNR of $\psi_1$ as a function of $\varepsilon$ (\myfigref{fig::words_SNR}, left). 
A SNR cut-off at 2, gives the bandwidth $\varepsilon=150$. 
\myfigref{fig::words_SNR}, right, shows the spectral fall-off 
for this choice of $\varepsilon$. For $m=3$ and $q=12$, we have that  
$(\lambda_{q}/\lambda_1)^m<0.1$, i.e we can obtain a 
dimensionality reduction of a factor of about $1/100$ 
by the eigenmap $e_y \in \bbR^p \ \mapsto (\lambda_1^{m} \psi_1(y), \lambda_2^{m}
\psi_2(y),\ldots, \lambda_q^{m} \psi_q(y)) \in \bbR^q$ 
without losing much accuracy. Finally, to reduce the number of variables $n$, 
we form a quantized 
matrix  $\widetilde{\mathbb{A}}_{\varepsilon,m}$ for a coarse-grained random walk 
on a graph with $k<n$ nodes. It can be shown~\citep{LafonLee2006},
that the spectral properties of $\mathbb{A}_{\varepsilon,m}$ and 
$\widetilde{\mathbb{A}}_{\varepsilon,m}$ are similar when the 
coarse-graining (quantization) corresponds to $k$-means clustering 
in diffusion space.

{\myfigref{fig::word_map} shows the first two diffusion coordinates 
of the cluster centers (the ``meta-words'') for $k=100$. These representative words
 have roughly been rearranged 
according to their semantics and can be used as conceptual indices for 
document representation and text retrieval. Starting to the left, moving counter-clockwise, 
we here have words that express concepts in medicine, biology, earth sciences, physics, 
astronomy, computer science and social sciences. Table~\ref{table:metawords} 
gives examples of words in a cluster and the corresponding word centers. 
\comment{ The diffusion centers or ``meta-words'' form a coarse-grained 
representation of the word graph and can, for example, be used as conceptual indices 
for document retrieval and document clustering. This will be discussed in later work.
}
\begin{figure}
\begin{center}
\includegraphics[width=3.2in]{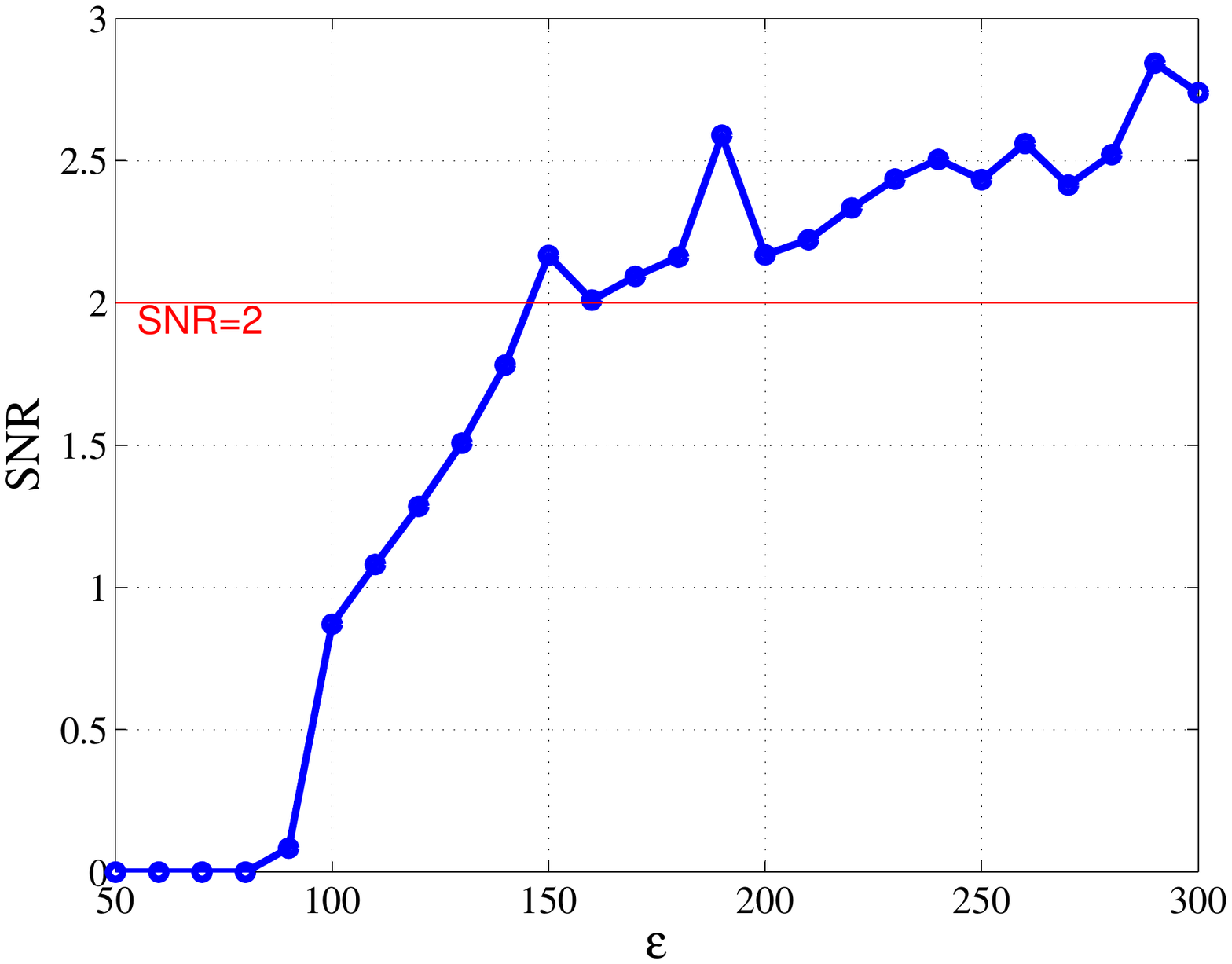}
\includegraphics[width=3.2in]{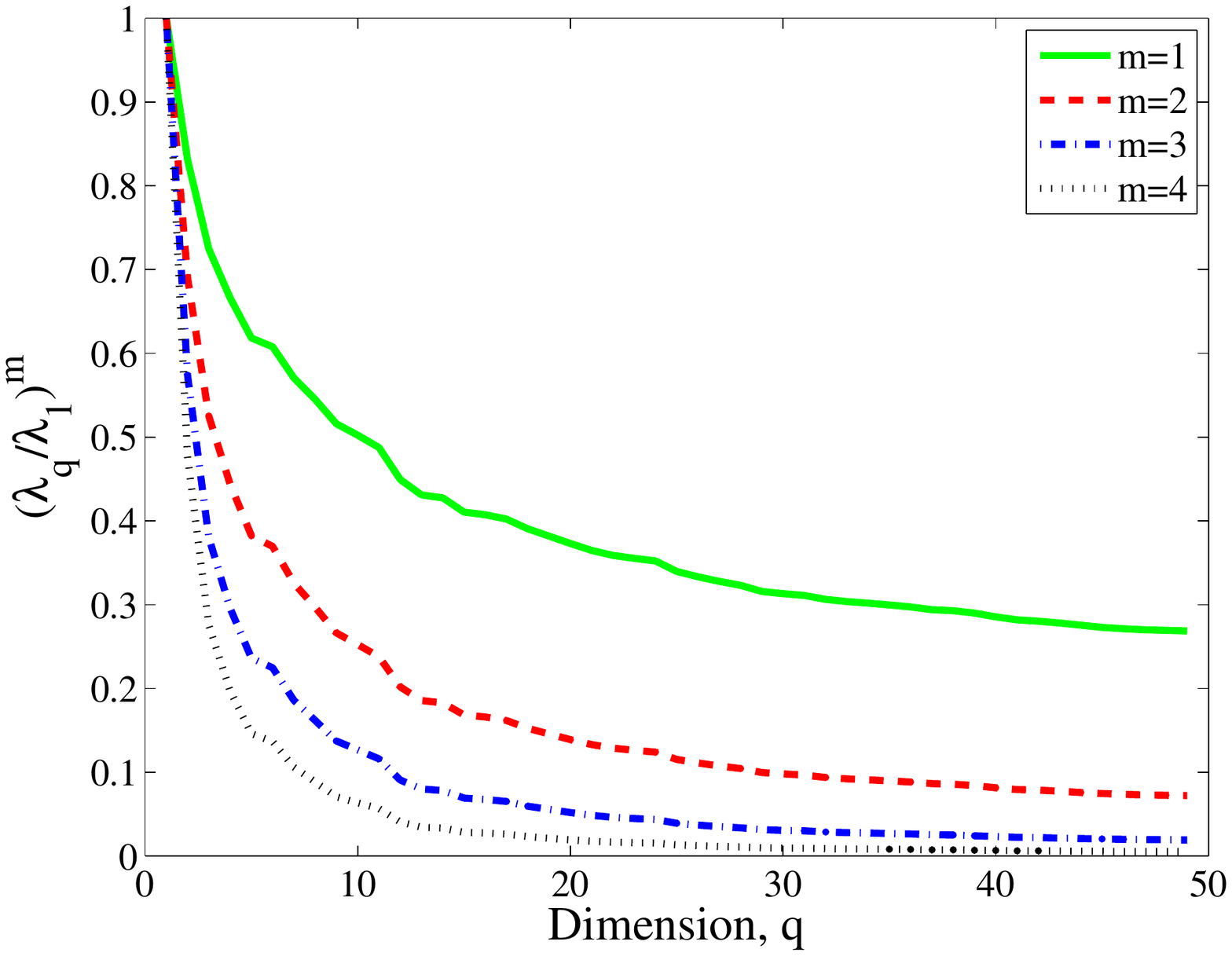}
\end{center}
\vspace{-3cm}
\caption{\footnotesize Left: Estimated SNR of $\psi_1$ as a function of $\varepsilon$. Right: Decay of 
the eigenvalues of $\mathbb{A}_{\varepsilon,m}$ for $\varepsilon=150$ and $m=1,2,3,4$.}
\label{fig::words_SNR}
\end{figure}

\begin{figure}
\begin{center}
\includegraphics[width=7in]{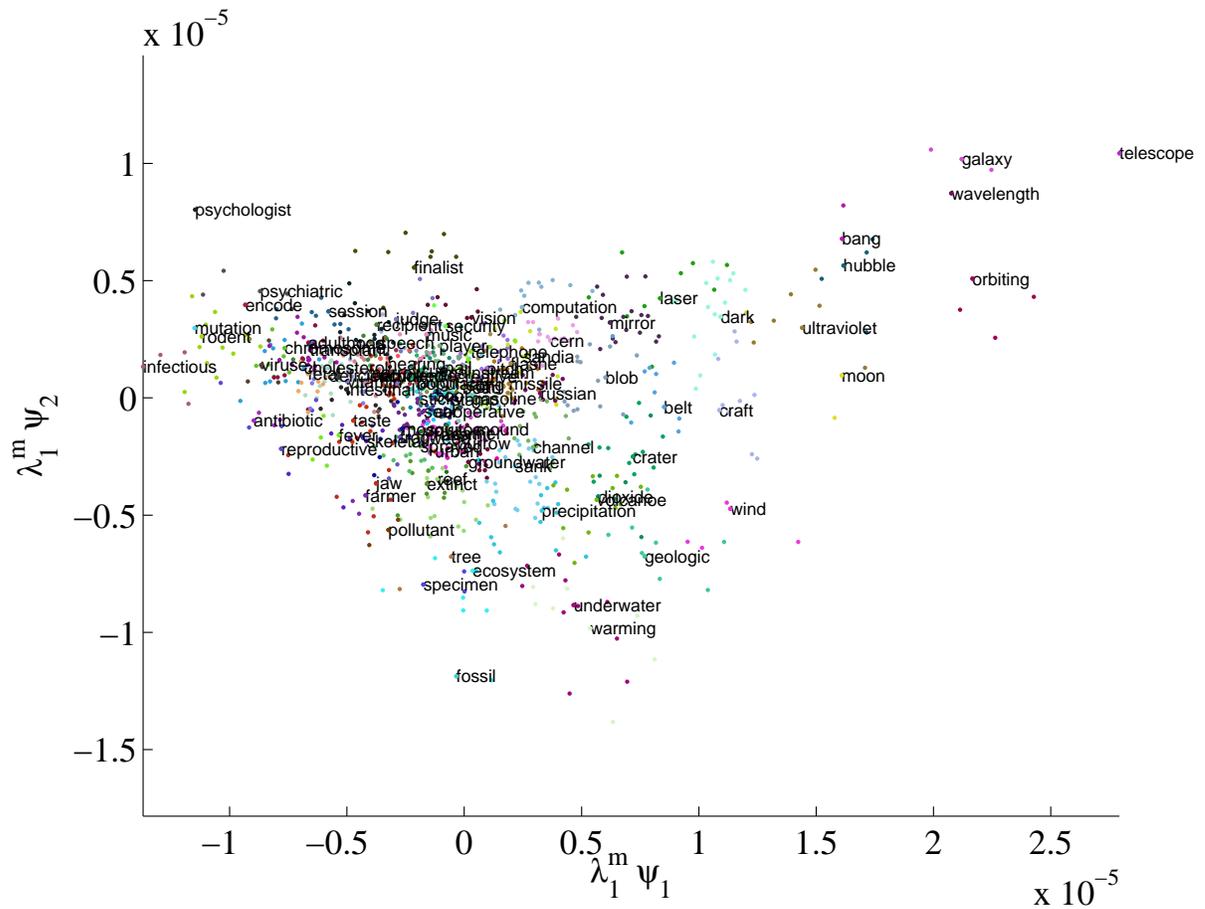}
\end{center}
\vspace{-6cm}
\caption{\footnotesize Parameterization and grouping of words using diffusion mapping. 
The text labels the representative word centers (meta-words) in each group. 
Note that the words 
are roughly arranged according to their semantic meaning.}
\label{fig::word_map}
\end{figure}

\begin{table}
\begin{center}
%\centering
{\caption{Examples of word groupings}\label{table:metawords}}
\begin{tabular}{|l|l|} \hline
Word center & Remaining words in group\\ \hline
 
{\footnotesize virus} & {\footnotesize aids, allergy, hiv, vaccine, viral} \\
{\footnotesize reproductive} & {\footnotesize fruit, male, offspring, reproductive, sex, sperm}\\
{\footnotesize vitamin} & {\footnotesize calory, drinking, fda, sugar, supplement, vegetable}\\
{\footnotesize fever} & {\footnotesize epidemic, lethal, outbreak, toxin}\\
{\footnotesize ecosystem} & {\footnotesize ecologist, fish, forest, marine, river, soil, tropical}\\
{\footnotesize warming}  & {\footnotesize climate, el, nino, forecast, pacific, rain, weather, winter}\\
{\footnotesize geologic} & {\footnotesize beneath, crust, depth, earthquake, plate, seismic, trapped, volcanic}\\
{\footnotesize laser} & {\footnotesize atomic, beam, crystal, nanometer, optical, photon, pulse, quantum, semiconductor}\\
{\footnotesize hubble} & {\footnotesize dust, gravitational, gravity, infrared}\\
{\footnotesize galaxy} & {\footnotesize cosmic, universe}\\
{\footnotesize finalist} & {\footnotesize award, competition, intel, prize, scholarship, student, talent, winner}
\\
 \hline
\end{tabular}

\end{center}
\end{table}

\section{Discussion}
\label{section::discussion}

Spectral methods are rapidly gaining popularity.
Their ability to reveal nonlinear structure
makes them ideal for complex, high-dimensional problems.
We have attempted to provide insight into these techniques
by identifying the population quantities that are being
estimated and studying their large sample properties.

Our analysis shows that spectral kernel methods in most cases 
have a convergence 
rate similar to classical non-parametric smoothing.
Laplacian-based kernel methods, for example, use the same smoothing 
operators as in traditional nonparametric regression. The end goal however 
is not smoothing, but data parameterization and structure definition of data. 
Spectral methods exploit the fact that the eigenvectors of
 local smoothing operators provide information on the underlying geometry and 
 connectivity of the data.
 
 We close by briefly mention how SCA and diffusion maps also 
can be used in clustering, density estimation and regression.
The full details of these applications will be reported in separate 
papers.

\subsection{Clustering}

One approach to clustering 
is spectral clustering. %~\citep{Luxburg:2007}.
The idea is to reparameterize the data
using the first few nontrivial eigenvectors
$\psi_1,\ldots,\psi_m$
and then apply a standard clustering algorithm such as
$k$-means clustering.
This can be quite effective for
finding non-spherical clusters.

Diffusion distances can also be used for this purpose.
Recall that ordinary $k$-means clustering
seeks to find points 
$C = \{c_1,\ldots, c_k\}$ to minimize
the empirical distortion
$$
\hat\Delta(C)=\frac{1}{n}\sum_{i=1}^n \|X_i - Q_C(X_i)\|^2
$$
where $Q_C$ is the quantization map that takes $X_i$
to the closest $c_j\in C$.
The empirical distortion estimates the population distortion
$$
\Delta(C) = \mathbb{E}\|X -  Q_C(X)\|^2.
$$
By using $k$-means in diffusion map coordinates %~\citep{LafonLee2006}
we instead minimize
$$
\Delta(C) = \mathbb{E} D_t(X,Q_t(X))
$$
where $Q_t(x) = {\rm argmin}_{c\in C}D_t(x,c)$.
The details are in~\cite{Lee:Wasserman:2008}.

\subsection{Density Estimation}

If $Q$ is a quanization map then
the quantized density estimator 
\citep{meinrit:2001} 
is
$$
\hat{p}(x)=\frac{1}{n}\sum_{i=1}^n \frac{1}{h^d} K_h(\|x - 
Q(X_i)\|).
$$
For highly clustered data, the quantized density estimator can 
have smaller mean squared error than the usual kernel density 
estimator.
Similarly, we can define  
the quantized diffusion density estimator as
$$
\hat{p}(x)=\frac{1}{n}\sum_{i=1}^n \frac{1}{h^d} K(\hat{D}_t(x,X_i))
$$
which can potentially have small mean squared error
for appropriately chosen $t$. See~\cite{Buchman:EtAl:2008} for an application 
to density estimation of hurricane tracks in the Atlantic Ocean.

\subsection{Regression}

A common method for nonparametric regression is to
expand the regression function
$m(x) = \mathbb{E}(Y|X=x)$ in a basis
and then estimate the coefficients of the expansion
from the data. Usually, the basis is chosen beforehand.
The diffusion map basis provides a natural data-adaptive basis
for doing nonparametric regression.
We expand $m(x) = \mathbb{E}(Y|X=x)$ as
$m(x) = \sum_j \beta_j \psi_j(x)$.
Let $\hat{m}(x) = \sum_{j=1}^q \hat\beta_j \hat\psi_{\varepsilon,j}(x)$
where $q$ and $\varepsilon$ are chosen by cross-validation. 
See~\cite{Richards:EtAl:2008} for an application to astronomy data 
and spectroscopic redshift prediction.

\section{Appendix}

\subsection{Spectral Decomposition and Euclidean Distances in 
Diffusion Space}

\label{appendix:diffusion}
In this section, we describe how a symmetric operator $\tilde{A}$, the 
stochastic differential operator $A$ and 
its adjoint (the Markov operator) $A^{*}$ are related, and how these 
relations lead to different normalization schemes for the 
corresponding eigenvectors. (For ease of notation, we have omitted the 
subindex ${\varepsilon}$, since we here consider a fixed $
\varepsilon>0$.) We also 
show that the diffusion metric 
corresponds to a weighted Euclidean distance in the embedding space 
induced by the diffusion map.

Suppose that $P$ is a probability measure with a compact support $
\mathcal{X}$. %$X_1, \ldots , X_n$ is an IID sample
Let $k:\mathcal{X} \times \mathcal{X}$ be a similarity function that
is symmetric, continuous, and positivity-preserving, i.e. $k(x,y)>0$
for all $x,y \in \mathcal{X}$. For simplicity, we assume in addition
that $k$ is positive semi-definite, i.e. for all bounded functions
$f$ on $\mathcal{X}$, $\int_\mathcal{X}\int_\mathcal{X} k(x,y) f(x)
f(y) dP(x) dP(y) \geq 0$. Consider two different normalization
schemes of $k$:
%\begin{equation}
$$\begin{array}{llll}
\tilde{a}(x,y) &=&\frac{k(x,y)}{\sqrt{\rho(x)} \sqrt{p(y)}}& \ \ \ \ \ 
\rm{(symmetric)}\\
a(x,y) &=&\frac{k(x,y)}{\rho(x)}& \ \ \ \ \ \rm{(stochastic)}
\end{array}
$$ %\end{equation}
where $\rho(x)=\int k(x,y)dP(y)$.

Define the {\em symmetric} integral operator $\tilde{A}$ by
%\begin{equation} %\tilde{A}: L^2(\mathcal{X}; dP) \rightarrow L^2(\mathcal{X}; dP), \ \ \ 
$$\tilde{A} f(x) = \int_{\mathcal{X}}
\tilde{a}(x,y) f(y) dP(y) .\label{eq:A_sym}$$%\end{equation} 
Under the
stated conditions, $k(x,y)$ is an $L^2$-kernel. It follows that
$\tilde{A}$ is a self-adjoint compact operator. The eigenvalues
$\{\lambda_{\ell}\}_{\ell \geq 0}$ of $\tilde{A}$ are real and the 
associated
eigenfunctions $\{v_{\ell}\}_{\ell \geq 0}$ form an orthonormal basis of
$L^2(\mathcal{X}; dP)$. According to Mercer's theorem, we have the
spectral decomposition
\begin{equation}
\tilde{a}(x,y) = \sum_{\ell \geq 0} \lambda_{\ell} v_{\ell}(x)
v_{\ell}(y) , \label{eq:Mercer} \end{equation} 
where the series on the
right converges uniformly and absolutely to $\tilde{a}(x,y)$.

Now consider the integral operator $A$ and its adjoint 
(the Markov operator) $A^{*}$:
%\begin{equation}
$$\begin{array}{lll}
%A: L^2(\mathcal{X}; p dP) \rightarrow L^2(\mathcal{X};p dP), \ \ \  & 
A f(x) &=& \int_{\mathcal{X}} a(x,y) f(y)  dP(y)\\
% A^{*}: L^2(\mathcal{X};  dP/p) \rightarrow L^2(\mathcal{X}; dP/p), \ \ \  & 
A^{*}f(x) &=& \int_{\mathcal{X}} f(y) a(y,x)  dP(y) ,\\
\end{array}$$
%\end{equation}
where $\langle Af,g \rangle_{L^2(\mathcal{X}; dP)}=\langle f,A^{*}g 
\rangle_{L^2(\mathcal{X}; dP)}$. Let $\ s(x)=\rho(x)/\int \rho(y) dP(y)$. 
If $\tilde{A} v_{\ell} = \lambda_{\ell} v_{\ell}$, then we have the 
corresponding eigenvalue equations
%$A \psi_{\ell} = \lambda_{\ell} \psi_{\ell}$ and $A^{*} \varphi_{\ell} = 
%\lambda_{\ell}  \varphi_{\ell}$, where
\begin{equation} 
A \psi_{\ell} = \lambda_{\ell} \psi_{\ell}, \ \  {\rm where} \ \
\psi_{\ell}(x)=v_{\ell}(x)/\sqrt{s(x)}
\label{eq:eigen_A} \end{equation}
and
\begin{equation} 
A^{*} \varphi_{\ell} = \lambda_{\ell}  \varphi_{\ell}, \ \  {\rm
where} \ \ \varphi_{\ell}(x)=v_{\ell}(x)\sqrt{s(x)}.
\label{eq:eigen_Aadjoint} \end{equation}
Moreover, if $\{v_{\ell}\}_{\ell
\geq 0}$ is an orthonormal basis of $L^2(\mathcal{X}; dP)$, then the
sets $\{\psi_{\ell}\}_{\ell \geq 0}$ and $\{\varphi_{\ell}\}_{\ell \geq 0}$ form
orthonormal bases of the {\em weighted} $L^2$-spaces 
$L^2(\mathcal{X}; s dP)$ and
$L^2(\mathcal{X}; dP/s)$, respectively.  The operator
$A$ preserves constant functions, i.e. $A 1=1$. One can also show that
the matrix norm $\|\tilde{A}\|=\sup_{f \in L^2(\mathcal{X}; dP)}
\frac{\|\tilde{A}f\|}{\|f\|}=1$. Thus, the eigenvalue
$\lambda_0=1$ is the largest eigenvalue of the operators $A$ and
$A^{*}$. The corresponding eigenvector of $A$ is $\psi_0=1$, 
and the corresponding eigenvector of $A^{*}$ is $\varphi_0=s$.
%is proportional to the approximate density function $\rho$.

%The corresponding eigenvector $\psi_0$ of $A$ is a constant
%function, and the corresponding eigenvector $\varphi_0$ of $A^{*}$
%is proportional to the approximate density function $\rho$.

From \myeqref{eq:Mercer}, it follows that
%\begin{equation} 
$$a(x,y) = \sum_{\ell \geq 0} \lambda_{\ell} \psi_{\ell}(x)
\varphi_{\ell}(y) , 
\label{eq:biorthogonal_decomp} 
$$%\end{equation} 
where 
$\|\varphi_{\ell}\|_{L^2(\mathcal{X}; dP/s)}=\|\psi_{\ell}\|
_{L^2(\mathcal{X}; s dP)}=1$ 
for all $\ell \geq 0$, and 
$\langle \varphi_{k}, \psi_{\ell} \rangle_{L^2(\mathcal{X}; dP)}=0$ for $k 
\neq \ell$. 
More generally, if $a_m(x,y)$ is the kernel of the $m^{\rm th}$ iterate
$A^{m}$, where $m$ is a positive integer, then
\begin{equation} 
a_m(x,y) = \sum_{\ell \geq 0} \lambda_{\ell}^{m} \psi_{\ell}(x)
\varphi_{\ell}(y) . 
\label{eq:biorthogonal_decomp_t}
\end{equation} 
\comment{For a
countable set $\mathcal{X}$, one can interpret $a(x,y)$ as the
transition kernel of a homogeneous Markov chain. Element
$a^{t}(x,y)$ then represents the probability of a transition from
state $x$ to $y$ in $t$ time steps, and the eigenvector $\varphi_0$
is the unique stationary distribution of the chain.
(Question: What is the interpretation for a general state space $
\mathcal{X}$?)}

We define a one-parametric family of diffusion distances between points 
$x$ and $z$ according to
\begin{equation}
D_m^2(x,z) \equiv\|a_m(x,\cdot)-a_m(z,\cdot) \|^2_{L^2(\mathcal{X}; dP/
s)} , 
\label{eq:diffusion_dist1}
\end{equation}
where the parameter $m$ determines the scale of the analysis. The 
diffusion metric measures the rate of connectivity between points on a 
data set. It will be 
small if there are many paths of lengths less than or equal to $2m$ 
between the two points, and it will be large if the number of connections 
is small. One 
can see this clearly by expanding the expression in 
\myeqref{eq:diffusion_dist1} so that \begin{equation}D_m^2(x,z)=\frac{a_{2m}(x,x)}
{s(x)} + \frac{a_{2m}(z,z)}
{s(z)}-\left(\frac{a_{2m}(x,z)}{s(z)} + \frac{a_{2m}(z,x)}{s(x)} 
\right). \label{eq:mod_polarization}\end{equation} The quantity $D_m^2(x,z)$ is small when the transition 
probability 
densities $a_{2m}(x,z)$ and  $a_{2m}(z,x)$ are large.

Finally, we look for an embedding where Euclidean distances reflect the 
above diffusion metric. The biorthogonal decomposition in 
\myeqref{eq:biorthogonal_decomp_t} can
be viewed as an orthogonal expansion of the functions
$a_m(x,\cdot)$ with respect to the orthonormal basis
$\{\varphi_{\ell}\}_{\ell \geq 0}$ of $L^2(\mathcal{X}; dP/s)$; the 
expansion coefficients are given by $\{\lambda_{\ell}^{m} \psi_{\ell}(x) 
\}_{\ell 
\geq 0}$. Hence,
%\begin{equation}
$$D_m^2(x,z)=\sum_{\ell \geq 0}
(\lambda_{\ell}^{m}\psi_{\ell}(x)-\lambda_{\ell}^{m}\psi_{\ell}(z))^2 =
\|\Psi_m(x)-\Psi_m(z) \|^2, \label{eq:diffusion_dist2}$$%\end{equation}
where $\Psi_m: x \mapsto (\lambda_1^{m} \psi_1(x), \lambda_2^{m}
\psi_2(x),\ldots)$ 
is the diffusion map of the data at time step $m$. \comment{In other 
words, the distance $D_t(x,z)$ in
(\ref{eq:diffusion_dist1})
is a {\em Euclidean distance} in the
eigenmap space defined by the eigenvalues and eigenfunctions of the
operator $A$.}

\subsection{Proofs}

{\bf Proof of Theorem \ref{thm::main}.}
From Theorem 2 below, we have that
$$
\|A_t(\varepsilon_n,\hat{P}_n) - \mathbf{A}_t \| =
\left(O_P\left(\sqrt{\frac{ \log(1/\varepsilon_n)}
                          {n \varepsilon_n^{(d+4)/2}}}\right) + 
O(\varepsilon_n)\right)\times  \rho(t).
$$
Hence,
\begin{eqnarray*}
\|A_t(\varepsilon_n,q,\hat{P}_n) - \mathbf{A}_t \| & \leq &
\|A_t(\varepsilon_n,q,\hat{P}_n) - A_t(\varepsilon_n,\hat{P}_n)\| + 
\|A_t(\varepsilon_n,\hat{P}_n) - \mathbf{A}_t \|\\
&=&
\|A_t(\varepsilon_n,q,\hat{P}_n) - A_t(\varepsilon_n,\hat{P}_n)\| + 
\left(O_P\left(\sqrt{\frac{ \log(1/\varepsilon_n)}
                          {n \varepsilon_n^{(d+4)/2}}}\right) + 
O(\varepsilon_n)\right)\times  \rho(t)\\
&=&
\| \sum_{q+1}^\infty \hat\lambda_{\varepsilon_n,\ell}^{t/\varepsilon_n}
\hat\Pi_{\varepsilon_n,\ell}\| +
\left(O_P\left(\sqrt{\frac{ \log(1/\varepsilon_n)}
                          {n \varepsilon_n^{(d+4)/2}}}\right) + 
O(\varepsilon_n)\right)\times  \rho(t)\\
& \leq &
\sum_{q+1}^\infty \hat\lambda_{\varepsilon_n,\ell}^{t/\varepsilon_n} +
\left(O_P\left(\sqrt{\frac{ \log(1/\varepsilon_n)}
                          {n \varepsilon_n^{(d+4)/2}}}\right) + 
O(\varepsilon_n)\right)\times  \rho(t).
\end{eqnarray*}
Now we bound the first sum.
Note that,
$$
\sup_\ell |\hat\nu_{\varepsilon_n,\ell}^2 - \nu_{\varepsilon_n,\ell}^2|  =
\sup_\ell \frac{|\hat\lambda_{\varepsilon_n,\ell} - \lambda_{\varepsilon_n,\ell}|}
{\varepsilon_n}
 \leq 
\frac{\|\hat{A}_{\varepsilon_n} - A_{\varepsilon_n}\|}{\varepsilon_n}=
O_P(\gamma_n)
$$
where
$$
\gamma_n =
\sqrt{\frac{ \log (1/\varepsilon_n)}{n  \varepsilon_n^{(d+4)/2}}}.
$$
By a Taylor series
expansion,
$$
G_{\varepsilon_n} f = \mathbf{G} f + O(\varepsilon_n)
$$
uniformly for $f\in {\cal F}$.
(This is the same calculation used to compute the bias in kernel 
regression.
See also, \cite{Gine:Koltchinskii:2006} and \cite{Singer:06}).
So,
$$
\sup_\ell |\nu_{\varepsilon_n,\ell}^2 - \nu_{\ell}^2| \leq
\|G_{\varepsilon_n} - \mathbf{G}\| = O(\varepsilon_n).
$$
Therefore,
\begin{eqnarray*}
\sum_{q+1}^\infty \hat\lambda_{\varepsilon_n,\ell}^{t/\varepsilon_n} &=&
\sum_{q+1}^\infty (1- \varepsilon_n \hat\nu_{\varepsilon_n,\ell}^2)^{t/\varepsilon_n}\\
&=& \sum_{q+1}^\infty 
\exp\left\{ \frac{t}{\varepsilon_n} \log (1- \varepsilon_n \hat\nu_{\varepsilon_n,\ell}^2)\right\}\\
&=& \sum_{q+1}^\infty 
\exp\left\{ \frac{t}{\varepsilon_n} 
\log (1- \varepsilon_n [O_P(\gamma_n) + O(\varepsilon_n) - \nu_\ell^2])\right\}\\
&=&
(1+ O_P(\gamma_n) + O(\varepsilon_n))
\sum_{q+1}^\infty e^{-\nu_\ell^2 t}.
\end{eqnarray*}
In conclusion,
\begin{eqnarray*}
\|A_t(\varepsilon_n,q,\hat{P}_n) - \mathbf{A}_t \| &=&
\left(O_P\left(\sqrt{\frac{ \log(1/\varepsilon_n)}
                          {n \varepsilon_n^{(d+4)/2}}}\right) +
O(\varepsilon_n)\right)\times  \rho(t) \\
&&\ \ \ +\ \ 
\Biggl(1+ O_P(\gamma_n) + O(\varepsilon_n)\Biggr)
\sum_{q+1}^\infty e^{-\nu_\ell^2 t}\\
& \leq &
\rho(t)
\left(O_P\left(\sqrt{\frac{ \log(1/\varepsilon_n)}
                          {n \varepsilon_n^{(d+4)/2}}}\right) +
O(\varepsilon_n)\right)\ + \ 
\sum_{q+1}^\infty e^{-\nu_\ell^2 t}.
\end{eqnarray*}

\vspace{1cm}

{\bf Proof of Theorem \ref{theorem::alt}.}
Recall that
$A_t(\varepsilon_n,\hat{P}_n)  = e^{t (\hat{A}_{\varepsilon_n} - I)/
\varepsilon_n}$.
From lemma 1,
$\|A_\varepsilon - \hat{A}_\varepsilon\| = \gamma(\varepsilon)$
where
$$
\gamma(\varepsilon) =
O_P\left(\sqrt{\frac{ \log(1/\varepsilon_n)}{n \varepsilon_n^{d/2}}}\right).
$$
Hence,
\begin{eqnarray*}
\frac{\hat{A}_\varepsilon - I}{\varepsilon} &=&
\frac{\hat{A}_\varepsilon - A_\varepsilon}{\varepsilon} +
\frac{A_\varepsilon - I}{\varepsilon} =
\frac{\gamma(\varepsilon)}{\varepsilon}  + \mathbf{G} + O(\varepsilon)
\end{eqnarray*}
and so
$$
A_t(\varepsilon,\hat{P}_n)  =
\mathbf{A}_t e^{t(\gamma(\varepsilon)+ O(\varepsilon^2)/\varepsilon} =
\mathbf{A}_t \Biggl[ I + t\bigl(\gamma(\varepsilon)+ O(\varepsilon^2)\bigr)/
\varepsilon + 
o\bigl(t(\gamma(\varepsilon)+ O(\varepsilon))\bigr)/\varepsilon\Biggr]
$$
Therefore,
\begin{eqnarray*}
\|\mathbf{A}_t - A_t(\varepsilon,\hat{P}_n) \| &=&
 \|\mathbf{A}_t\| 
\left(O_P\left(
\sqrt{\frac{ \log(1/\varepsilon_n)}{n \varepsilon^{(d+4)/2}}}\right)
+O(\varepsilon)\right)\\
& \leq &
\left(O_P\left(\sqrt{\frac{ \log(1/\varepsilon_n)}{n \varepsilon^{(d+4)/2}}}
\right) +
O(\varepsilon)\right)\ 
\sum_{\ell=1}^\infty e^{-\nu_\ell^2 t}.\ \Box
\end{eqnarray*}

\begin{lemma}\label{lemma::sup}
Let $\varepsilon_n\to 0$ and $n \varepsilon_n^{d/2}/
\log(1/\varepsilon_n) \to \infty$.
Then
$$
\|A_\varepsilon  - \hat{A}_\varepsilon \| =
O_P\left(\sqrt{\frac{\log(1/\varepsilon_n)}{n \varepsilon_n^{d/2}}}\right).
$$
\end{lemma}

{\bf Proof.}
Uniformly, for all $f \in {\cal F}$, and all $x$ in the support of $P$,
$$
|A_\varepsilon f(x) - \hat{A}_\varepsilon f(x)| \leq
|A_\varepsilon f(x) - \tilde{A}_\varepsilon f(x)| +
|\tilde{A}_\varepsilon f(x) - \hat{A}_\varepsilon f(x)|
$$
where
$\tilde{A}_\varepsilon f(x) = \int \hat{a}_\varepsilon(x,y)f(y) dP(y).$
\relax From \cite{Gine:Guillou:02},
$$
\sup_x \frac{| \hat{p}_\varepsilon(x) - p_\varepsilon(x)|}
            {| \hat{p}_\varepsilon(x)p_\varepsilon(x)|} =
  O_P\left(\sqrt{\frac{\log (1/\varepsilon_n)}{n \varepsilon_n^{d/2}}}\right).
$$
Hence,
\begin{eqnarray*}
|A_\varepsilon f(x) - \tilde{A}_\varepsilon f(x)| & \leq &
\frac{| \hat{p}_\varepsilon(x) - p_\varepsilon(x)|}
{| \hat{p}_\varepsilon(x)p_\varepsilon(x)|}\int |f(y)| k_
\varepsilon(x,y)dP(y) \\
&= &  O_P\left(\sqrt{\frac{\log (1/\varepsilon_n)}{n \varepsilon_n^{d/2}}}
\right) 
\int |f(y)| k_\varepsilon(x,y)dP(y)\\
& = &
O_P\left(\sqrt{\frac{\log (1/\varepsilon_n)}{n \varepsilon_n^{d/2}}}\right). 
\end{eqnarray*}
Next, we bound
$\tilde{A}_\varepsilon f(x) - \hat{A}_\varepsilon f(x)$.
We have
\begin{eqnarray*}
\tilde{A}_\varepsilon f(x) - \hat{A}_\varepsilon f(x) & = &
\int f(y) \hat{a}_\varepsilon(x,y) (d\hat{P}_n(y) - dP(y))\\
&=& \frac{1}{p(x)+o_P(1)}\int f(y) k_\varepsilon(x,y)(d\hat{P}_n(y) - 
dP(y)).
\end{eqnarray*}
Now, expand
$f(y) = f(x) + r_n(y)$
where $r_n(y) = (y-x)^T\nabla f(u_y)$ and
$u_y$ is between $y$ and $x$.
So,
$$
\int f(y) k_\varepsilon(x,y)(d\hat{P}_n(y) - dP(y)) =
f(x) \int  k_\varepsilon(x,y)(d\hat{P}_n(y) - dP(y))  +
 \int r_n(y) k_\varepsilon(x,y)(d\hat{P}_n(y) - dP(y)) .
$$
By an application of Talagrand's inequality
to each term, as in
Theorem 5.1 of Gin\'e and Koltchinskii (2006),
we have
$$
\int f(y) k_\varepsilon(x,y)(d\hat{P}_n(y) - dP(y)) =
O_P\left(\sqrt{\frac{\log (1/\varepsilon_n)}{n \varepsilon_n^{d/2}}}\right) .
$$
Thus,
$$
\sup_{f\in {\cal F}}\| \hat{A}_\varepsilon f - A_\varepsilon f \|_\infty = 
O_P\left(\sqrt{\frac{\log (1/\varepsilon_n)}{n \varepsilon_n^{d/2}}}\right)
$$
This also holds uniformly over
$\{f\in {\cal F}: \|f\|=1\}$.
Moreover,
$\| \hat{A}_\varepsilon f - A_\varepsilon f \|_2 \leq 
C \| \hat{A}_\varepsilon f - A_\varepsilon f \|_\infty$
for some $C$
since $P$ has compact support.
Hence,
\begin{eqnarray*}
\sup_{f\in {\cal F}} \frac{\| \hat{A}_\varepsilon f - A_\varepsilon f \|_2}{\|f\|} 
=
\sup_{f\in {\cal F}, \|f\|=1} \| \hat{A}_\varepsilon f - A_\varepsilon f \|_2 =
O_P\left(\sqrt{\frac{\log (1/\varepsilon_n)}{n \varepsilon_n^{d/2}}}\right). 
\Box
\end{eqnarray*}

\vspace{1cm}

%\begin{lemma}\label{lemma::var}
%$$
%\left\|\sum_{\ell=0}^{q} (\hat{\Pi}_{\varepsilon,\ell} - {\Pi}_{\varepsilon,
%\ell} )\right\| = 
%O_P\left(\sqrt{\frac{\log (1/\varepsilon_n)}{n \delta(q) \varepsilon_n^{(d
%+4)/2}}}\right)
%$$
%where $\delta(q)=\frac{1}{2}(\nu_{q+1}^2-\nu_q^2)$ and
%$$
%\max_{0\leq \ell \leq q}\|\hat\Pi_{\varepsilon,\ell} - \Pi_{\varepsilon,\ell}\| = 
%O_P\left(\sqrt{\frac{q^2 \log (1/\varepsilon_n)}
%                   {n \underline{\delta}(q) \varepsilon_n^{(d+4)/2}}}\right)
%$$ 
%where $\underline{\delta}(q)=\min_{0\leq \ell \leq q}\frac{1}{2}(\nu_{\ell
%+1}^2-\nu_\ell^2)$.
%%For $\ell=0,1,\ldots$,
%%$$
%%\|\hat\psi_{\varepsilon,\ell} - \psi_{\varepsilon,\ell}\| = 
%%O_P\left(\sqrt{\frac{1}{n^{4/(4+d)}}}\right)
%%$$
%%and so
%%$$
%%\max_{0\leq \ell \leq q_n}\|\hat\psi_{\varepsilon,\ell} - \psi_{\varepsilon,\ell}\| = 
%%O_P\left(\sqrt{\frac{q_n^2}{n^{4/(4+d)}}}\right).
%%$$
%\end{lemma}
%
%
%
%{\bf Proof.}
%We will use Theorem 3 of Zwald and Blanchard (2005)
%which states that, for operators $A$ and $B$,
%\begin{equation}\label{eq::zwald}
%\| \Pi^q(A) - \Pi^q(A+B)\| \leq \frac{\|B\|}{\delta}
%\end{equation}
%where
%$\Pi^q$ is the projector on the subspace spanned by the first $q$ 
%eigenvectors
%and $\delta= (\lambda_q-\lambda_{q+1})/2$. 
%The conclusion follows from Lemma \ref{lemma::sup},
%(\ref{eq::zwald}) and 
%the perturbation expansion 
%$\lambda_{\varepsilon,q} = 
%1-\varepsilon \nu_{\varepsilon,q}^2 =1-\varepsilon \nu_{q}^2 + 
%O(\varepsilon^2)$. 
%$\Box$ 
%
%

\vspace{1cm}

{\bf Proof of Theorem \ref{theorem::lownoise}.}
Let
$A_n = \{ |\psi_1(X)| \leq \delta_n\}$.
Then
$$
A_n^c \bigcap 
\Bigl\{ \hat{H}(X) \neq H(X)\Bigr\}\ \ \ 
{\rm implies\ that\ \ }\ 
\Bigl\{|\hat\psi_{\varepsilon,1}(X) - \psi_1(X)| > \delta_n \Bigr\}.
$$
Also,
$\sup_x | \psi_1(x) - \psi_{\varepsilon,1}(x)| \leq c \varepsilon_n$
for some $c>0$.
Hence,
\begin{eqnarray*}
\mathbb{P}\left(\hat{H}(X) \neq H(X)\right)&=&
\mathbb{P}\left(\hat{H}(X) \neq H(X),A_n\right) + 
\mathbb{P}\left(\hat{H}(X) \neq H(X),A_n^c\right)\\
& \leq &
\mathbb{P}(A_n) + 
\mathbb{P}\left(\hat{H}(X) \neq H(X),A_n^c\right)\\
& \leq &
C \delta_n^\alpha + 
\mathbb{P}\left(|\psi_1(X) - \hat\psi_{\varepsilon,1}(X)| > \delta_n\right)\\
& \leq &
C \delta_n^\alpha + 
\mathbb{P}\left(|\psi_1(X) - \psi_{\varepsilon,1}(X)| + 
|\psi_{\varepsilon,1}(X) -\hat\psi_{\varepsilon,1}(X)| > \delta_n\right)\\
& \leq &
C \delta_n^\alpha + 
\mathbb{P}\left(|\hat\psi_{\varepsilon,1}(X)-\psi_{\varepsilon,1}(X)|+c 
\varepsilon_n  > 
\delta_n\right)\\
& = &
C \delta_n^\alpha + 
\mathbb{P}\left(|\hat\psi_{\varepsilon,1}(X)-\psi_{\varepsilon,1}(X)|  > 
\delta_n- c\varepsilon_n \right)\\
& \leq &
C \delta_n^\alpha + 
\frac{\mathbb{E}\|\hat\psi_{\varepsilon,1}(X)-\psi_{\varepsilon,1}(X)\|}
{\delta_n- c\varepsilon_n}\\
& \leq &
C \delta_n^\alpha + 
O_P\left(\sqrt{\frac{\log(1/\varepsilon_n)}{n \varepsilon_n^{(d+4)/2}}}
\right)
\frac{1}{\delta_n- c\varepsilon_n}\\
\end{eqnarray*}
Set
$\delta = 2c\varepsilon_n$
and
$\varepsilon_n = n^{-2/(4\alpha + d + 8)}$
and so
$$
\mathbb{P}\left(\hat{H}(X) \neq H(X)\right) \leq
n^{ - \frac{2\alpha}{4\alpha +8 + d }}.\ \ \Box
$$

\
%& \leq &
%2 C \int_{S_j} |\omega_{\epsilon,m}(u|x) - s_{\epsilon}(u))| du\\
%& \leq & 2 C \gamma < \delta.
%\end{eqnarray*}
%Hence, $D_{\epsilon,m}^2(x,y)  < \delta$.
%
%Now let $x \in S_j$ and $y\in S_k$.
%Then,
%\begin{eqnarray*}
%D_{\epsilon,m}^2(x,y) &=&
%\int_{S_j} \omega_{\epsilon,m}^2(u|x)  \frac{p(u)}{s_\epsilon(u)} du + 
%\int_{S_k} \omega_{\epsilon,m}^2(u|y)  \frac{p(u)}{s_\epsilon(u)} du \\
%& \geq &
%\int_{S_j} \omega_{\epsilon,1}^2(u|x)  \frac{p(u)}{s_\epsilon(u)} du + 
%\int_{S_k} \omega_{\epsilon,1}^2(u|y)  \frac{p(u)}{s_\epsilon(u)} du \\
%& \geq &
%2 \min_j \int_{S_j} \omega_{\epsilon,1}^2(u|x)  \frac{p(u)}{s_
%\epsilon(u)} du  \equiv M.
%\end{eqnarray*}
%
%

\bibliographystyle{chicago}
\footnotesize 
\bibliography{paper}

\begin{thebibliography}{}

\bibitem[\protect\citeauthoryear{Audibert and Tsybakov}{Audibert and
  Tsybakov}{2007}]{Audibert:Tsybakov:2007}
Audibert, J.-Y. and A.~B. Tsybakov (2007).
\newblock Fast learning rates for plug-in classifiers.
\newblock {\em Annals of Statistics\/}~{\em 35\/}(2), 608--633.

\bibitem[\protect\citeauthoryear{Belkin and Niyogi}{Belkin and
  Niyogi}{2003}]{BelkinNiyogi03}
Belkin, M. and P.~Niyogi (2003).
\newblock Laplacian eigenmaps for dimensionality reduction and data
  representation.
\newblock {\em Neural Computation\/}~{\em 6\/}(15), 1373--1396.

\bibitem[\protect\citeauthoryear{Belkin and Niyogi}{Belkin and
  Niyogi}{2005}]{Belkin:Niyogi:05}
Belkin, M. and P.~Niyogi (2005).
\newblock Towards a theoretical foundation for {L}aplacian-based manifold
  methods.
\newblock In {\em Proc. COLT}, Volume 3559, pp.\  486--500.

\bibitem[\protect\citeauthoryear{Ben-David, Luxburg, and P\'al}{Ben-David
  et~al.}{2006}]{Ben-David:EtAl:06}
Ben-David, S., U.~V. Luxburg, and D.~P\'al (2006).
\newblock A sober look at clustering stability.
\newblock In {\em COLT}, pp.\  5--19. Springer.

\bibitem[\protect\citeauthoryear{Bengio, Delalleau, LeRoux, Paiement, Vincent,
  and Ouimet}{Bengio et~al.}{2004}]{Bengio:EtAl:2004}
Bengio, Y., O.~Delalleau, N.~LeRoux, J.-F. Paiement, P.~Vincent, and M.~Ouimet
  (2004).
\newblock Learning eigenfunctions links spectral embedding and kernel {PCA}.
\newblock {\em Neural Comput.\/}~{\em 16\/}(10), 2197--2219.

\bibitem[\protect\citeauthoryear{Bernstein, {d}e Silva, Langford, and
  Tenenbaum}{Bernstein et~al.}{2000}]{Bernstein:EtAl:00}
Bernstein, M., V.~{d}e Silva, J.~C. Langford, and J.~B. Tenenbaum (2000).
\newblock Graph approximations to geodesics on embedded manifolds.
\newblock Technical report, Department of Mathematics, Stanford University.

\bibitem[\protect\citeauthoryear{Bickel and Levina}{Bickel and
  Levina}{2004}]{Bickel:Levina:04}
Bickel, P.~J. and E.~Levina (2004).
\newblock Maximum likelihood estimation of instrinsic dimension.
\newblock {\em NIPS\/}.

\bibitem[\protect\citeauthoryear{Bousquet, Chapelle, and Hein}{Bousquet
  et~al.}{2003}]{BousquetCH03}
Bousquet, O., O.~Chapelle, and M.~Hein (2003).
\newblock Measure based regularization.
\newblock In {\em NIPS}.

\bibitem[\protect\citeauthoryear{Buchman, Lee, and Schafer}{Buchman
  et~al.}{2008}]{Buchman:EtAl:2008}
Buchman, S., A.~B. Lee, and C.~M. Schafer (2008).
\newblock Density estimation of hurricane trajectories in the {A}tlantic
  {O}cean by spectral connectivity analysis.
\newblock In preparation.

\bibitem[\protect\citeauthoryear{Coifman and Lafon}{Coifman and
  Lafon}{2006}]{Coifman:Lafon:06}
Coifman, R. and S.~Lafon (2006).
\newblock Diffusion maps.
\newblock {\em Applied and Computational Harmonic Analysis\/}~{\em 21}, 5--30.

\bibitem[\protect\citeauthoryear{Coifman, Lafon, Lee, Maggioni, Nadler, Warner,
  and Zucker}{Coifman et~al.}{2005a}]{PNAS1}
Coifman, R., S.~Lafon, A.~Lee, M.~Maggioni, B.~Nadler, F.~Warner, and S.~Zucker
  (2005a).
\newblock Geometric diffusions as a tool for harmonics analysis and structure
  definition of data: Diffusion maps.
\newblock {\em Proceedings of the National Academy of Sciences\/}~{\em
  102\/}(21), 7426--7431.

\bibitem[\protect\citeauthoryear{Coifman, Lafon, Lee, Maggioni, Nadler, Warner,
  and Zucker}{Coifman et~al.}{2005b}]{PNAS2}
Coifman, R., S.~Lafon, A.~Lee, M.~Maggioni, B.~Nadler, F.~Warner, and S.~Zucker
  (2005b).
\newblock Geometric diffusions as a tool for harmonics analysis and structure
  definition of data: Multiscale methods.
\newblock {\em Proceedings of the National Academy of Sciences\/}~{\em
  102\/}(21), 7432--7437.

\bibitem[\protect\citeauthoryear{Coifman and Maggioni}{Coifman and
  Maggioni}{2006}]{CoifmanMauro05}
Coifman, R. and M.~Maggioni (2006).
\newblock Diffusion wavelets.
\newblock {\em Applied and Computational Harmonic Analysis\/}~{\em 21}, 53--94.

\bibitem[\protect\citeauthoryear{Donoho and Grimes}{Donoho and
  Grimes}{2003}]{DonohoGrimes03}
Donoho, D. and C.~Grimes (2003, May).
\newblock Hessian eigenmaps: new locally linear embedding techniques for
  high-dimensional data.
\newblock {\em Proceedings of the National Academy of Sciences\/}~{\em
  100\/}(10), 5591--5596.

\bibitem[\protect\citeauthoryear{Fan}{Fan}{1993}]{Fan:1993}
Fan, J. (1993).
\newblock Local linear regression smoothers and their minimax efficiencies.
\newblock {\em The Annals of Statistics\/}~{\em 21}, 196--216.

\bibitem[\protect\citeauthoryear{Fouss, Pirotte, and Saerens}{Fouss
  et~al.}{2005}]{FoussEtAl:05}
Fouss, F., A.~Pirotte, and M.~Saerens (2005).
\newblock A novel way of computing similarities between nodes of a graph, with
  application to collaborative recommendation.
\newblock In {\em Proc. of the 2005 IEEE/WIC/ACM International Joint Conference
  on Web Intelligence}, pp.\  550--556.

\bibitem[\protect\citeauthoryear{Gin\'e and Guillou}{Gin\'e and
  Guillou}{2002}]{Gine:Guillou:02}
Gin\'e, E. and A.~Guillou (2002).
\newblock Rates of strong uniform consistency for multivariate kernel density
  estimators.
\newblock {\em Ann Inst. H. Poincar\/}~{\em 38}, 907--921.

\bibitem[\protect\citeauthoryear{Gin\'e and Koltchinskii}{Gin\'e and
  Koltchinskii}{2006}]{Gine:Koltchinskii:2006}
Gin\'e, E. and V.~Koltchinskii (2006).
\newblock Empirical graph {L}aplacian approximation of {L}aplace-{B}eltrami
  operators: Large sample results.
\newblock In {\em High Dimensional Probability: Proceedings of the Fourth
  International Conference}, IMS Lecture Notes, pp.\  1--22.

\bibitem[\protect\citeauthoryear{Grigor'yan}{Grigor'yan}{2006}]{Grigoryan:06}
Grigor'yan, A. (2006).
\newblock Heat kernels on weighted manifolds and applications.
\newblock {\em Cont. Math.\/}~{\em 398}, 93--191.

\bibitem[\protect\citeauthoryear{Hastie and Stuetzle}{Hastie and
  Stuetzle}{1989}]{Hastie:Stuetzle:1989}
Hastie, T. and W.~Stuetzle (1989).
\newblock Principal curves.
\newblock {\em Journal of the American Statistical Association\/}~{\em 84},
  502--516.

\bibitem[\protect\citeauthoryear{Hein, Audibert, and {v}on Luxburg}{Hein
  et~al.}{2005}]{Hein:EtAl:05}
Hein, M., J.-Y. Audibert, and U.~{v}on Luxburg (2005).
\newblock From graphs to manifolds --- weak and strong pointwise consistency of
  graph {L}aplacians.
\newblock In {\em Proc. COLT}.

\bibitem[\protect\citeauthoryear{Kambhatla and Leen}{Kambhatla and
  Leen}{1997}]{Kambhatla:Leen:1997}
Kambhatla, N. and T.~K. Leen (1997).
\newblock Dimension reduction by local principal component analysis.
\newblock {\em Neural Computation\/}~{\em 9}, 1493--1516.

\bibitem[\protect\citeauthoryear{Kohler and Krzyzak}{Kohler and
  Krzyzak}{2007}]{Kohler:Krzyzak:07}
Kohler, M. and A.~Krzyzak (2007).
\newblock On the rate of convergence of local averaging plug-in classification
  rules under a margin condition.
\newblock {\em IEEE Transactions on Information Theory\/}~{\em 53}, 1735--1742.

\bibitem[\protect\citeauthoryear{Lafferty and Wasserman}{Lafferty and
  Wasserman}{2007}]{Lafferty:Wasserman:07}
Lafferty, J. and L.~Wasserman (2007).
\newblock Statistical analysis of semi-supervised regression.
\newblock In {\em NIPS}.

\bibitem[\protect\citeauthoryear{Lafon}{Lafon}{2004}]{Lafon:2004}
Lafon, S. (2004).
\newblock {\em Diffusion Maps and Geometric Harmonics}.
\newblock Ph.\ D. thesis, Yale University.

\bibitem[\protect\citeauthoryear{Lafon and Lee}{Lafon and
  Lee}{2006}]{LafonLee2006}
Lafon, S. and A.~Lee (2006).
\newblock Diffusion maps and coarse-graining: A unified framework for
  dimensionality reduction, graph partitioning, and data set parameterization.
\newblock {\em IEEE Trans. Pattern Anal. and Mach. Intel.\/}~{\em 28},
  1393--1403.

\bibitem[\protect\citeauthoryear{Lange, Roth, Braun, and Buhmann}{Lange
  et~al.}{2004}]{Lange:EtAl:04}
Lange, T., V.~Roth, M.~L. Braun, and J.~M. Buhmann (2004).
\newblock Stability-based validation of clustering solutions.
\newblock {\em Neural Computation\/}~{\em 16\/}(6), 1299--1323.

\bibitem[\protect\citeauthoryear{Lasota and Mackey}{Lasota and
  Mackey}{1994}]{Lasota:Mackey:94}
Lasota, A. and M.~C. Mackey (1994).
\newblock {\em Chaos, Fractals, and Noise: Stochastic Aspects of Dynamics\/}
  (Second ed.).
\newblock Springer.

\bibitem[\protect\citeauthoryear{Lee and Wasserman}{Lee and
  Wasserman}{2008}]{Lee:Wasserman:2008}
Lee, A.~B. and L.~Wasserman (2008).
\newblock Data quantization and density estimation via spectral connectivity
  analysis.
\newblock In preparation.

\bibitem[\protect\citeauthoryear{Mammen and Tsybakov}{Mammen and
  Tsybakov}{1999}]{Mammen99smoothdiscrimination}
Mammen, E. and A.~B. Tsybakov (1999).
\newblock Smooth discrimination analysis.
\newblock {\em Ann. Statist\/}~{\em 27}, 1808--1829.

\bibitem[\protect\citeauthoryear{Mardia, Kent, and Bibby}{Mardia
  et~al.}{1980}]{Mardia:1980}
Mardia, K.~V., J.~T. Kent, and J.~M. Bibby (1980).
\newblock {\em Multivariate Analysis}.
\newblock Academic Press.

\bibitem[\protect\citeauthoryear{Meinicke and Ritter}{Meinicke and
  Ritter}{2002}]{meinrit:2001}
Meinicke, P. and H.~Ritter (2002).
\newblock Quantizing density estimators.
\newblock {\em Advances in Neural Information Processing Systems\/}~{\em 14},
  825--832.

\bibitem[\protect\citeauthoryear{Page, Brin, Motwani, and Winograd}{Page
  et~al.}{1998}]{Page:EtAl:98}
Page, L., S.~Brin, R.~Motwani, and T.~Winograd (1998).
\newblock The pagerank citation ranking: Bringing order to the web.
\newblock Technical report, Stanford University.

\bibitem[\protect\citeauthoryear{Richards, Freeman, Lee, and Schafer}{Richards
  et~al.}{2009}]{Richards:EtAl:2008}
Richards, J.~W., P.~E. Freeman, A.~B. Lee, and C.~M. Schafer (2009).
\newblock Exploiting low-dimensional structure in astronomical spectra.
\newblock {\em Astrophysical Journal\/}.
\newblock To appear.

\bibitem[\protect\citeauthoryear{Roweis and Saul}{Roweis and
  Saul}{2000}]{RoweisSaul00}
Roweis, S. and L.~Saul (2000).
\newblock Nonlinear dimensionality reduction by locally linear embedding.
\newblock {\em Science\/}~{\em 290}, 2323--2326.

\bibitem[\protect\citeauthoryear{Sch\"olkopf, Smola, and M\"uller}{Sch\"olkopf
  et~al.}{1998}]{Scholkopf98}
Sch\"olkopf, B., A.~Smola, and K.-R. M\"uller (1998).
\newblock Nonlinear component analysis as a kernel eigenvalue problem.
\newblock {\em Neural Computation\/}~{\em 10\/}(5), 1299--1319.

\bibitem[\protect\citeauthoryear{Singer}{Singer}{2006}]{Singer:06}
Singer, A. (2006).
\newblock From graph to manifold {L}aplacian: The convergence rate.
\newblock {\em Applied and Computational Harmonic Analysis\/}~{\em 21},
  128--134.

\bibitem[\protect\citeauthoryear{Stewart}{Stewart}{1991}]{Stewart:1991}
Stewart, G. (1991).
\newblock Perturbation theory for the singular value decomposition.
\newblock {\em Svd and Signal Processing, II\/}.

\bibitem[\protect\citeauthoryear{Szummer and Jaakkola}{Szummer and
  Jaakkola}{2001}]{SzummerJaakkola01}
Szummer, M. and T.~Jaakkola (2001).
\newblock Partially labeled classification with markov random walks.
\newblock In {\em Advances in Neural Information Processing Systems},
  Volume~14.

\bibitem[\protect\citeauthoryear{Tenenbaum, {d}e Silva, and Langford}{Tenenbaum
  et~al.}{2000}]{Tenenbaum:2000}
Tenenbaum, J.~B., V.~{d}e Silva, and J.~C. Langford (2000).
\newblock {A Global Geometric Framework for Nonlinear Dimensionality
  Reduction}.
\newblock {\em Science\/}~{\em 290\/}(5500), 2319--2323.

\bibitem[\protect\citeauthoryear{{v}on Luxburg}{{v}on
  Luxburg}{2007}]{Luxburg:2007}
{v}on Luxburg, U. (2007).
\newblock A tutorial on spectral clustering.
\newblock {\em Statistics and Computing\/}~{\em 17\/}(4), 395--416.

\bibitem[\protect\citeauthoryear{von Luxburg, Belkin, and Bousquet}{von Luxburg
  et~al.}{2008}]{Luxburg:EtAl:2008}
von Luxburg, U., M.~Belkin, and O.~Bousquet (2008).
\newblock Consistency of spectral clustering.
\newblock {\em Annals of Statistics\/}~{\em 36\/}(2), 555--586.

\bibitem[\protect\citeauthoryear{Zwald and Blanchard}{Zwald and
  Blanchard}{2006}]{Zwald:Blanchard:2006}
Zwald, L. and G.~Blanchard (2006).
\newblock On the convergence of eigenspaces in kernel principal component
  analysis.
\newblock In {\em Advances in Neural Inf. Proc. Systems (NIPS 05)}, Volume~18,
  pp.\  1649--1656. MIT Press.

\end{thebibliography}

\end{document}